\documentclass[aps,prd,onecolumn,nofootinbib,superscriptaddress]{revtex4}
\usepackage{float}
\usepackage{graphicx}
\usepackage{amsmath}
\usepackage{amsfonts}
\usepackage{amssymb,ulem}
\usepackage{color}%
\usepackage{dcolumn}
\usepackage{subfigure}

\usepackage{MnSymbol,wasysym}
\usepackage{braket,diagbox}
\usepackage{eurosym}
\usepackage{calrsfs}
\usepackage[usenames,dvipsnames,svgnames]{xcolor}

\newcommand{\RNum}[1]{\uppercase\expandafter{\romannumeral #1\relax}}
\usepackage[colorlinks=true,linkcolor=blue,urlcolor=blue,filecolor=black,citecolor=red,
pdfstartview=FitV,pdftitle={},pdfsubject={},pdfkeywords={},pdfpagemode=None,bookmarksopen=true]{hyperref}

\begin{document}
\baselineskip=0.4 cm

\title{Precession and Lense-Thirring effect of hairy Kerr spacetimes}
\author{Meng-He Wu}
\email{mhwu@sues.edu.cn}
\affiliation{School of Mathematics, Physics and Statistics, Shanghai University of Engineering Science, Shanghai 201620, China}
\affiliation{Center of Application and Research of Computational Physics, Shanghai University of Engineering Science, Shanghai 201620, China}

\author{Hong Guo}
\email{gravhguo@gmail.com}
\affiliation{School of Physics and Astronomy, Shanghai Jiao Tong University, Shanghai 200240, China}

\author{Xiao-Mei Kuang}
\email{xmeikuang@yzu.edu.cn (corresponding author)}
\affiliation{Center for Gravitation and Cosmology, College of Physical Science and Technology, Yangzhou University, Yangzhou, 225009, China}

\date{\today}

\begin{abstract}
\baselineskip=0.4 cm
We investigate the frame-dragging effect of the hairy Kerr spacetimes  on the spin of a test gyro and accretion disk. Firstly, we analyze Lense-Thirring (LT) precession frequency, geodetic precession frequency, and the general spin precession frequency of a test gyro attached to a stationary observer in the spacetime. We find that the black hole hair suppresses those precession frequencies in comparison  with that occurs in Kerr spacetime in general relativity.  Moreover, using those frequencies as probe,  we differentiate the hairy Kerr black hole (BH) from naked singularity (NS).  Specifically, as the observer approaches the central source along any direction,  the frequencies grow sharply  for the hairy Kerr BH,  while for the hairy NS they
are finite except at the ring singularity.  Then,  we investigate  the quasiperiodic oscillations (QPOs) phenomena as the accretion disk approaches the hairy Kerr BH or NS. To this end, we analyze  the bound circular orbits and their perturbations. We find that as the orbits approach the corresponding  inner-most stable circular orbit (ISCO),  both LT precession frequency and periastron precession frequency behave differently in the hairy Kerr BH and NS. Additionally,  the hairy parameters have significant effects on the two frequencies.  We expect that our theoretical studies could shed light on astrophysical observations in distinguishing hairy theories from Einstein's gravity, and also in distinguishing  BH  from NS in spacetime with hair.

\end{abstract}


\maketitle

\tableofcontents

\newpage
\section{Introduction}
Einstein's general relativity (GR) is a successful theory in modern physics and passes plenty of test in astrophysics as well as astronomy. The existence of black holes is a prediction of GR, and black holes provide natural laboratories to test  gravity in the strong field regime.  Recent observations on gravitational waves \cite{LIGOScientific:2016aoc,LIGOScientific:2018mvr,LIGOScientific:2020aai} and the supermassive black hole images \cite{EventHorizonTelescope:2019dse,EventHorizonTelescope:2019ths,EventHorizonTelescope:2019pgp,EventHorizonTelescope:2022xnr,EventHorizonTelescope:2022xqj} agree well with the predictions of Kerr black hole described by the Kerr metric from GR.
More  observations, such as those from  Next Generation Very Large Array \cite{2019clrp.2020...32D} and Thirty Meter Telescope \cite{TMT:2015pvw}, also provide significant properties in the regime of strong gravity of  black hole spacetimes. In particular, these  observations open a valuable window to explore, distinguish, or constrain physically viable black hole solutions that exhibit small deviations from the Kerr metric. On the other hand, due to the additional
surrounding sources, the black holes in our Universe could obtain an extra global charge dubbed `hair'
and the spacetime may deviate from the Kerr metric.
Recently, a rotating hairy Kerr black hole was constructed with the use of  the gravitational
decoupling (GD) approach \cite{Ovalle:2020kpd,Contreras:2021yxe}, which is
designed for describing deformations of known solutions
of GR due to the inclusion of additional sources. The  hairy Kerr black hole  attracts lots of attentions.  Plenty of  theoretical and observational investigations have been done in this hairy Kerr black hole spacetime, for examples,  thermodynamics \cite{Mahapatra:2022xea}, quasinormal modes and (in)stability \cite{Cavalcanti:2022cga,Yang:2022ifo,Li:2022hkq},  strong gravitational lensing and parameter constraint from Event Horizon Telescope observations \cite{Islam:2021dyk,Afrin:2021imp}.

There are many theoretical scenarios in which a
visible singularity exists, especially after it was found in \cite{Joshi:2011zm} that a spacetime with a central naked singularity can be formed as an equilibrium
end state of the gravitational collapse of general matter cloud. However,  whether naked singularity (NS) really exists in nature and what is its physical signature distinguished from black hole (BH) are still open questions.   In Mathematics, the Kerr spacetime metric in GR, as a solution of Einstein field equations, describes Kerr BH and the Kerr  singularity is contained in the event horizon, otherwise, if the event horizon disappears, the metric describes a NS. In shadow scenario,  the shadow cast by NS spacetime was found to have similar size to that an equally massive Schwarzschild
black hole can cast \cite{Shaikh:2018lcc}. More elaborate discussions on shadow from NS have been carried out  in \cite{Joshi:2020tlq,Dey:2020haf, Dey:2020bgo},  which are important theoretical studies and match  recent observations on the shadow of M87* and  SgrA*. In addition, the orbital dynamics of particles or stars around a horizonless ultra-compact object could be another important physical signature which could be distinguishable  from that around  black holes, since the nature of timelike geodesics in a spacetime closely depends upon the geometrical essence of the spacetime. In this scenario, the timelike
geodesics around different types of compact objects and the  trajectory of massive particles or stars
are widely studied \cite{Pugliese:2011xn,Martinez:2019nor,Hackmann:2014tga,Potashov:2019kxq,Bhattacharya:2017chr,Bambhaniya:2019pbr,Deng:2020hxw,Lin:2021noq,Bambhaniya:2021jum,Ota:2021mub} and reference therein. People expect that the nature of the precession of
the timelike orbits  can provide more  information about the central compact object, as
GRAVITY \cite{GRAVITY:2020lpa} and SINFONI \cite{Eisenhauer:2005cv} are continuously tracking the dynamics
of  `S' stars  orbiting around SgrA*.

For the  timelike orbits around a rotating compact body, the geodetic precession \cite{deSitter:1916zz} and  Lense-Thirring (LT) precession \cite{Lense:1918LT} are two extraordinary effects
predicted by GR.  The former effect is also known as  de Sitter precessions and it is due to the spacetime curvature of the central body, while the latter effect is due to the rotation of the central body which causes the dragging of locally inertial frames. One can examine  the dragging effects by considering a test gyro based on the fact that a gyroscope tends to keep its spinning axis rigidly pointed in a fixed direction relative to a reference star. The LT precession and geodetic effects have been measured in the Earth’s gravitational field by the Gravity Probe B experiment, in which the satellite consists of four gyroscopes and a telescope orbiting $642 km$  above the Earth \cite{Everitt:2011hp}. The geodetic precession in Schwarzschild BH and the Kerr BH have been studied in \cite{sakina1979parallel}. The LT precession is more complex and usually some approximations should be involved.
In the weak field approximations, 
the magnitude of LT precession frequency is proportional to the spin parameter of the central body and decreases in the order of $r^{-3}$ with $r$ the distance between the test gyro from the central rotating body \cite{2009Hartle}. In the strong gravity limit, the LT precession frequency were studied in various rotating compact objects, for instances,
in Kerr black hole \cite{Chakraborty:2013naa,Bini:2016iym} and its generalizations \cite{Chakraborty:2012wv}, in rotating traversable wormhole \cite{Chakraborty:2016oja} and in rotating neutron star \cite{Chakraborty:2014qba}. Those studies further show that the behavior of LT frequency in the strong gravity regime closely depends on the nature of the central rotating bodies. In particular, it was proposed in \cite{Chakraborty:2016mhx} that the spin precession of a test gyro  can be used to distinguish Kerr naked singularities from black holes, which was then extended in modified theories of gravity \cite{Rizwan:2018lht,Rizwan:2018rgs,Pradhan:2020nno,Solanki:2021mkt}.

Thus, one of the main aims of this paper is to investigate the spin precession frequency, including the LT frequency and geodetic frequency, of the test gyro in the hairy Kerr spacetime, and differentiate the hairy Kerr BH from hairy NS. We find a clear difference on the precession frequencies  between hairy rotating BH and NS, and the effects of the hairy parameters are also systematically explored.
 
 Additionally, we also investigate the accretion disk physics in the hairy Kerr spacetime,  which is realized by studying  the orbital precession of a test timelike particle around the hairy Kerr BH and NS. The motivation stems from the followings.  Accretion in Low-Mass X-ray Binaries (LMXBs) occurs  in the strong gravity regime around compact bodies, in which the  quasiperiodic oscillations (QPOs) phenomena involves and their frequencies in the $Hz$ to $kHz$ range have been detected \cite{Torok:2011qy,Bambi:2012ku,Tripathi:2019bya}. Plenty of efforts have been made to  explain the
QPOs phenomena, see  \cite{Klisin2006} for a review. It  shows that the geodesic models of QPOs are related with three characteristic
frequencies of the massive particles orbiting around the compact bodies, namely the orbital epicyclic frequency and the radial and vertical
epicyclic frequencies.  Therefore, since the QPOs provide a way to testify the strong gravity,  these three frequencies could be used  as a tool to study the crucial differences among the central compact objects \cite{Rizwan:2018rgs} or test/constrain alternative theories of gravity \cite{Aliev:2013jqz,Doneva:2014uma,Azreg-Ainou:2020bfl,Jusufi:2020odz,Ghasemi-Nodehi:2020oiz}.
In particular, more recently it was proposed in \cite{Motta:2017ijc}  that LT effect
may have connection with QPOs phenomena and perhaps be used to explain the QPOs of accretion disks around rotating black holes. Thus,
Along with using the  spin precession frequency and LT precession frequency of  the test gyro to indicate the differences between hairy Kerr BH and NS,  we further study the  three characteristic
frequencies of massive particles orbiting around the corresponding central rotating objects. It is found that the LT precession  and periastron precession  of massive particle behave crucial differences between hairy Kerr BH and NS,  as a support of the results from the test gyro in strong gravity regime of compact bodies.

Our paper is organized as follows. In section \ref{sec:SGL}, we briefly review the hairy Kerr spacetime derived from GD approach, and analyze the parameter spaces for the corresponding black hole and naked singularity. Then we derive the timelike geodesic equations in the spacetime. In section \ref{sec:test gyroscope}, we derive the spin precession frequency of a test gyro in the hairy Kerr spacetime. Then we compare the LT  frequency as well as the spin frequency between the hairy Kerr BH and NS cases. In section \ref{sec:test particle}, we  study the accretion disk physics by analyzing  inner-most stable circular orbit (ISCO), and LT precession  and periastron precession  in terms of the three characteristic
frequencies,  of a timelike particle around the hairy Kerr BH and NS, respectively. The last section is our conclusion and discussion.
Throughout the paper, we  use $G_{\rm N}=c=\hbar=1$  and all quantities are rescaled to be dimensionless by the parameter $M$.

\section{Hairy Kerr spacetime and the timelike geodesic equations}\label{sec:SGL}

In this section, firstly, we will show a brief review on the idea of gravitational decoupling (GD) approach and  the hairy spacetime constructed from GD approach by  Ovalle \cite{Ovalle:2020kpd}. Then we derive the timelike geodesic equations in the hairy Kerr spacetime. 

The no-hair theorem in classical general relativity states that black holes are only described by mass, electric charge and spin \cite{Israel:1967wq,Hawking:1971vc,Mazur:1982db}. But it is possible that the interaction between black hole spacetime and matters brings in other charge, such that the black hole could carry hairs. The physical effect of  these hairs can modify the spacetime of the background of black hole, namely hairy black holes may form. Recently, Ovalle et.al used the GD approach to obtain a spherically symmetric metric with hair \cite{Ovalle:2020kpd}. The hairy black hole in this scenario has great generality because  there is no certain matter fields in the GD approach, in which the corresponding Einstein equation is expressed by
\begin{equation}\label{eq-EE}
R_{\mu\nu}-\frac{1}{2}Rg_{\mu\nu}=8\pi\Tilde{T}_{\mu\nu}.
\end{equation}
Here $\Tilde{T}_{\mu\nu}$ is the total energy momentum tensor written as $\Tilde{T}_{\mu\nu}=T_{\mu\nu}+\vartheta_{\mu\nu}$ where $T_{\mu\nu}$ and $\vartheta_{\mu\nu}$ are  energy momentum tensor in GR and  energy momentum tensor introduced by matter fields or others, respectively. $\nabla_\mu \Tilde{T}_{\mu\nu}=0$ is satisfied because of the Bianchi indentity. It is direct to prove that when $\vartheta_{\mu\nu}=0$, the solution to \eqref{eq-EE} degenerates into Schwarzschild metric. The hairy solution  with proper treatment (strong energy condition) of $\vartheta_{\mu\nu}$ was constructed and the detailed algebra calculations are shown in \cite{Ovalle:2020kpd,Contreras:2021yxe}. Here we will not re-show their steps, but directly refer to  the formula of the hairy metric 
\begin{equation}\label{eq-static}
ds^2=-f(r)dt^2+\frac{dr^2}{f(r)}+r^2(d\theta^2+\sin^2\theta d\phi^2)
~~\mathrm{with}~~ f(r)=1-\frac{2M}{r}+\alpha e^{-r/(M-l_0/2)}.
\end{equation}
In this solution, $M$ is the black hole mass, $\alpha$ is the deformation parameter due to the introduction of surrounding matters and it describes the physics related with the strength of hairs, and $l_0=\alpha l$ with $l$ a parameter with length dimension corresponds to primary hair which should satisfy $l_0\leq 2M$ to guarantee the asymptotic flatness. The metric \eqref{eq-static} reproduces the Schwarzchild spacetime in the absence of the matters, i.e., $\alpha=0$. 

Later, considering that astrophysical black holes in our universe usually have rotation, the authors of \cite{Afrin:2021imp} induced the rotating counterpart of the static solution \eqref{eq-static}, which is stationary and axisymmetric, and in Boyer-Lindquist coordinates it reads as
\begin{eqnarray}\label{eq-metric}
ds^2=&&g_{tt} dt^2+g_{rr} dr^2+g_{\theta\theta} d\theta^2+g_{\phi\phi} d\phi^2+2g_{t\phi} dtd\phi\nonumber\\
=&&-\left(\frac{\bigtriangleup-a^2\sin^2\theta}{\Sigma}\right)dt^2+\sin^2\theta\left(\Sigma+a^2\sin^2\theta\left(2-\frac{\bigtriangleup-a^2\sin^2\theta}{\Sigma}\right)\right)
d\phi^2+\frac{\Sigma}{\bigtriangleup}dr^2+\Sigma d\theta^2\nonumber\\ &&-2a\sin^2\theta
\left(1-\frac{\bigtriangleup-a^2\sin^2\theta}{\Sigma}\right)dtd\phi
\end{eqnarray}
with
\begin{eqnarray}
\Sigma=r^2+a^2\cos^2\theta,~~~\bigtriangleup=r^2+a^2-2Mr+\alpha r^2 e^{-r/(M-\frac{l_0}{2})}.
\end{eqnarray}
It is noticed that the above metric is also proved to satisfy the equations of motion in the GD approach \cite{Ovalle:2020kpd}.
The metric describes certain deformation of the Kerr solution due to the introduction of additional material sources (such as dark energy or dark matter). In the metric, $a$ is the spin parameter and $M$ is the black hole mass parameter. Similar to those in static case, $\alpha$ is the deviation parameter from GR and $l_0$ is the primary hair which is required to be $l_0\leq 2M$ for asymptotic flatness. When $\alpha=0$, the metric reduces to standard Kerr metric in GR, namely no surrounding matters.

\begin{figure}[H]
{\centering
\includegraphics[scale=0.8]{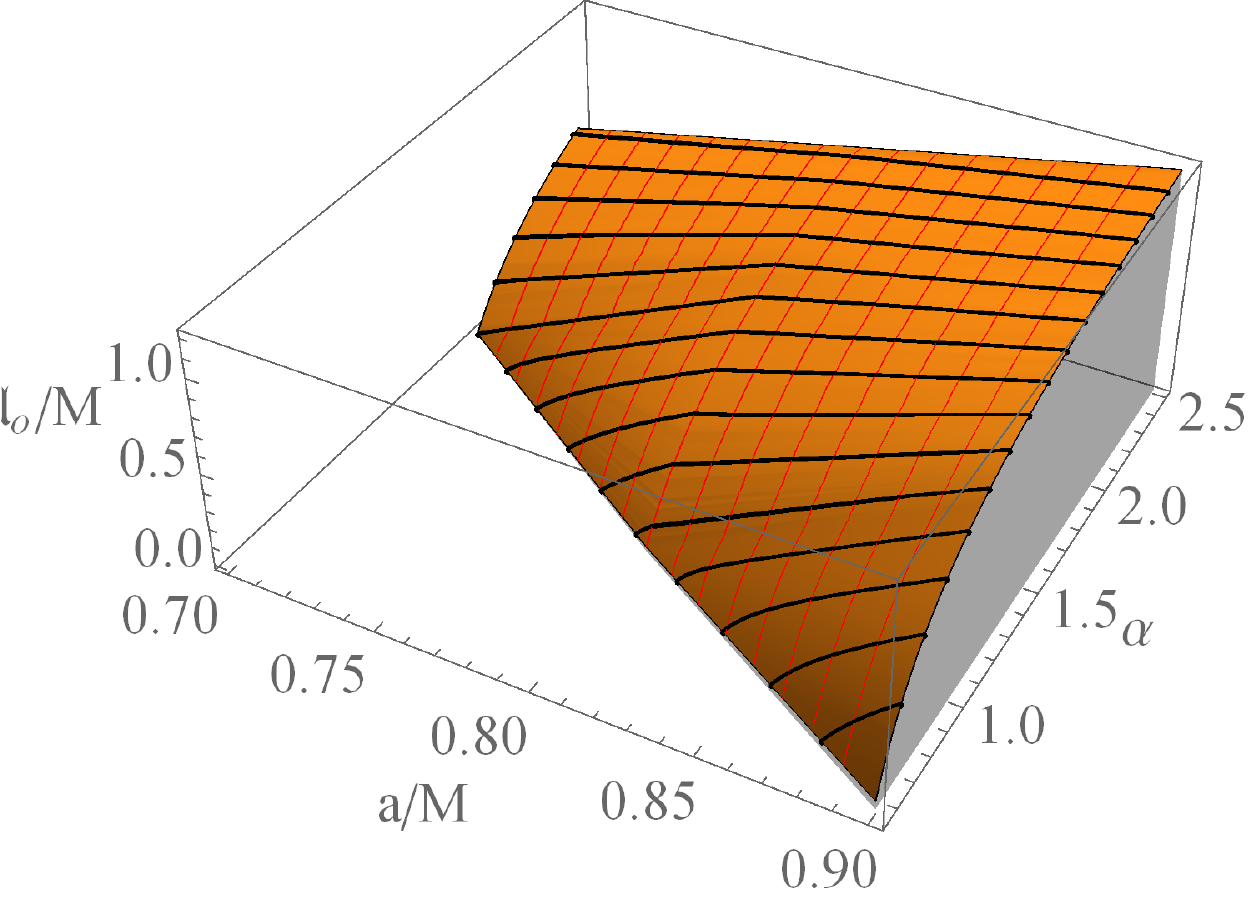}\hspace{1cm}
\caption{The parameters space $(l_0,a,\alpha)$ for non-extremal  hairy black hole (the white region), extremal black hole (the orange surface) and naked singularity (the shaded region).}\label{fig:geometry}	}	
\end{figure}

 The metric \eqref{eq-metric} could describe non-extremal hairy Kerr black hole, extremal hairy Kerr black hole and hairy naked singularity which correspond to two distinct, two equal and no  real positive roots of $g^{rr}=\bigtriangleup=0$ as well as $\Sigma\neq 0$,  meaning they have two horizons, single horizon and no horizon, respectively.  The parameter space $(l_0-a)$ of the related  geometries for some discrete $\alpha$ has been show in \cite{Afrin:2021imp}. Here in FIG. \ref{fig:geometry}, we show a 3D plot in the parameters space $(l_0,a,\alpha)$ for various geometries. In the figure, for the parameters in the white region, the spacetime with the metric \eqref{eq-metric} is a non-extremal hairy Kerr black hole with two horizons, while it is a naked singularity without horizon for the parameters in the shaded region; and for the parameters on the orange surface, it is an extremal hairy Kerr black hole with single horizon. Besides, how the hairy parameters affect the static limit surface and the ergoregion of the hairy Kerr black hole has also been explored in \cite{Afrin:2021imp}. Moreover, as we aforementioned, some theoretical and observational properties of the hairy Kerr black hole have been carried out, such as the thermodynamics \cite{Mahapatra:2022xea}, quasinormal modes and (in)stability \cite{Cavalcanti:2022cga,Yang:2022ifo,Li:2022hkq},  strong gravitational lensing and parameter constraint from Event Horizon Telescope observations \cite{Islam:2021dyk,Afrin:2021imp}.

In the following sections, we shall partly study the physical signatures in the strong field regime of those different central objects in the hairy Kerr spacetime  \eqref{eq-metric}. We will analyze the inertial frame dragging effects on a test gyro and on the accretion physics respectively, which is expected to show the difference between black hole and naked singularity. 
To proceed, we should analyze the timelike geodesic equations in the hairy Kerr spacetime. Since for the spacetime, we have two Killing vector fields $\partial_t$ and  $\partial_\phi$, so we can define the conserved energy $\mathcal{E}$ and axial-component of the angular momentum $L$ of which the expressions are 
\begin{equation}\label{eq-conservation}
\mathcal{E}=-g_{tt}\dot{t}-g_{t\phi}\dot{\phi},~~~~~~L=g_{t\phi}\dot{t} +g_{\phi\phi}\dot{\phi},
\end{equation}
where the dot represent the derivative with respect to the affine parameter $\lambda$. We employ the Hamilton-Jacobi method \cite{Carter:1968rr} and introduce the Hamilton-Jacobi equation for the particle with rest mass $\mu$ 
\begin{equation}
\mathcal{H}=-\frac{\partial S}{\partial\lambda}=\frac{1}{2}g_{\mu\nu}\frac{\partial S}{\partial x^\mu}\frac{\partial S}{\partial x^\nu}=\mu^2
\end{equation}
where $\mathcal{H}$ and $S$ are the canonical Hamiltonian and Jacobi action, respectively, and $x^\mu$ denote the coordinates $t,r,\theta,\phi$ in the metric  \eqref{eq-metric}. Then after variation, we find that we can separate the Jacobi action as $S=\frac{1}{2}\mu^2\lambda-\mathcal{E}t+L\phi+S_r(r)+S_{\theta}(\theta)$ and define the constant $\mathcal{C}$ via
\begin{equation}\label{eq-defK}
\left(\frac{dS_{\theta}}{d\theta}\right)^2+\frac{(L-a\mathcal{E}\sin^2\theta)^2}{\sin^2\theta}=
-\Delta\left(\frac{dS_{r}}{dr}\right)^2+\frac{\left((r^2+a^2)\mathcal{E}^2-aL\right)^2}{\Delta}=\mathcal{C}
\end{equation}
 to  separate the outcome into four first-order differential equations describing the geodesic motion
\begin{eqnarray}
\Sigma\dot{t}&=&a(L-a\mathcal{E}\sin^2\theta)+\frac{r^2+a^2}{\Delta}\left((r^2+a^2)\mathcal{E}-aL\right),\label{eq-motion3}\\
\Sigma\dot{\phi}&=&\frac{L}{\sin^2\vartheta}-a\mathcal{E}+\frac{a}{\Delta}\left((r^2+a^2)\mathcal{E}-aL\right),\label{Requation}\\
\Sigma\dot{r}&=&\Delta\left(\frac{dS_r}{dr}\right)=\sqrt{\left(r^2+a^2)\mathcal{E}-aL\right)^2-\Delta\left(Q+(L-a\mathcal{E})^2+\mu^2r^2\right)},\\
\Sigma\dot{\theta}&=&\frac{dS_{\theta}}{d\theta}=\sqrt{Q+\cos^2\theta\biggl(a^2(\mathcal{E}^2-\mu^2)-\frac{L^2}{\sin^2\theta}\biggr)}, \label{eq-geodesic}
\end{eqnarray}
where $Q\equiv\mathcal{C}-(L-a\mathcal{E})^2$ with $\mathcal{C}$ defined in \eqref{eq-defK} is the Carter constant. Thus, the geodesics equations in the hairy Kerr spacetime \eqref{eq-metric} are separable as in Kerr case, and the modification from the hair appears in the metric function $\Delta$. Now we can proceed to investigate the influence of hair on the orbital precession. For convenience, we will assume that the gyros/stars/particles we will consider are minimally coupled to the metric, and also there is no direct coupling between them and the surrounding matters, i.e., the modified gravity effect or print of hair would only be reflected in the metric functions.

\section{Spin precession of test gyroscope in hairy Kerr spacetime}\label{sec:test gyroscope}
In this section, we will study the spin precession frequency of a test gyroscope  attached to a stationary observer, dubbed stationary gyroscopes for brevity in the hairy Kerr spacetime, for whom  the $r$ and $\theta$ coordinates are remaining fixed with respect to  infinity.  Such a stationary observer has a four-velocity  $u_{stationary}^\mu=u_{stationary}^t(1,0,0,\Omega)$, where $t$ is the time coordinate and $\Omega=d\phi/dt$ is the angular velocity of the observer. 

Let us consider stationary gyroscopes moving along Killing trajectory, whose spin undergoes Fermi-Walker transport along $u=(-K^2)^{-1/2}K$ with $K$ the timelike Killing vector field.
Since the hairy Kerr spacetime \eqref{eq-metric} has two Killing vectors which are  the time translation Killing vector $\partial_t$ and the azimuthal Killing vector $\partial_\phi$, so a more general Killing vector is  $K=\partial_t+\Omega\partial_\phi$. In this case, the general spin precession frequency of a test stationary gyroscope is the rescaled vorticity filed of the observer congruence, and its one-form  is given by \cite{2004graa.book}
\begin{equation}
\widetilde{\Omega}_p=\frac{1}{2K^2}*(\widetilde{K}\wedge d\widetilde{K})
\end{equation}
 where  $\widetilde{K}$ is the covector of $K$, the denotations $*$ and $\wedge$ represent the Hodge dual and wedge product, respectively. According to the  pedagogical steps of \cite{Chakraborty:2016mhx},  the corresponding vector of the covector $\widetilde{\Omega}_p$ is
\begin{equation}
\vec{\Omega}_{p}=\frac{\varepsilon_{ckl}{ }}{2\sqrt{-g}%
	\left( 1+2\Omega \frac{g_{0c}}{g_{00}}+\Omega ^{2}\frac{g_{cc}}{g_{00}}\right) }%
\left[ \left( g_{0c,k}-\frac{g_{0c}}{g_{00}}g_{00,k}\right) +\Omega \left(
g_{cc,k}-\frac{g_{cc}}{g_{00}}g_{00,k}\right) +\Omega ^{2}\left( \frac{g_{0c}%
}{g_{00}}g_{cc,k}-\frac{g_{cc}}{g_{00}}g_{0c,k}\right) \right]\partial_l,
\end{equation}
where $g$ is the determinant of the metric $g_{\mu\nu}  (\mu,\nu=0,1,2,3)$ and $\varepsilon _{ckl} (c,k,l=1,2,3)$ is the Levi-Civita symbol.  Then in the hairy Kerr spacetime \eqref{eq-metric} which is stationary and axisymmetric spacetime, the above expression can be reduced as
\begin{eqnarray}\label{eq:Omega_p}
\vec{\Omega}_{p}=&&\frac{\left(C_1\hat{r}  + C_2 \hat{\theta}\right)}{2\sqrt{-g}
	\left( 1+2\Omega \frac{g_{t\phi}}{g_{tt}}+\Omega ^{2}\frac{g_{\phi\phi}}{g_{tt}}\right) } ,
\end{eqnarray}
with
\begin{equation}\label{eq:Omega_HF}
\begin{aligned}
C_1=&-\sqrt{g_{r r}}\left[\left(g_{t \phi, \theta}-\frac{g_{t \phi}}{g_{t t}} g_{t t, \theta}\right)+\Omega\left(g_{\phi \phi, \theta}-\frac{g_{\phi \phi}}{g_{t t}} g_{t t, \theta}\right)+\Omega^2\left(\frac{g_{t \phi}}{g_{t t}} g_{\phi \phi, \theta}-\frac{g_{\phi \phi}}{g_{t t}} g_{t \phi, \theta}\right)\right], \\
C_2=&\sqrt{g_{\theta\theta}}\left[\left(g_{t\phi,r}-\frac{g_{t\phi}}{g_{tt}}g_{tt,r}\right)
+\Omega\left(g_{\phi\phi,r}-\frac{g_{\phi\phi}}{g_{tt}}g_{tt,r}\right)
+\Omega^2\left(\frac{g_{t\phi}}{g_{tt}}g_{\phi\phi,r}-\frac{g_{\phi\phi}}{g_{tt}}g_{t\phi,r}\right)
\right].
\end{aligned}
\end{equation}
 \begin{figure}[H]
 \resizebox{\linewidth}{!}{\begin{tabular}{cc}
\includegraphics[scale=0.3]{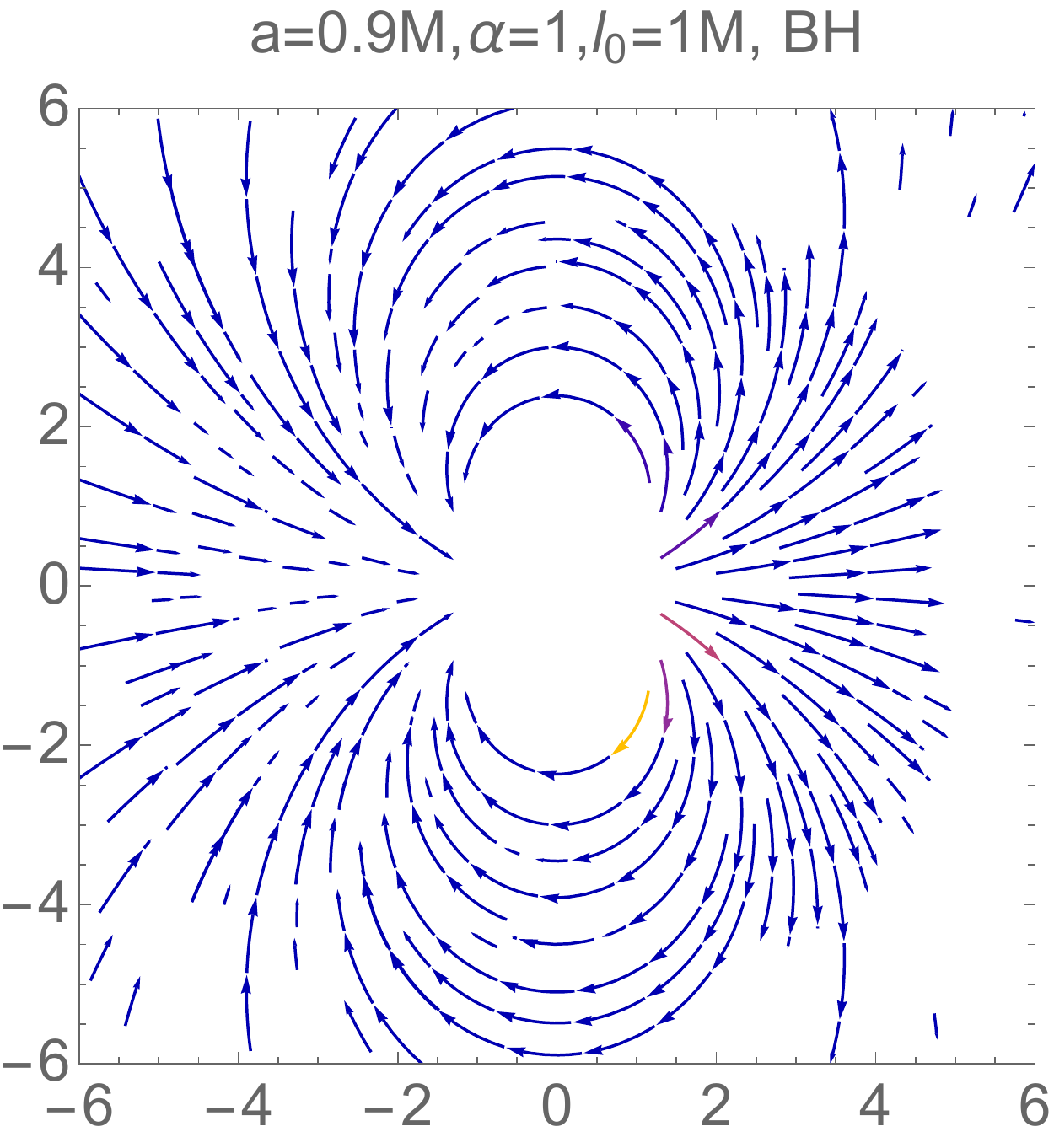}~~~~&
\includegraphics[scale=0.3]{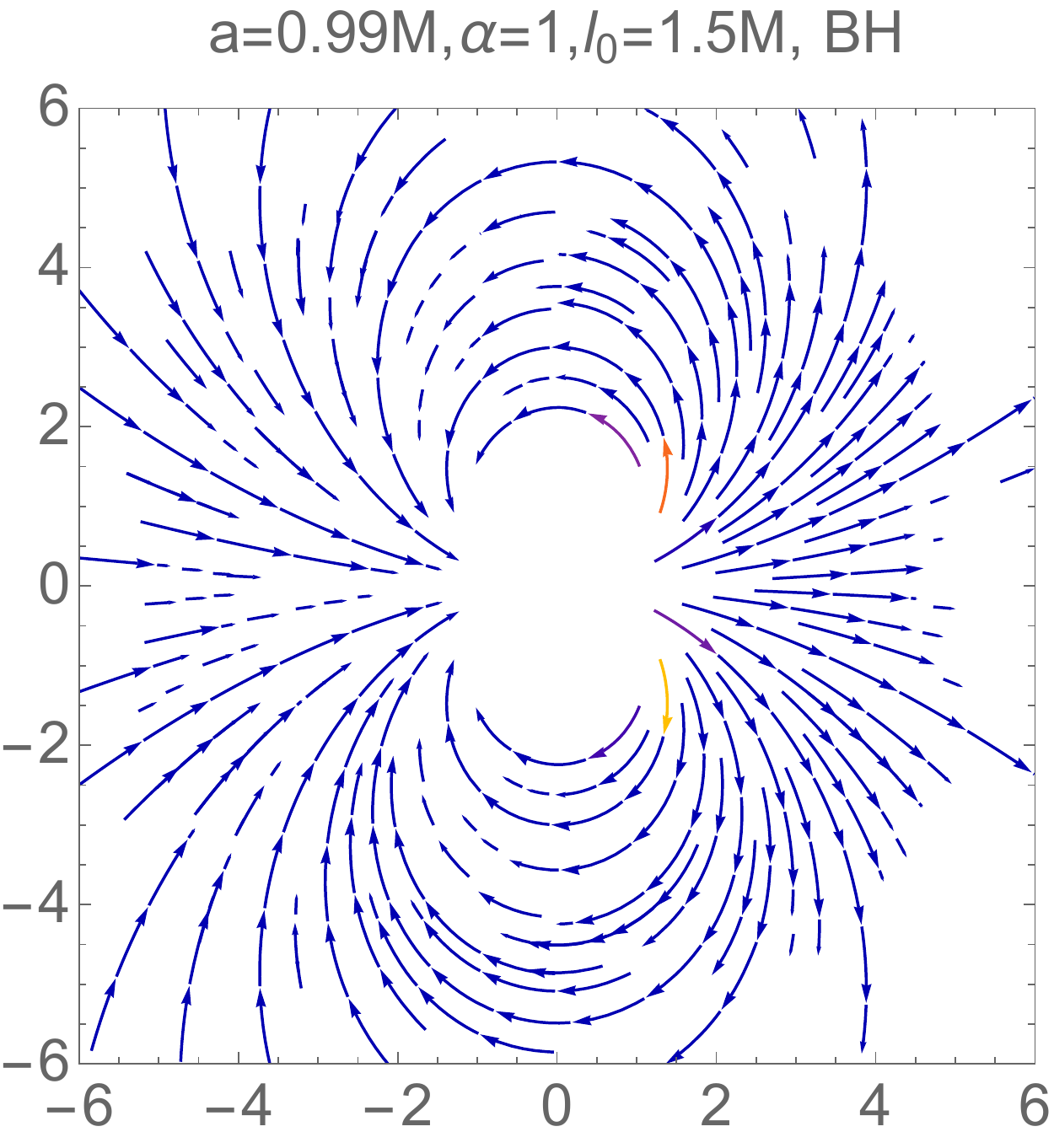}\\
\tiny (a) & \tiny (b) \\[6pt]
\\
\includegraphics[scale=0.3]{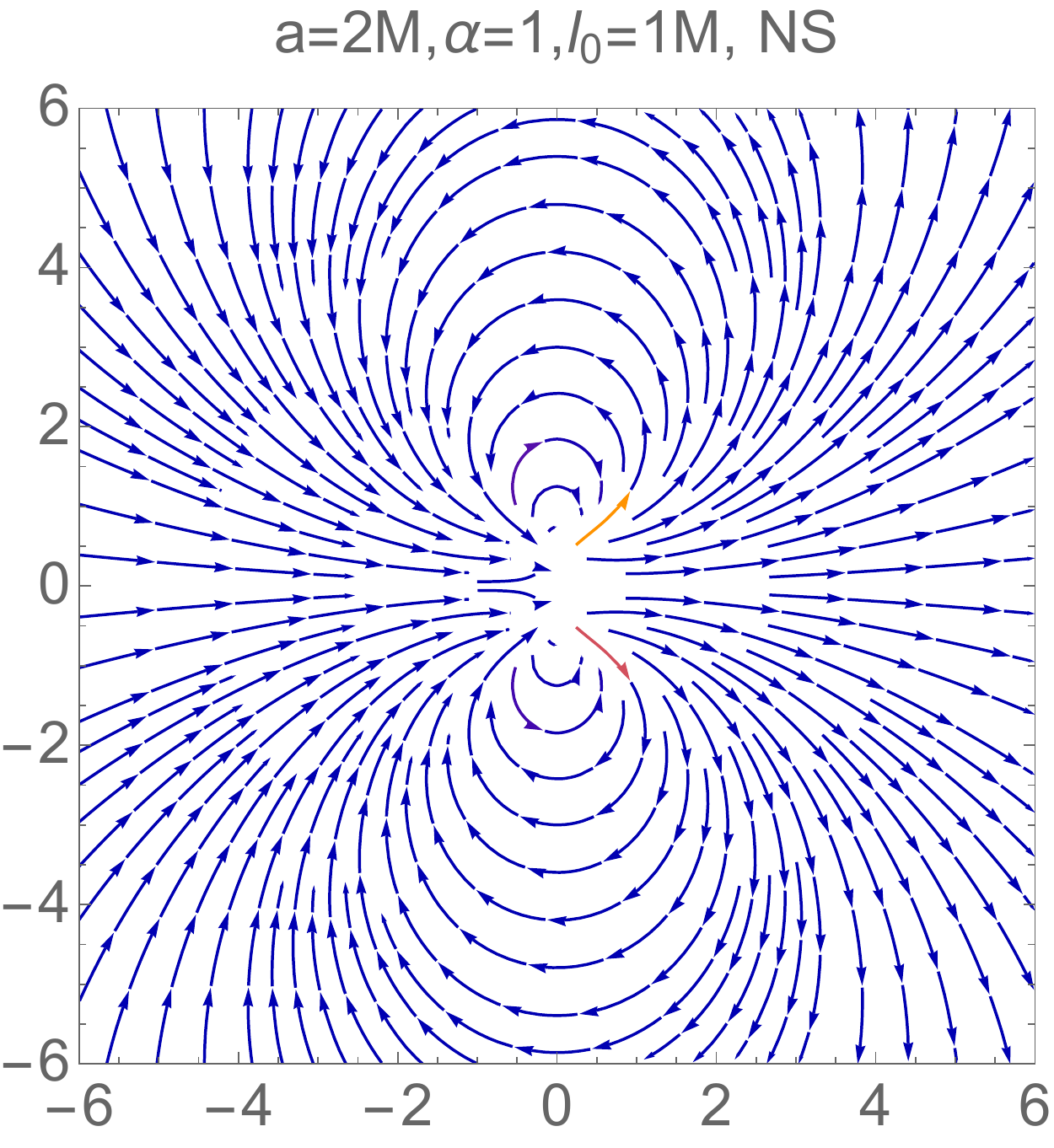}~~~~&
\includegraphics[scale=0.3]{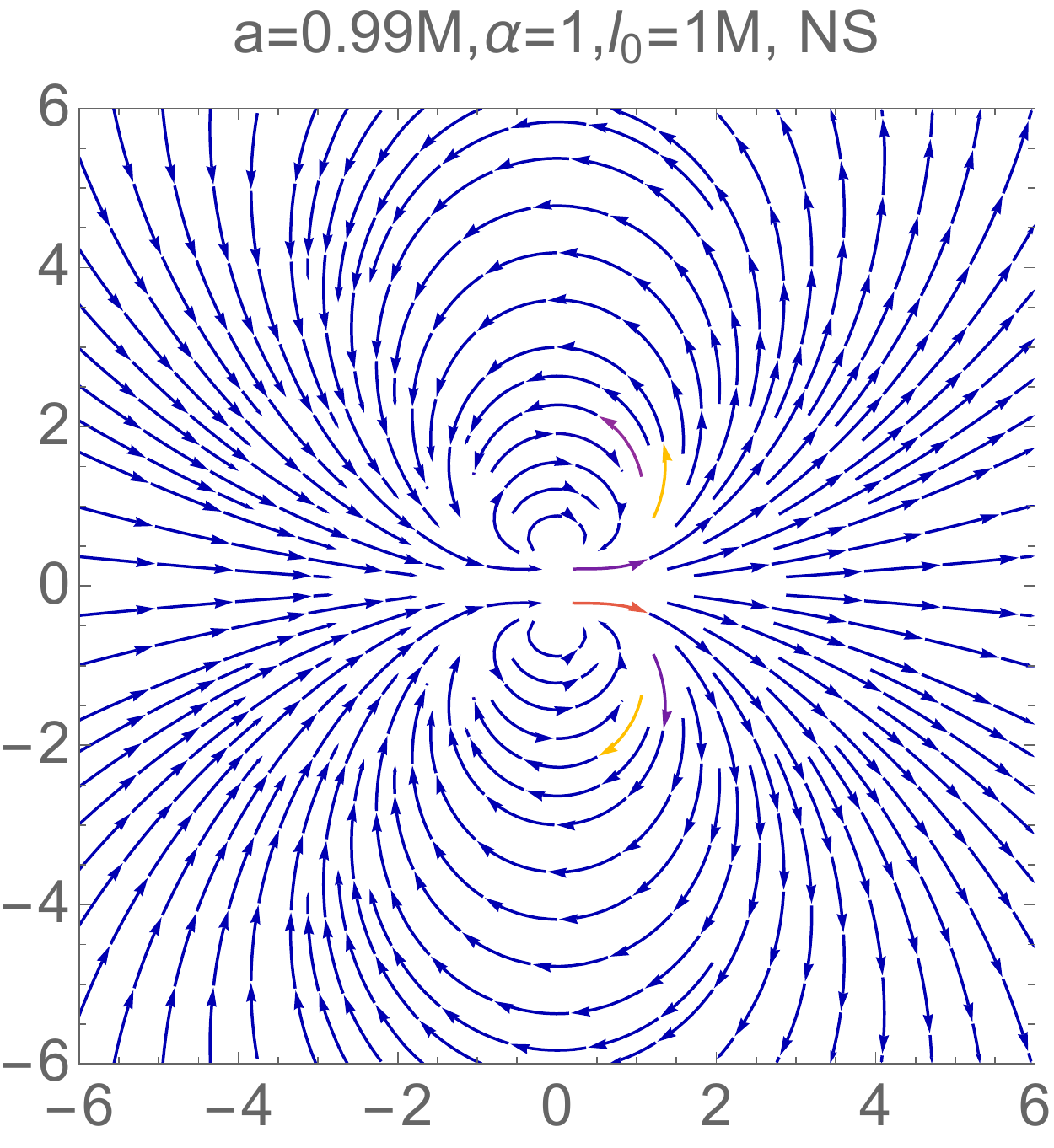}\\
\tiny (c) & \tiny (d) \\[6pt]
\end{tabular}}
\caption{The vector field of the LT precession frequency for the hairy Kerr spacetime. The plots in (a) with $a=0.9M,~\alpha=1,~l_0=1M$ and (b) with $a=0.99M,~\alpha=1,~l_0=1.5M$ are for hairy Kerr black hole , while in (c) with $a=2M,~\alpha=1,~l_0=1M$ and (d) with $a=0.99M,~\alpha=1,~l_0=1M$ are for hairy  naked singularity.}
\label{fig:Vector-OmegaLT}		
\end{figure}
It is worth to emphasize that  this expression is valid for observers both inside and outside of the ergosphere for a restricted range of $\Omega$
\begin{equation}\label{eq:OmegaRange}
\Omega _{-}(r,\theta )<\Omega (r,\theta )<\Omega _{+}(r,\theta ),
\end{equation}
with $\Omega _{\pm}=(-g_{t\phi}\pm\sqrt{g^2_{t\phi}-g_{\phi\phi}g_{tt}})/g_{\phi\phi}$,
which could ensure that the observer at fixed $r$ and $\theta$ is timelike.

Since the gyroscope has an angular velocity $\Omega$, the general precession frequency \eqref{eq:Omega_p} stems from two aspects: spacetime rotation (LT precession) and curvature (geodetic precession).  To disclose the properties of precession of gyroscope in hairy Kerr spacetime,  we will study the influence of hairy parameters on
the LT precession, geodetic precession and the general spin precession of the hairy Kerr black holes and naked singularities, respectively.

\subsection{Case with $\Omega=0$: LT precession frequency}
When the angular velocity vanishes, saying $\Omega=0$, it means that the gyroscope is attached to a static observer
in a stationary spacetime. Such static observers do not change their location with respect to infinity only outside the ergoregion. It corresponds to a four-velocity  $u_{static}^\mu=u_{static}^t(1,0,0,0)$ and the Killing vector $K=\partial_t$. Thus, $\vec{\Omega}_{p}$ reduces to the LT precession frequency, $\vec{\Omega}_{LT}$, of the gyroscope attached to a static observer outside the ergosphere \cite{Chakraborty:2013naa}. Subsequently,  the LT precession frequency in the hairy Kerr spacetime \eqref{eq-metric} takes the form
\begin{eqnarray}\label{eq:Omega_LT}
\vec{\Omega}_{LT}=&&\frac{1}{2\sqrt{-g} }\Big\{
-\sqrt{g_{rr}}\left(g_{t\phi,\theta}-\frac{g_{t\phi}}{g_{tt}}g_{tt,\theta}\right)
\hat{r}+
\sqrt{g_{\theta\theta}}\left(g_{t\phi,r}-\frac{g_{t\phi}}{g_{tt}}g_{tt,r}\right)
\hat{\theta}
\Big\}.
\end{eqnarray}

Samples of the above LT-precession frequency vector for hairy Kerr black holes and naked singularities are plotted in Cartesian plane shown in FIG.\ref{fig:Vector-OmegaLT}. The upper row exhibits the vector for hairy Kerr BH, giving that the  LT precession frequency is finite outside the ergoshpere and it diverges when the observer approach the ergosphere. In the bottom row for hairy NS, the  LT precession frequency is regular in the entire spacetime region except the ring singularity $r=0$ and $\theta=\pi/2$.

\begin{figure}[H]
 \resizebox{\linewidth}{!}{\begin{tabular}{cc}
\includegraphics[scale=0.6]{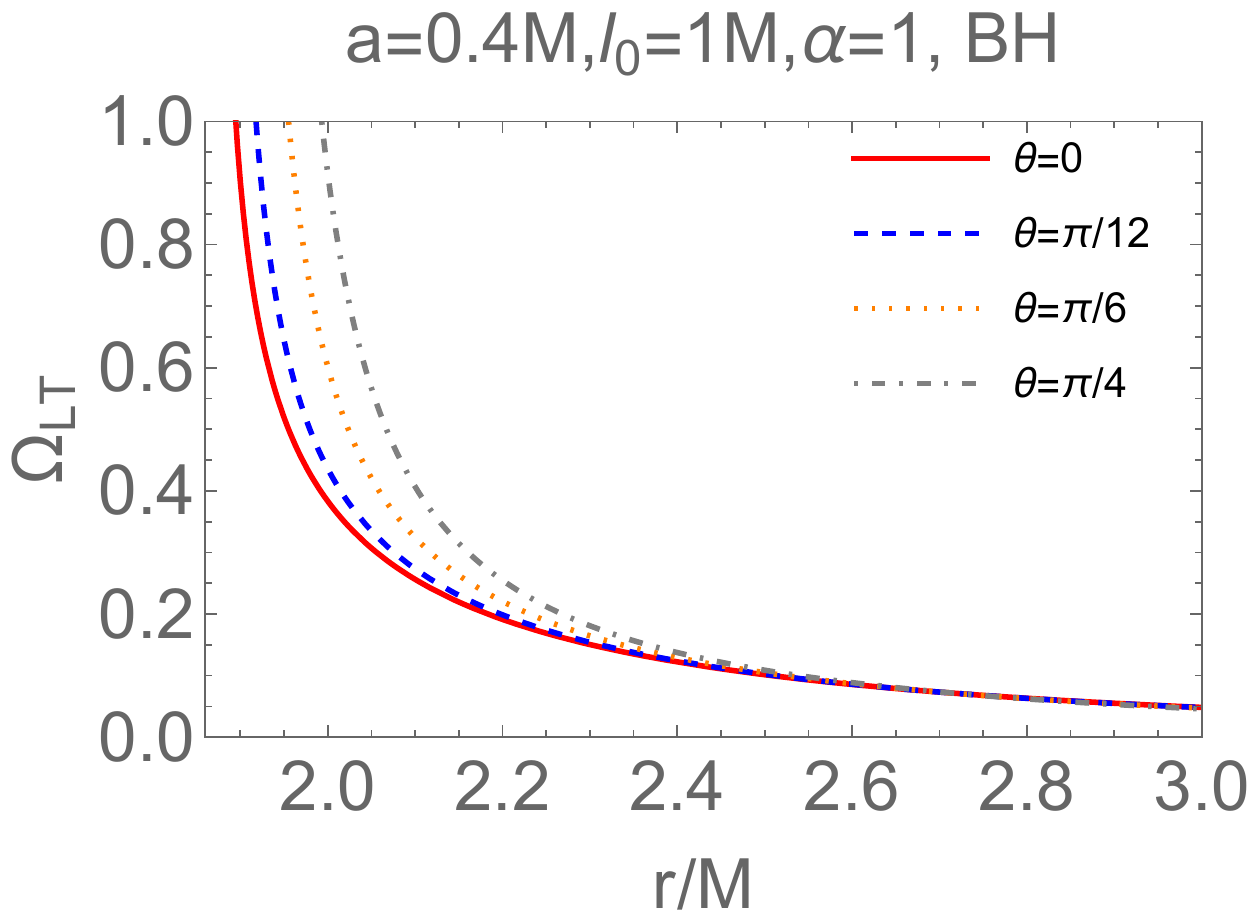}~~~~&
\includegraphics[scale=0.6]{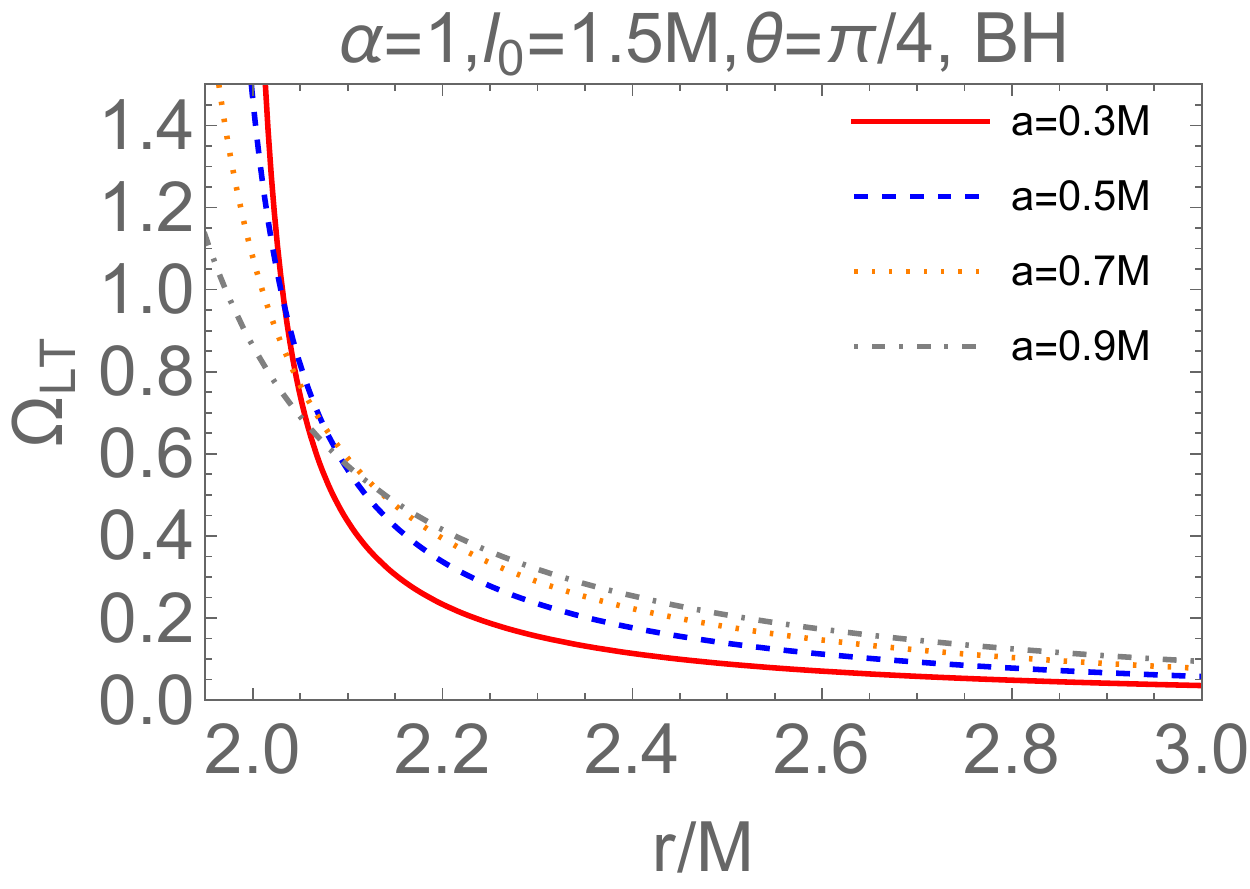}\\
(a) & (b) \\[6pt]
\\
\includegraphics[scale=0.6]{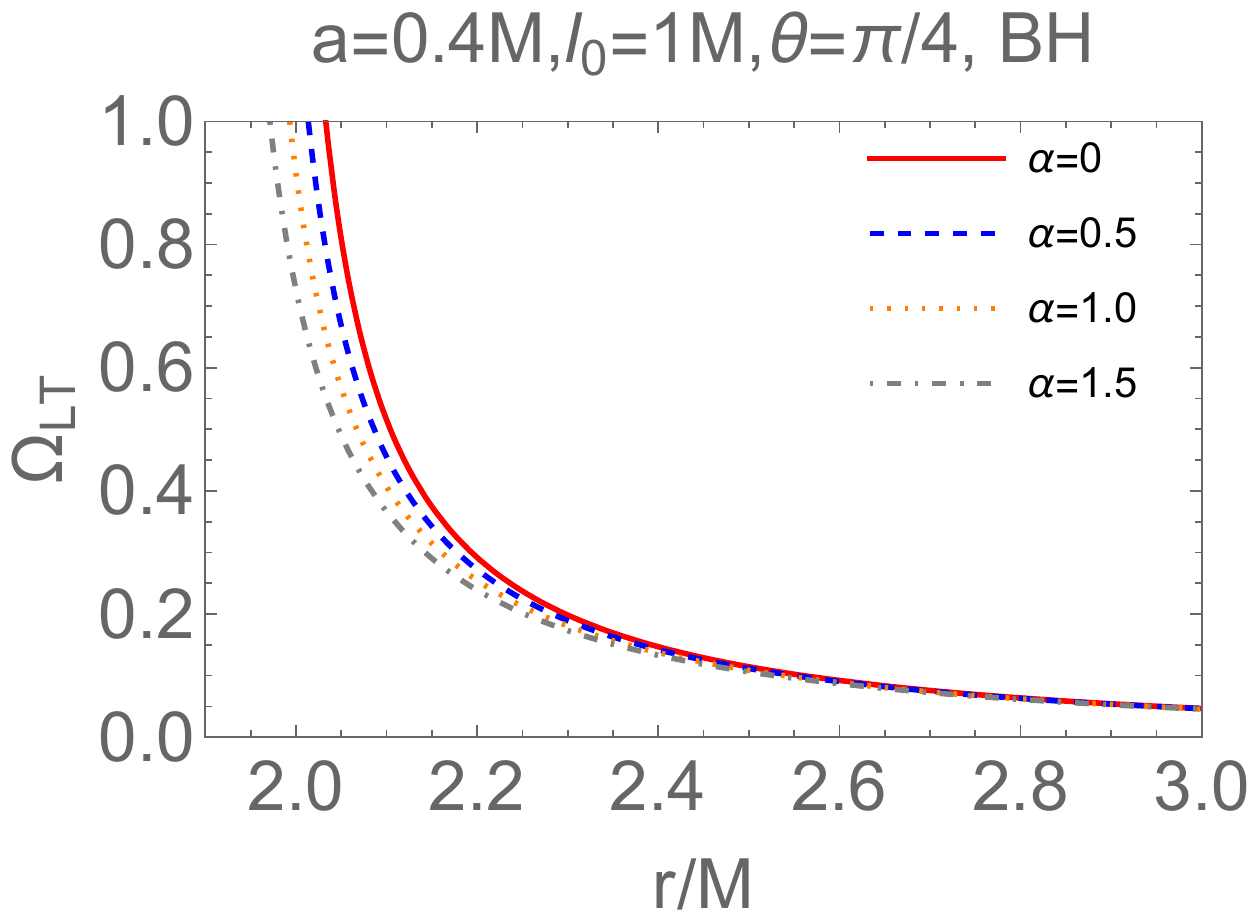}~~~~&
\includegraphics[scale=0.4]{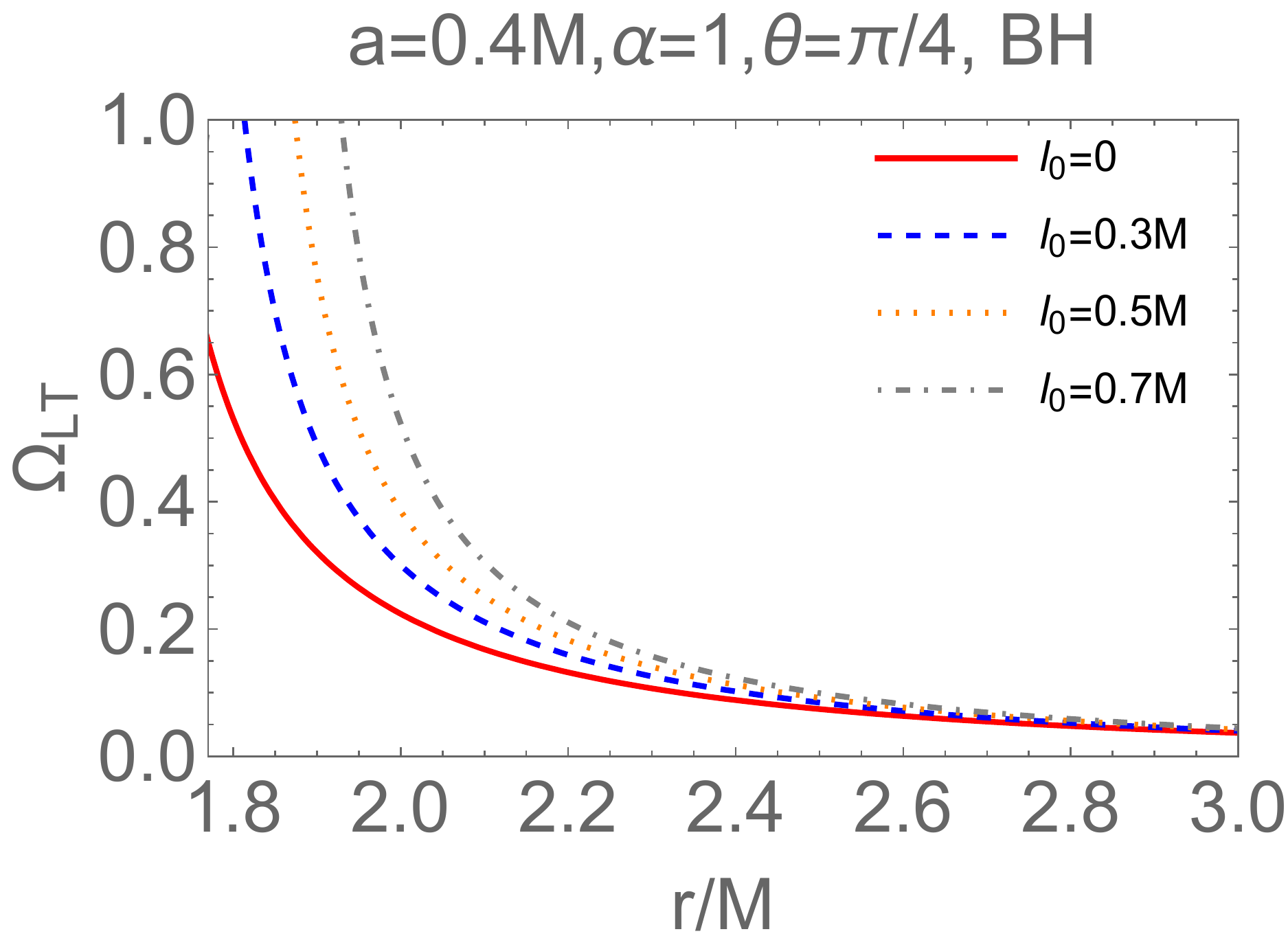}\\
(c) & (d) \\[6pt]
\end{tabular}}
\caption{Lense-Thirling precession frequency as a function of radial coordinate for hairy Kerr black hole. In plot (a), we focus the effect of the angle with fixed $a=0.4M,~l_0=1M $ and $\alpha=1$;  In plot (b), we focus on the effect of the spining parameter $a$ with fixed $\alpha=1,~l_0=1.5M$ and $\theta=\pi/4$; In plot (c), we focus on the effect of $\alpha$ with fixed $a=0.4M,~l_0=1M$ and $\theta=\pi/4$ ; In plot (d), we focus on the effect of $l_0$ with fixed $a=0.4M,~\alpha=1$ and $\theta=\pi/4$.}\label{fig:OmegaLT-BH}	
\end{figure}

\begin{figure}[H]
\resizebox{\linewidth}{!}{\begin{tabular}{cc}
\includegraphics[scale=0.6]{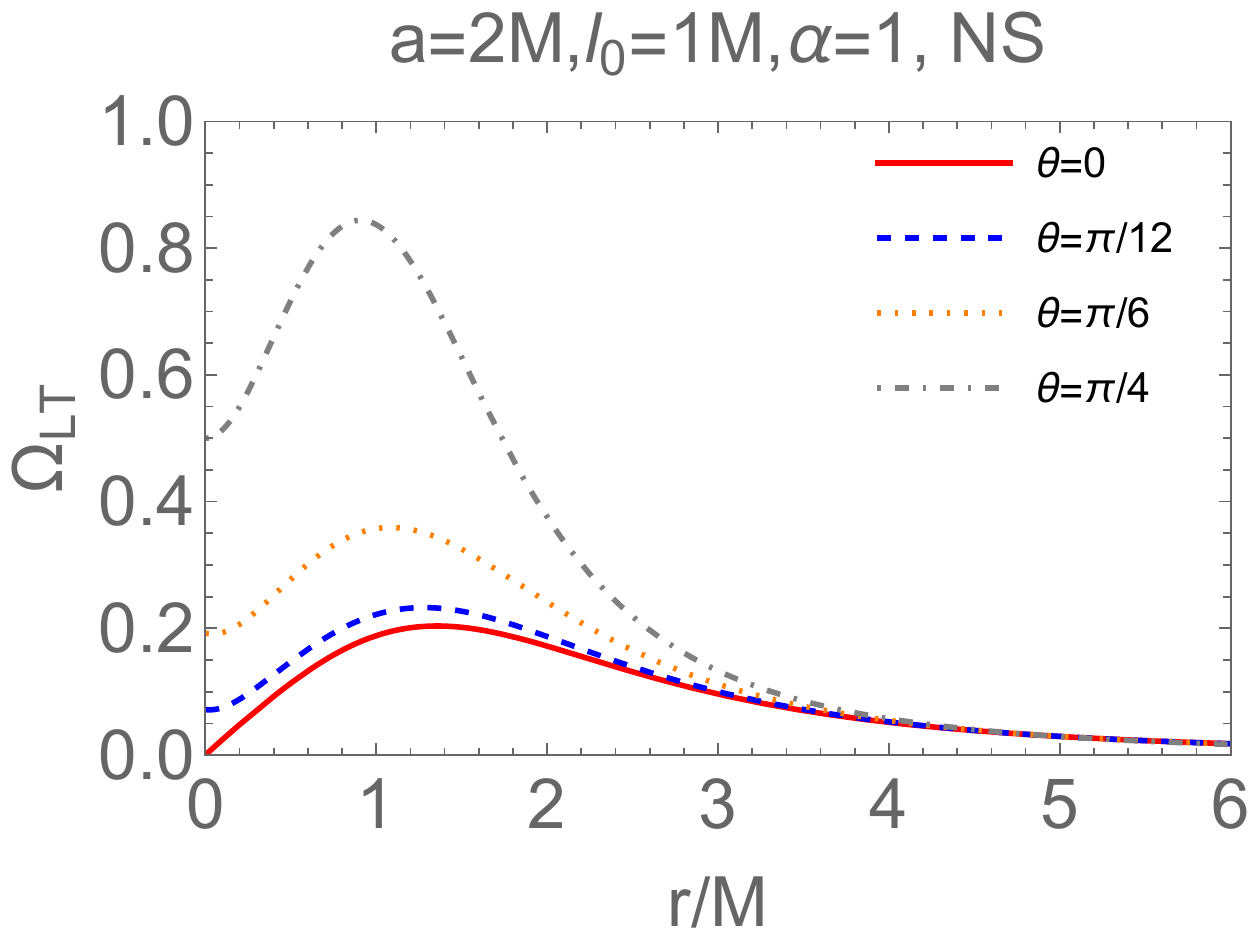}~~~~&
\includegraphics[scale=0.6]{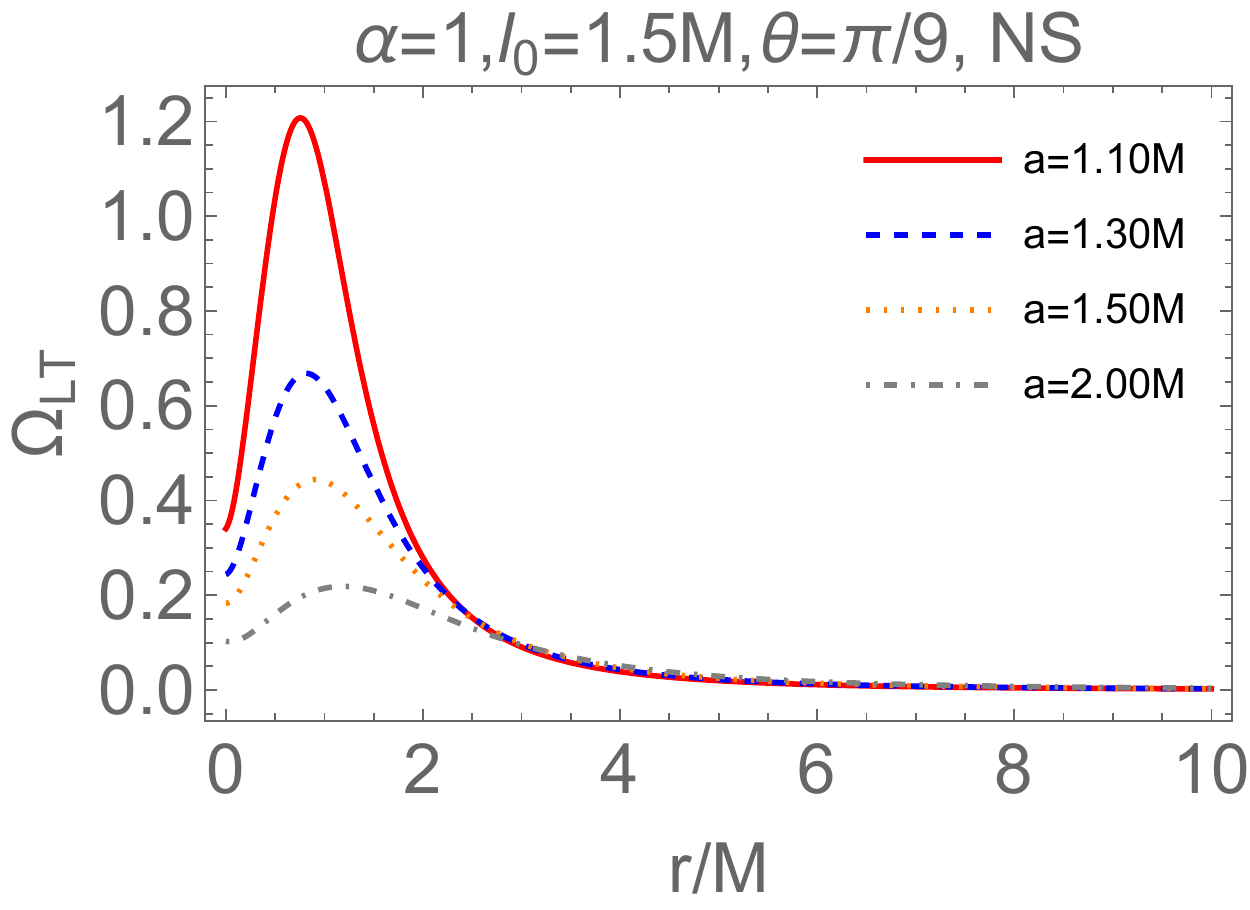}\\
(a)&(b)\\[6pt]
\\
\includegraphics[scale=0.6]{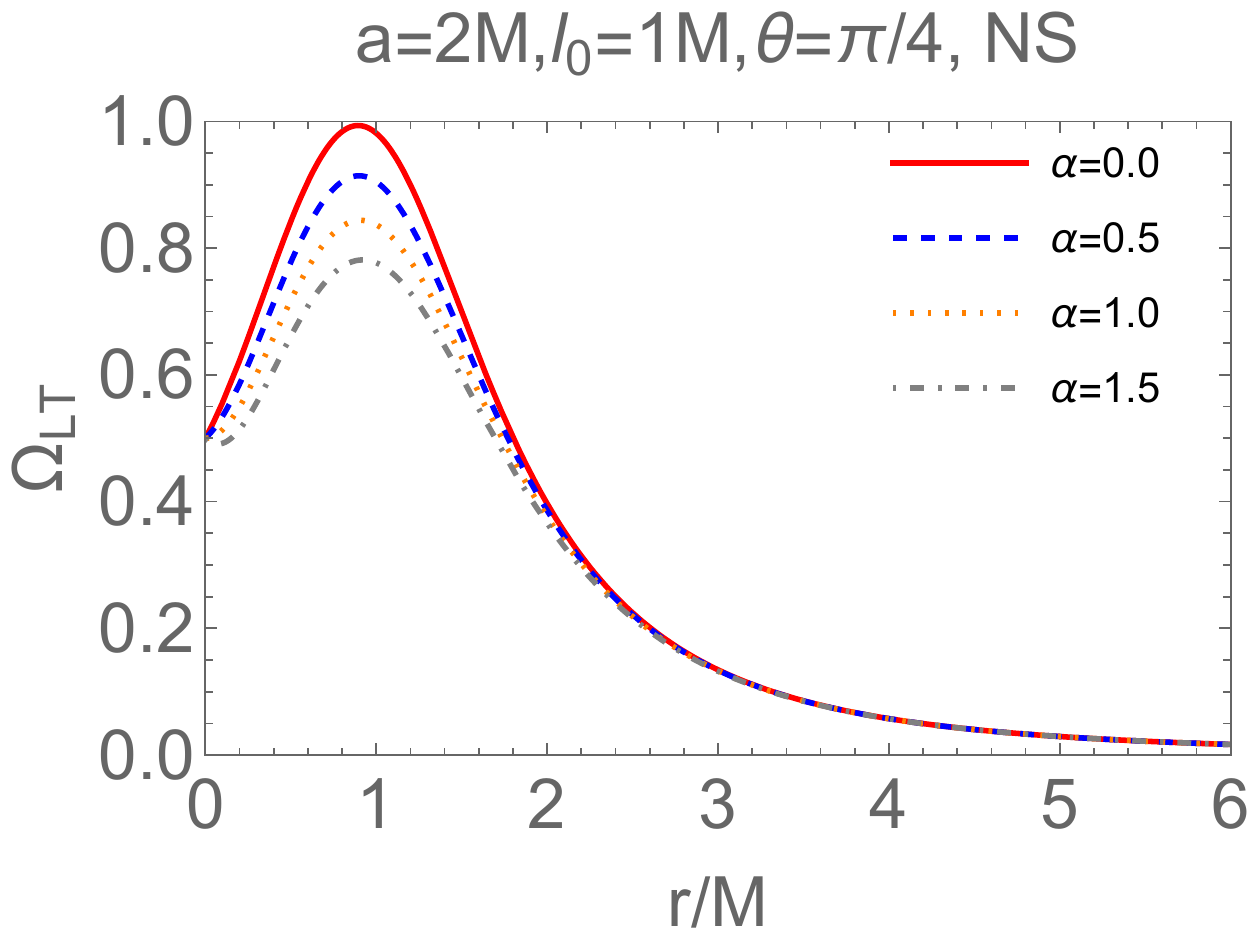}~~~~&
\includegraphics[scale=0.6]{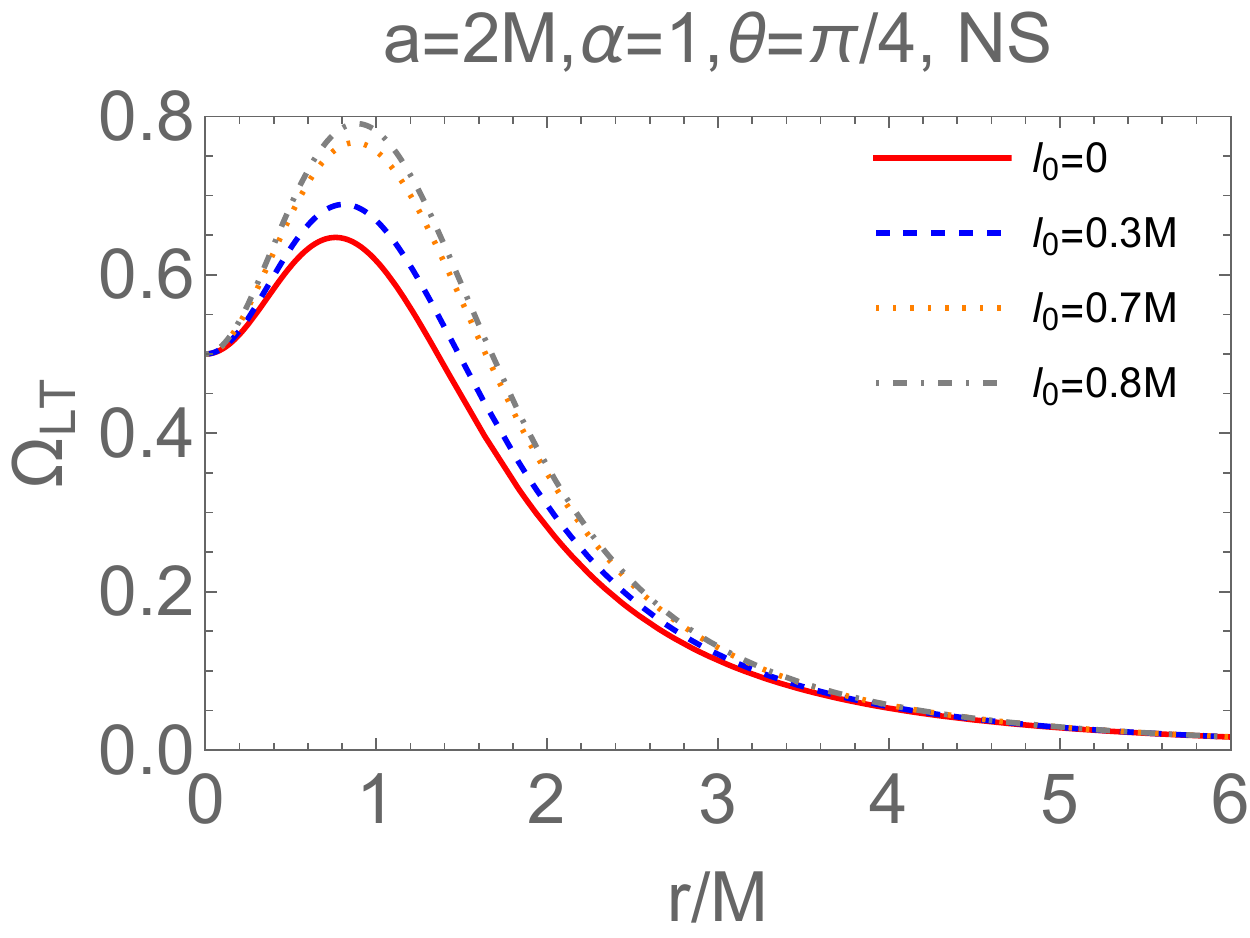}\\
(c)&(d)\\[6pt]
\end{tabular}}
\caption{Lense-Thirling precession frequency  as a function of radial coordinate for hairy naked singularity. In plot (a), we check the effect of the angle with fixed $a=2M,~l_0=1M$ and $\alpha=1$;  In plot (b), we check the effect of the spining parameter $a$ with fixed $\alpha=1,~l_0=1.5M$ and $\theta=\pi/9$; In plot (c), we check the effect of $\alpha$ with fixed $a=2M,~l_0=1M$ and $\theta=\pi/4$ ; In plot (d), we check the effect of $l_0$ with fixed $a=2M,~\alpha=1$ and $\theta=\pi/4$.}\label{fig:OmegaLT-NS}	
\end{figure}

The magnitude of LT precession frequency is given by
\begin{eqnarray}\label{eqM:Omega_LT}
\Omega_{LT}=&&\frac{1}{2\sqrt{-g} }\sqrt{g_{rr} \left(g_{t\phi,\theta}-\frac{g_{t\phi}}{g_{tt}}g_{tt,\theta}\right)^{2}
+
g_{\theta\theta}\left(g_{t\phi,r}-\frac{g_{t\phi}}{g_{tt}}g_{tt,r}\right)^{2}
}
\end{eqnarray}
which  are shown in FIG. \ref{fig:OmegaLT-BH} and FIG. \ref{fig:OmegaLT-NS} for samples of parameters.  In FIG. \ref{fig:OmegaLT-BH} for hairy Kerr BH, the LT precession frequency diverges at the ergosphere. In the region far away from the static limit, the LT precession frequency hardly affects by various parameters as expected. When the observer goes near the static limit,  the LT precession frequency increases as the angle and spin parameter increase, similar to that in the case of Kerr BH \cite{Chakraborty:2016ipk}; in addition, it becomes smaller for larger $\alpha$ or smaller $l_0$. 

 In FIG. \ref{fig:OmegaLT-NS} for hairy NS, the LT precession frequency is always finite and possesses a peak.  Similar to that in Kerr case  \cite{Chakraborty:2016ipk}, as the angle increases, both the peak and frequency near the ring singularity increase, while the spin parameter has the opposite influence. Moreover, the hairy parameter $l_0$ also enhances the peak but $\alpha$ suppresses it; while the hairy parameters have no imprint on the frequency at the ring singularity. Though the concrete expression of the LT precession frequency are complex, we can analytically reduce the behaviors at $r\rightarrow 0$ as
\begin{equation}
    \Omega_{LT}|_{r\rightarrow 0}=\frac{
    \sec^2(\theta)\tan(\theta)}{a^2}.
\end{equation}
Therefore, it is obvious  that LT precession frequency at $r\rightarrow 0$ is independent of the hairy parameters $l_0$ and $\alpha$, and it is finite unless $\theta=\pi/2$, which are consistent with what we see in FIG.\ref{fig:OmegaLT-NS}.

\subsection{Case with $a=0$: Geodetic precession frequency}
The metric \eqref{eq-metric} with $a=0$ describes hairy Schwarzschild  spacetime. In this case, the precession frequency \eqref{eq:Omega_p} does not vanish, though the LT-precession frequency \eqref{eq:Omega_LT} vanishes. The non-vanishing sector of the spin precession is known as the geodetic precession due to the curvature of the spacetime, and its formula is 
\begin{eqnarray}\label{eq:Omega_a0v1}
\vec{\Omega}_{p}\mid_{a=0}=&&\frac{1}{2\sqrt{-g}
	\left( 1+\Omega ^{2}\frac{g_{\phi\phi}}{g_{tt}}\right) }
\left[
-\Omega\sqrt{g_{rr}}\left(g_{\phi\phi,\theta}-\frac{g_{\phi\phi}}{g_{tt}}g_{tt,\theta}\right)\hat{r}+
\Omega\sqrt{g_{\theta\theta}}\left(g_{\phi\phi,r}-\frac{g_{\phi\phi}}{g_{tt}}g_{tt,r}\right)
\hat{\theta}
\right].
\end{eqnarray}

\begin{figure}[H]
\center{
\includegraphics[scale=0.6]{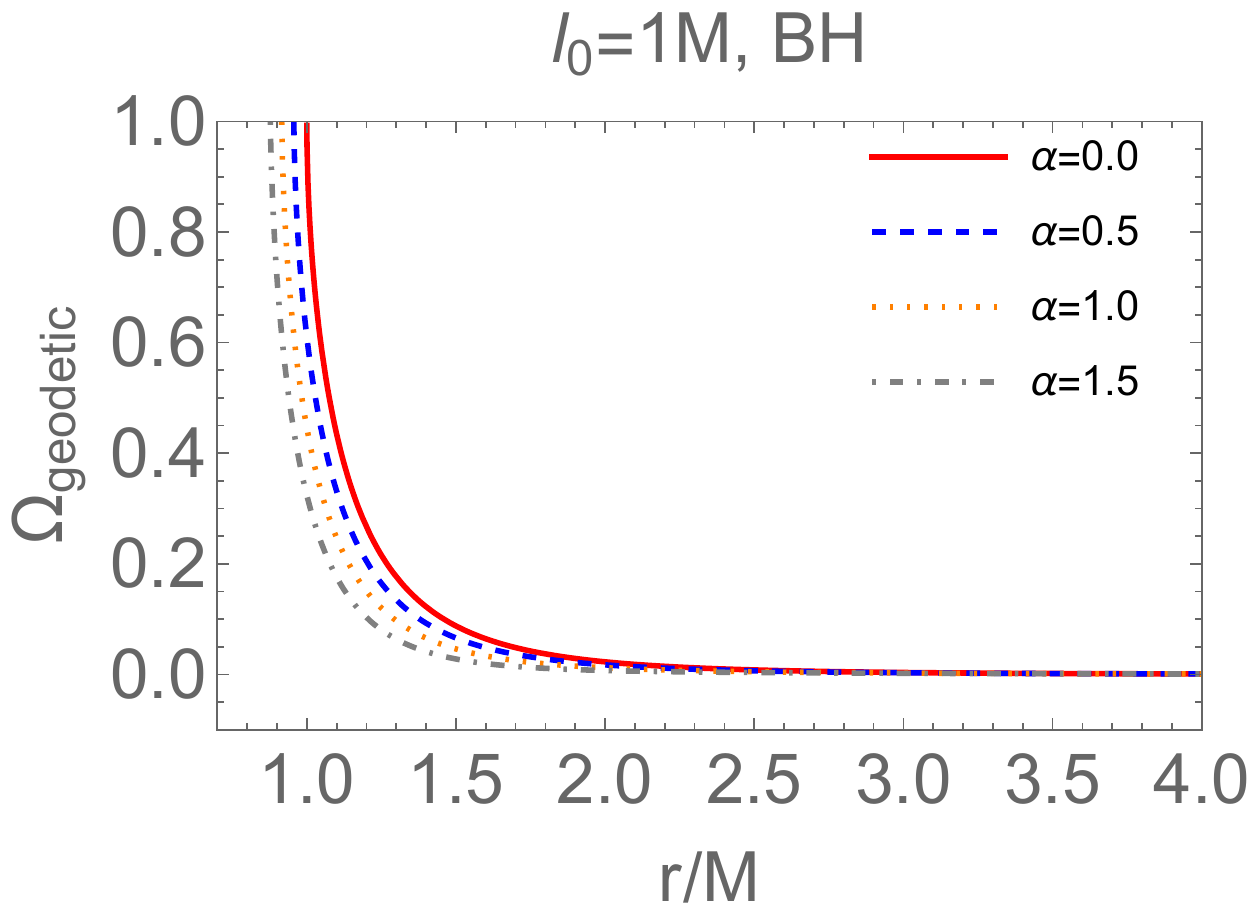}\hspace{0.5cm}
\includegraphics[scale=0.6]{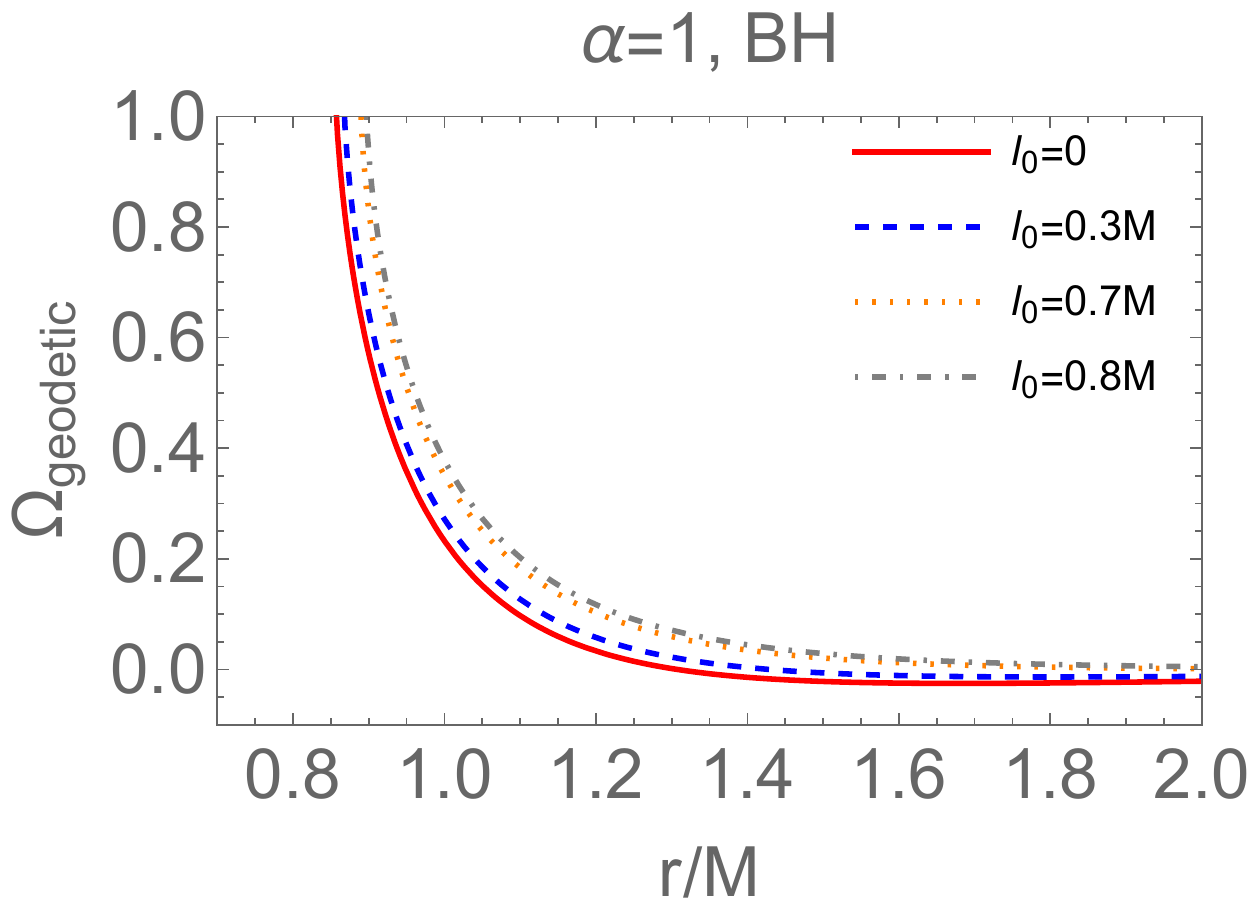}
\caption{The  geodetic precession   as a function of radial coordinate for the hairy Schwarszchild black hole. In the left panel, we tune  the deviation parameter $\alpha$ with fixed $l_0=1 M$, while in the right panel, we tune $l_0$ with fixed $\alpha=1$.}\label{fig:Omegageodetic}	}	
\end{figure}

To proceed, we can choose the observer in the equatorial plane by setting $\theta=\pi/2$ without loss of generality due to the spherical symmetry. Then we have
\begin{eqnarray}\label{eq:Omega_a0v2}
\Omega_{p}\mid_{a=0}=\frac{\Omega\sqrt{g_{\theta\theta}}\left(g_{\phi\phi,r}
-\frac{g_{\phi\phi}}{g_{tt}}g_{tt,r}\right)}{2\sqrt{-g}
	\left( 1+\Omega ^{2}\frac{g_{\phi\phi}}{g_{tt}}\right) }=\frac{(3l_0 M-6M^2+e^{\frac{2r}{l_0-2M}}r^2\alpha-l_0 r(1-e^{\frac{2r}{l_0-2M}}\alpha)+2M r(1+e^{\frac{2r}{l_0-2M}}\alpha))\Omega}{(l_0-2M)(2M+r(-1-e^{\frac{2r}{l_0-2M}}\alpha+r^2\Omega^2))},
\end{eqnarray}
meaning that the gyroscope moving in the hairy Schwarzschild spacetime will also precess. Therefore, in this case, the frequency  for the gyro moving along a circular geodesic could be  the Kepler frequency \cite{Glendenning:1993di} $\Omega_{Kep}=\sqrt{(M-\alpha r^2e^{-r/(M-l_0/2)}/(2M-l_0))/r^3}$ , with which we can induce 
\begin{eqnarray}\label{eq:Omega_a0v3}
\Omega_{p}\mid_{a=0,\Omega=\Omega_{Kep}}=\Omega_{Kep}=\sqrt{\frac{M}{r^3}-\alpha \frac{ e^{-r/(M-l_0/2)}}{(2M-l_0)r}}.
\end{eqnarray}
The above expression gives the precession frequency in the Copernican frame, computed with respect to the proper time $\tau$, which is related to the coordinate time via
\begin{equation}
d\tau=\sqrt{1-M/r^3+\alpha(1+r/(2M-l_0))e^{-r/(M-l_0/2)}} dt.
\end{equation}
Therefore, in the coordinate basis, the geodetic precession frequency is
\begin{eqnarray}\label{eq:Omega_a0'}
\Omega_{geodetic}=\left(1-\sqrt{1-\frac{M}{r^3}+\alpha(1+\frac{r}{2M-l_0})e^{-\frac{r}{M-l_0/2}}}\right)\sqrt{\frac{M}{r^3}-\alpha \frac{ e^{-r/(M-l_0/2)}}{(2M-l_0)r}}.
\end{eqnarray}
It is obvious that when the deviation parameter, $\alpha$, vanishes, the above expression reproduces the geodetic precession of the Schwarzschild black hole found in \cite{sakina1979parallel}. How the model parameters $\alpha$ and $l_0$ affect the geodetic precession of the hairy Schwarzschild BH is shown in FIG.\ref{fig:Omegageodetic}. We see that the  geodetic precession is suppressed by the hair comparing to that in GR, and its deviation from that in  Schwarzschild black hole is more significant as the $\alpha$ increases. Moreover, the  geodetic precession frequency is enhanced by larger $l_0$.

\subsection{Spin precession frequency in hairy Kerr black hole and naked singularity}
We move on to study the general spin precession frequency \eqref{eq:Omega_p} in hairy Kerr spacetime, and  analyze its difference between hairy Kerr BH and NS. 
As we aforementioned,  the angular velocity has a restricted range for the timelike stationary observers . So, we introduce a parameter $0<k<1$ so that the angular velocity $\Omega$ can be rewritten  in terms of $\Omega_{\pm}$  as
\begin{equation}\label{eq:O-rede}
\Omega =k\Omega _{+}+(1-k)\Omega _{-}=\frac{(2k-1)\sqrt{g^2_{t\phi}-g_{\phi\phi}g_{tt}}-g_{t\phi}}{g_{\phi\phi}}.
\end{equation}
It is obvious that for $k=1/2$, this expression can be reduced as 
\begin{equation}
\Omega \mid_{k=1/2}=-\frac{g_{t\phi}}{g_{\phi\phi}},
\end{equation}
with which  the observer  is called the zero-angular-momentum observer (ZAMO) \cite{Bardeen:1972fi}. It was addressed in \cite{Chakraborty:2016mhx} that the precession frequency of the gyroscope attached to ZAMO in the Kerr black hole spacetime has different behavior from a gyroscope attached to other observers with other angular velocities, because these gyros has no rotation with respect to the local geometry and stationary observer.

Substituting \eqref{eq:O-rede} into \eqref{eq:Omega_p}, we can obtain the general spin precession frequency in terms of the parameter $k$, thus the magnitude of spin precession frequency is written as
\begin{eqnarray}\label{eq:Omega_p_amg}
\Omega_{p}=&&\frac{\sqrt{C_1^{2}  + C_2^{2} }}{2\sqrt{-g}
	\left( 1+2\Omega \frac{g_{t\phi}}{g_{tt}}+\Omega ^{2}\frac{g_{\phi\phi}}{g_{tt}}\right) } ,
\end{eqnarray}
where $C_1$ and $C_2$ are defined in \eqref{eq:Omega_HF}.

\begin{figure}[H]
\resizebox{\linewidth}{!}{\begin{tabular}{ccc}
\includegraphics[scale=0.4]{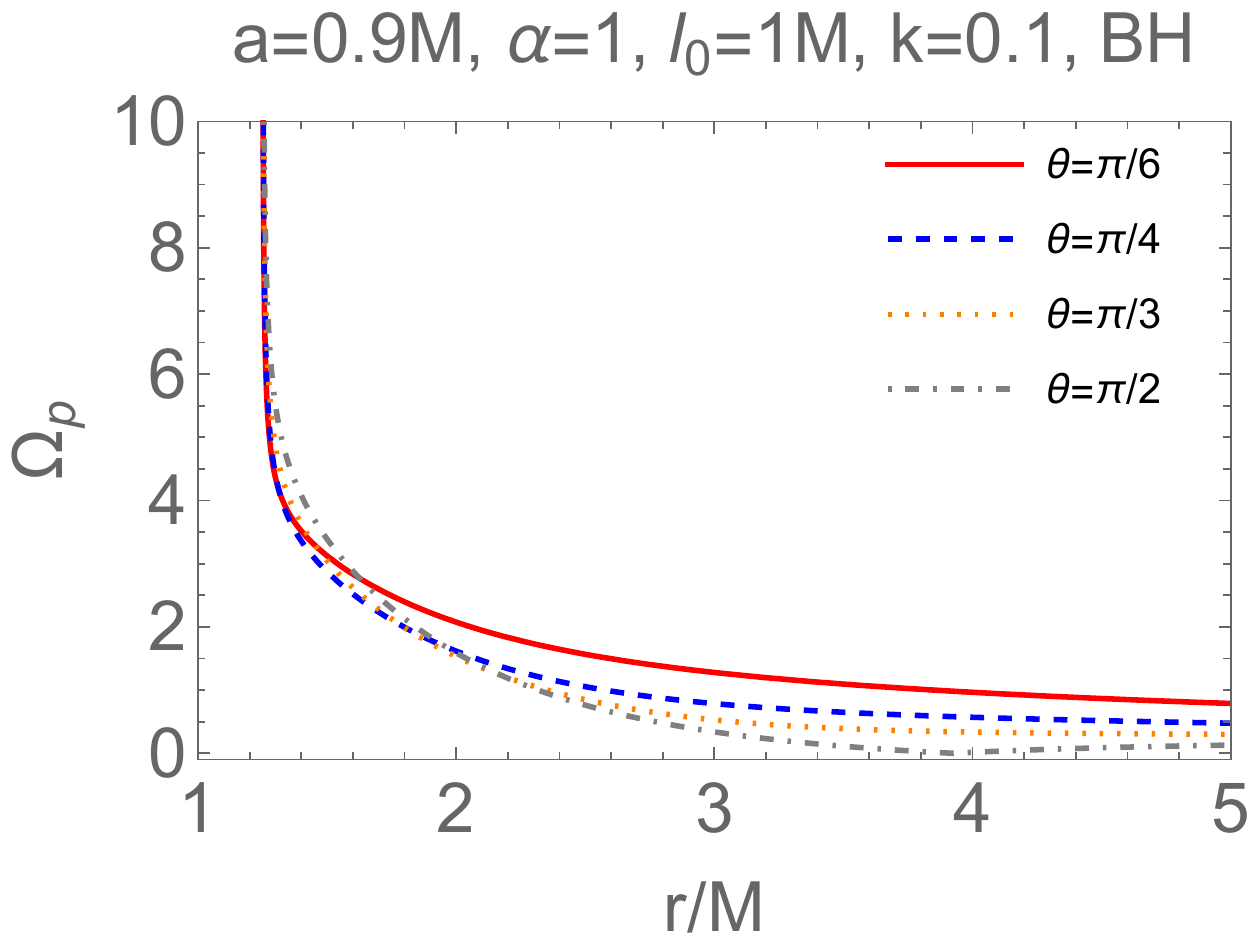}&
\includegraphics[scale=0.4]{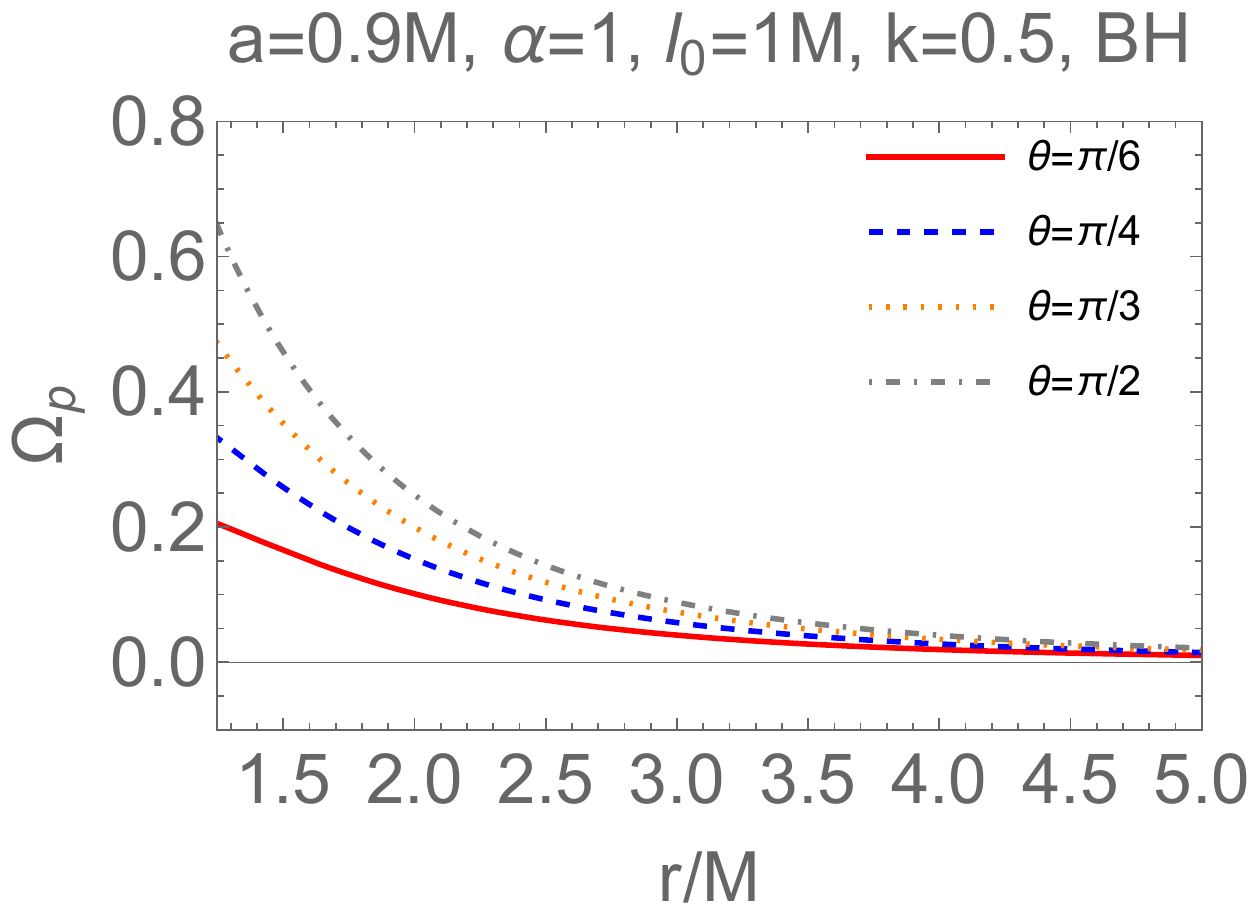}&
\includegraphics[scale=0.4]{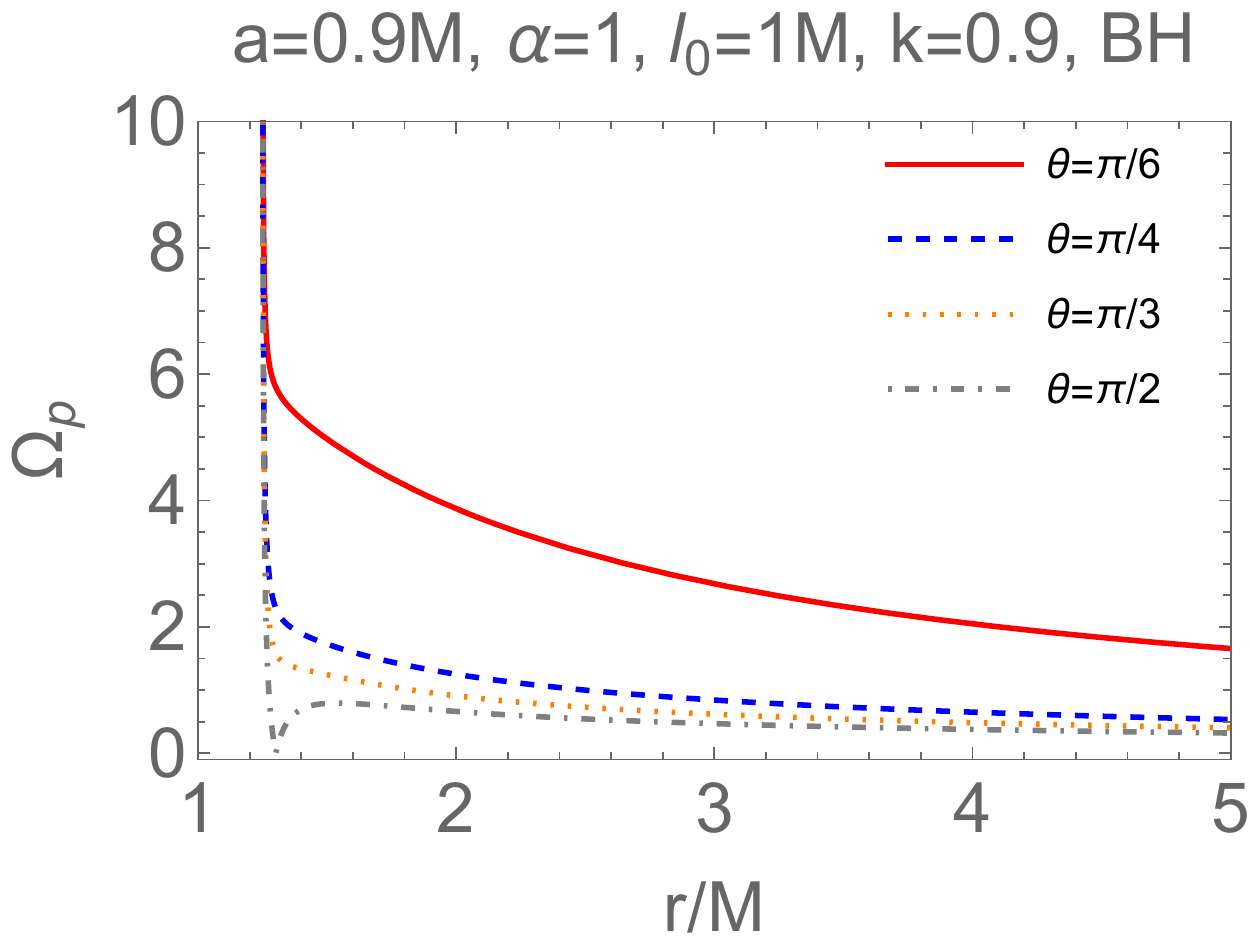}\\
(a)&(b)&(c)\\
\\
\includegraphics[scale=0.4]{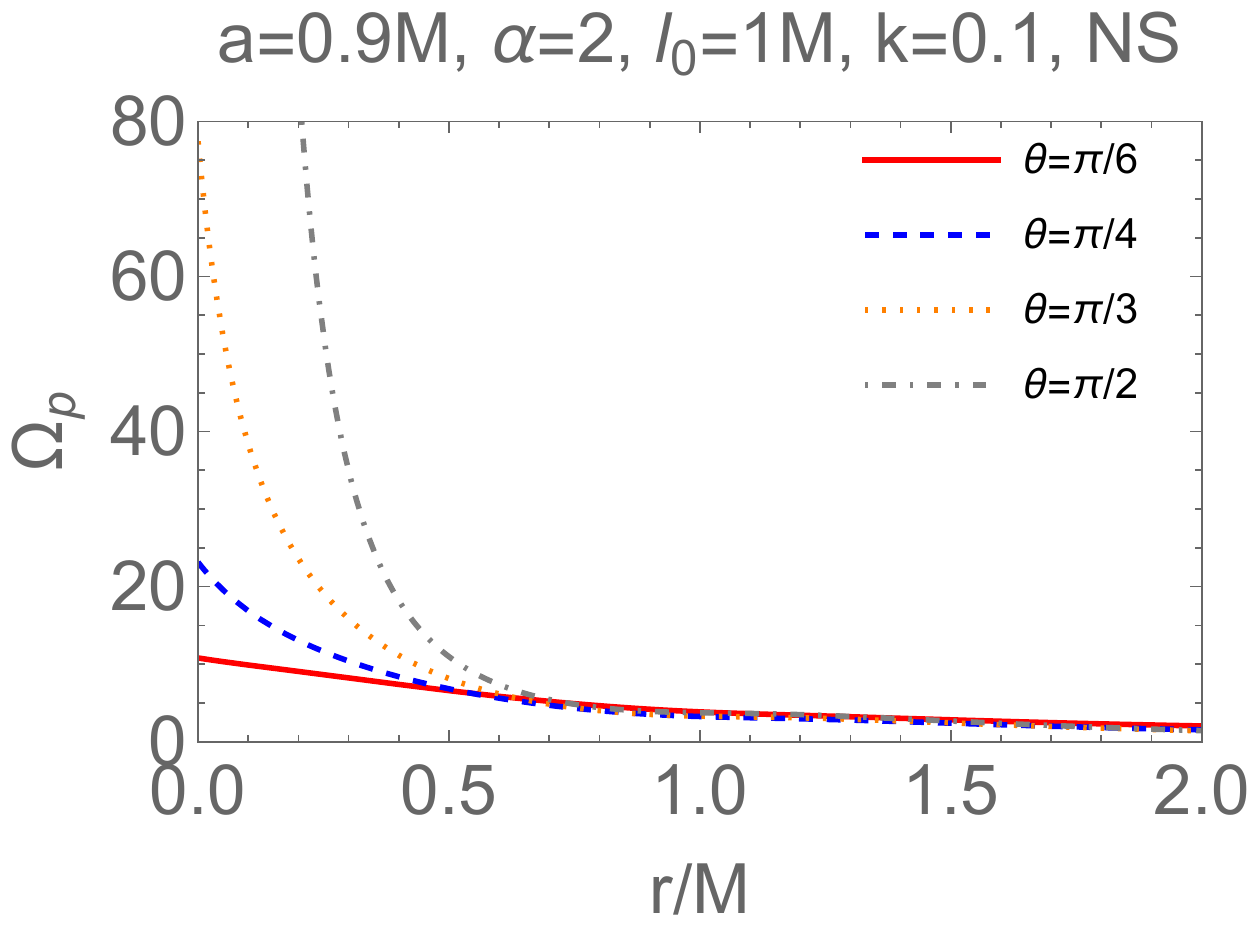} &
\includegraphics[scale=0.28]{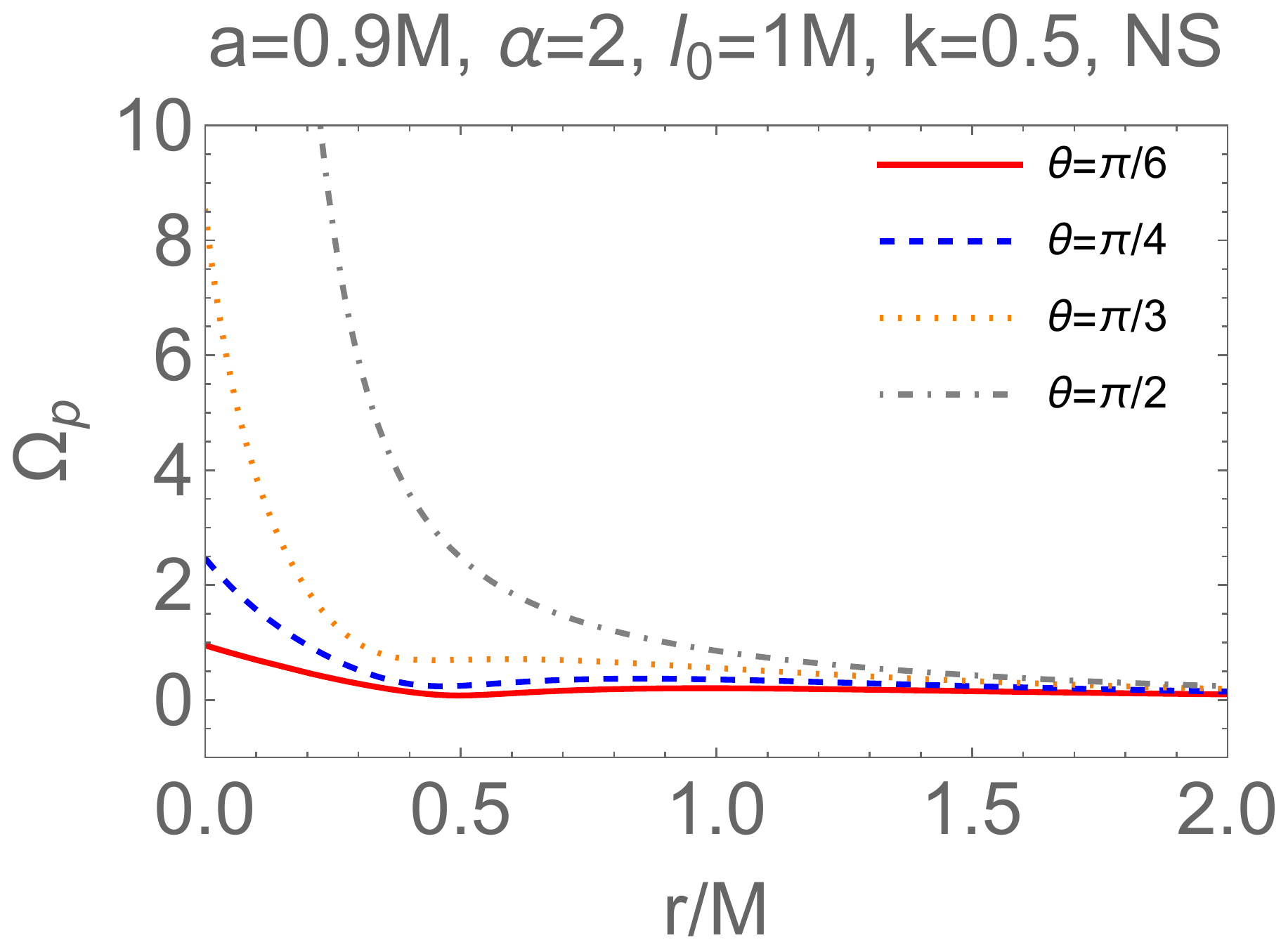} &
\includegraphics[scale=0.4]{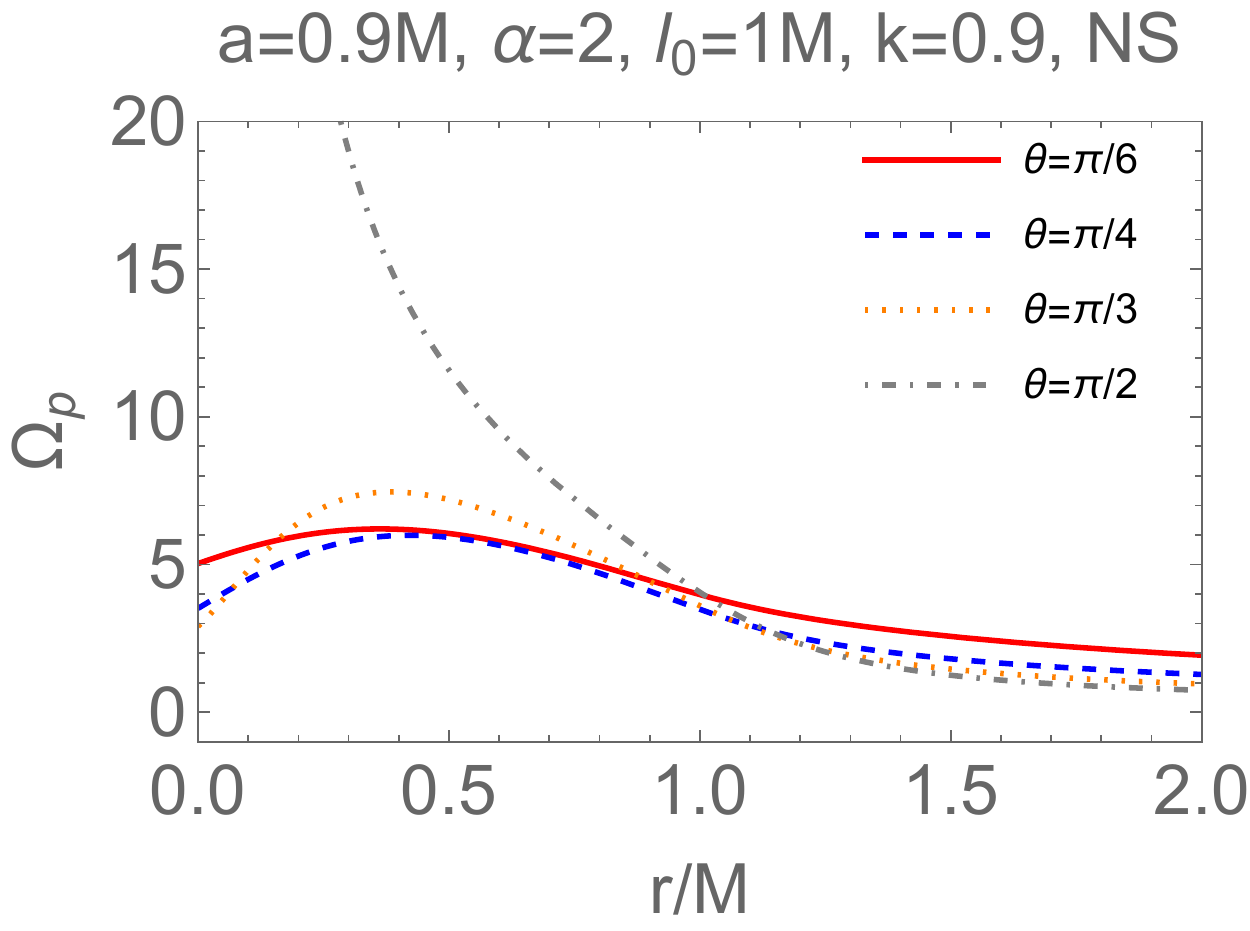} \\
(d)&(e)&(f)\\\\[6pt]
\end{tabular}}
\caption{The effect of $\theta$ on the spin precession frequency $\Omega_{p}$ for different $k$ in hairy Kerr spacetime. Here we fix $a=0.9M$ and $l_0=1M$, then $\alpha=1$ corresponds to hairy Kerr black hole (a-c)  while $\alpha=2$ corresponds to naked singularity (d-f). From left to right, $k=0.1$ (a, d), $0.5$ (b,e) and $0.9$ (c,f), respectively.}
\label{fig:OmegaP-k}
\end{figure}

\begin{figure}[H]
\center{
\includegraphics[scale=0.28]{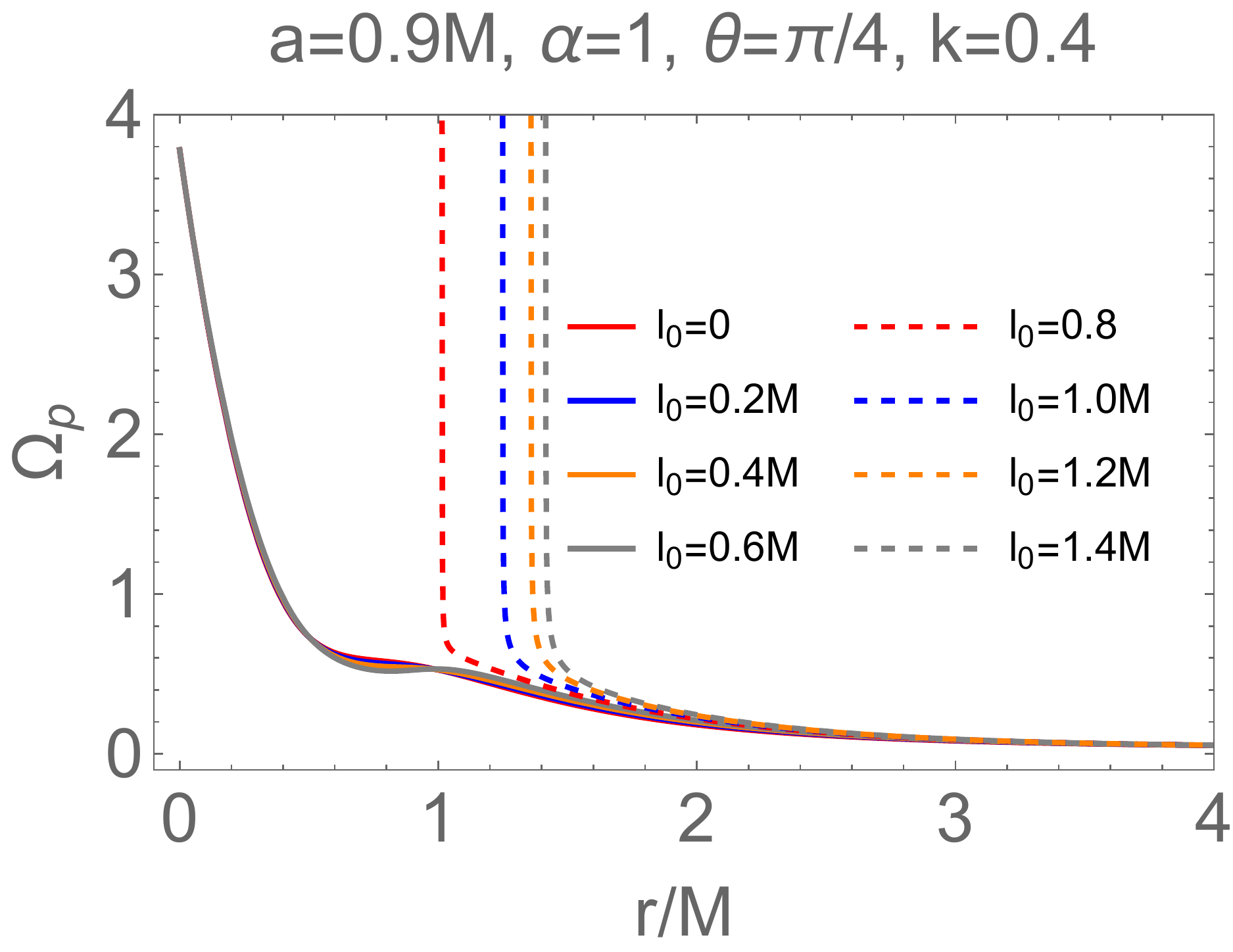}\hspace{0.5cm}
\includegraphics[scale=0.28]{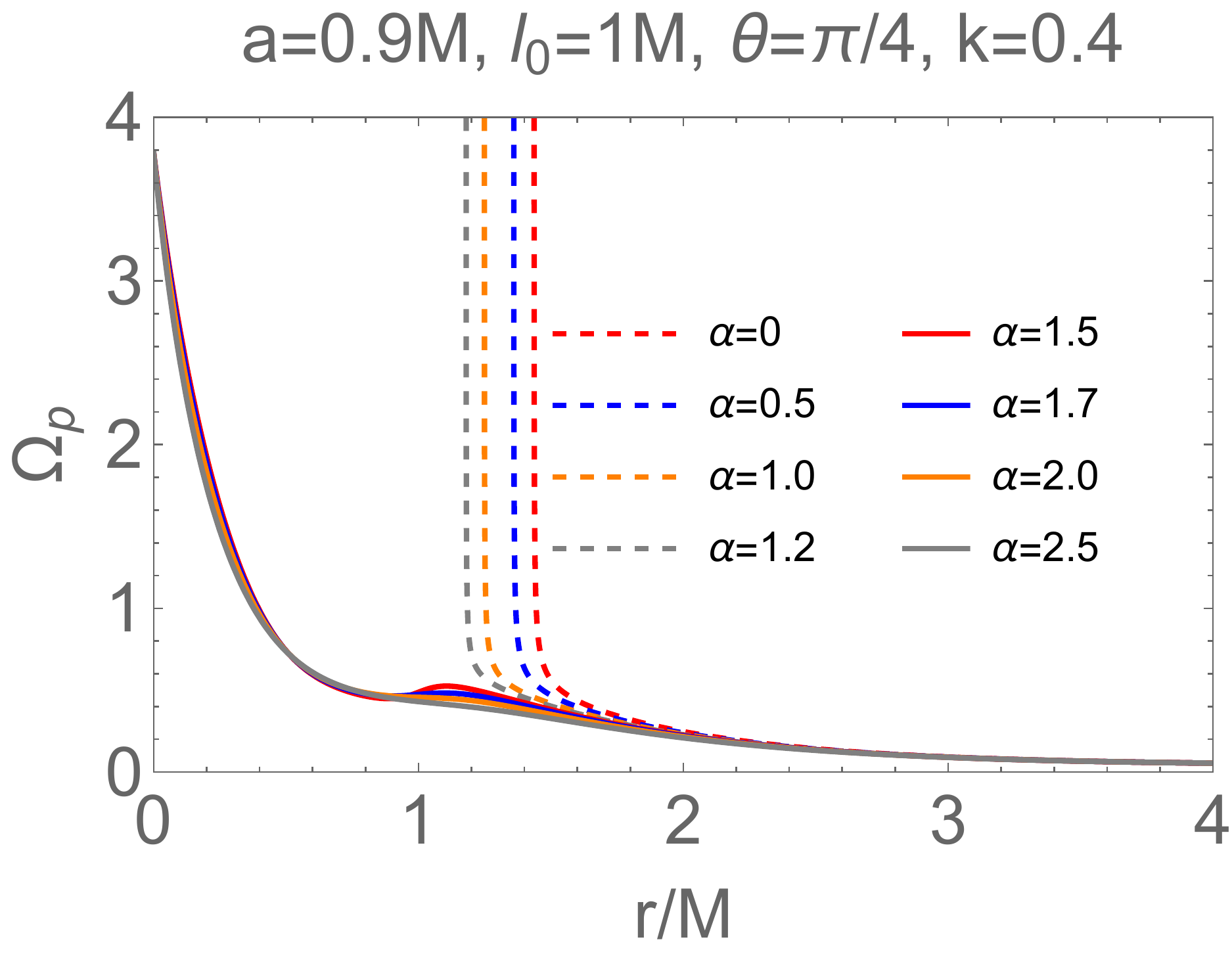}\hspace{0.5cm}
\includegraphics[scale=0.28]{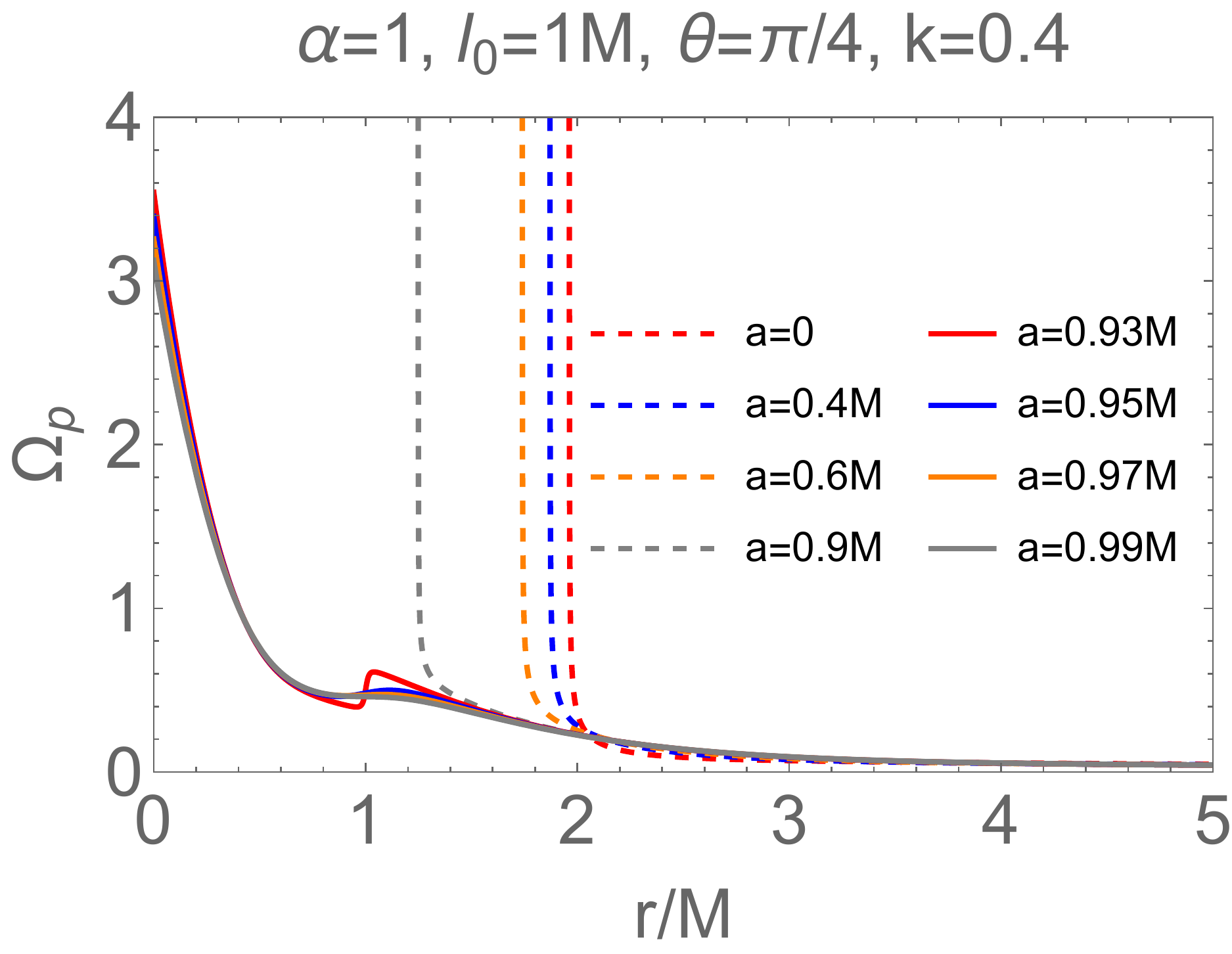}
\caption{The dependence of the spin precession frequency $\Omega_{p}$ on the model parameters. From left to right, we tune $l_0$, $\alpha$ and $a$, respectively. In each plot, the dashed curves correspond to  hairy Kerr black hole while the solid curves correspond to  naked singularity. }\label{fig:OmegaP-BH}	}	
\end{figure}

The magnitude of the general spin precession frequencies with different $k$ for the hairy Kerr BH and NS are depicted in FIG. \ref{fig:OmegaP-k}, where $k=0.1, 0.5, 0.9$ from the left to right column.  In the first row, we show the results for black hole with $a=0.9M$, $\alpha=1$ and $l_0=1M$. It is obvious that for hairy Kerr BH, the spin precession frequency of a gyroscope attached to
any observer beyond ZAMO, always diverges whenever it approaches the horizons along any direction. However, $\Omega_p$ remains finite for ZAMO observer everywhere including at the horizon.  In the second row, we show the results for hairy naked NS with $a=0.9M$, $\alpha=2$ and $l_0=1M$. It shows that the spin precession frequency for NS is finite even as the observer approach $r = 0$ along any direction except from the direction $\theta=\pi/2$. More results for different model parameters are shown in FIG.\ref{fig:OmegaP-BH} where we have set $k=0.4$ and $\theta=\pi/4$. Again, it is convenient to verity that
\begin{eqnarray}
    \Omega_p|_{r\rightarrow 0, \theta=\pi/4}=\sqrt{\frac{a^2(1-2k)^2+(17+4(k-1)k(17+8(k-1)k))M^2+12\sqrt{2}(1-2k(2+k(2k-3)))M^2}{8a^4(k-1)^2k^2}}
\end{eqnarray}
which is independent of $\alpha$ and $l_0$.

\section{LT precession and periastron precession of accretion disk physics around hairy Kerr spacetime} \label{sec:test particle}

In this section, we will study the LT precession frequency  and the periastron precession frequency  of accretion disk physics around different geometries  in the hairy Kerr spacetime, as the related physics could help to test the strong gravity of the central compact bodies. We expect that this study could further 
 differentiate the hairy Kerr black hole and naked singularity as the central objects. To this end, we should study the geodesic motion of a test massive particle around the hairy Kerr spacetime. To study the accretion disk physic via the orbits of test massive particle, we should fix the stable circular orbit around the central bodies and then perturb it. One important stable circular orbit is the inner-most stable circular orbit (ISCO) which is the last or smallest stable circular orbit of the particle. 
Therefore, we will first recall the procedure of deriving an orbit equation to introduce  the bound orbits, circular orbits and ISCO in the hairy Kerr spacetime. Then we explore the LT precession frequency  and the periastron precession frequency by perturbing the circular orbit in equatorial plane.
For convenience,
we shall focus on the orbits in  the equatorial plane of the spacetime.

\subsection{Bound orbit and ISCO}

For a test particle moving along the timelike geodesic with the four velocity $p^\mu$ in hairy Kerr spacetime \eqref{eq-metric}, we have two conserved quantities
for the massive particle {defined in \eqref{eq-conservation}.} So for the orbit in the equatorial plane with $\theta=\pi/2$, we can solve out 
\begin{eqnarray}
p^t&=&\frac{g_{\phi\phi}\mathcal{E}+g_{t\phi}L}{g_{t\phi}^2-g_{tt}g_{\phi\phi}}\mid_{\theta=\pi/2}=\frac{r^3\mathcal{E}+a^2 \mathcal{E}\left(2 M+r-e^{\frac{2 r}{l_0-2 M}} r \alpha\right)+a L\left(-2 M+e^{\frac{2 r}{l_0-2 M}} r \alpha\right)}{r\left(a^2+r\left(-2 M+r+e^{\frac{2 r}{l_0-2 M}} r \alpha\right)\right)},\\
p^\phi&=&-\frac{g_{t\phi}\mathcal{E}+g_{tt}L}{g_{t\phi}^2-g_{tt}g_{\phi\phi}}\mid_{\theta=\pi/2}=\frac{a\mathcal{E}\left(2 M-e^{\frac{2 r}{l_0-2 M}} r \alpha\right)+ L\left(-2 M+r+e^{\frac{2 r}{l_0-2 M}} r \alpha\right)}{r\left(a^2+r\left(-2 M+r+e^{\frac{2 r}{l_0-2 M}} r \alpha\right)\right)}.
\end{eqnarray} 
Inserting the above formulas into the normalization condition $p^\mu p_\mu=-1$ for timelike geodesic, we can obtain the radial velocity as 
\begin{eqnarray}
p^r&=&\pm\sqrt{\frac{-1-g_{tt}(p^t)^2-2g_{t\phi}p^tp^\phi-g_{\phi\phi}(p^\phi)^2}{g_{rr}}}\mid_{\theta=\pi/2}\nonumber\\
&=&\pm\sqrt{\left(\mathcal{E}^2-1-e^{\frac{2 r}{l_0-2 M}}\alpha\right)+\frac{2M}{r}-\frac{L^2-a^2(\mathcal{E}^2-1)+e^{\frac{2 r}{l_0-2 M}}\alpha(L-a\mathcal{E})^2}{r^2}+\frac{2M(L-a\mathcal{E})^2}{r^3}}
\end{eqnarray}
where $\pm$ corresponds to the radially outgoing and incoming cases, respectively. 
\begin{figure}[H]
\center{
\includegraphics[scale=0.5]{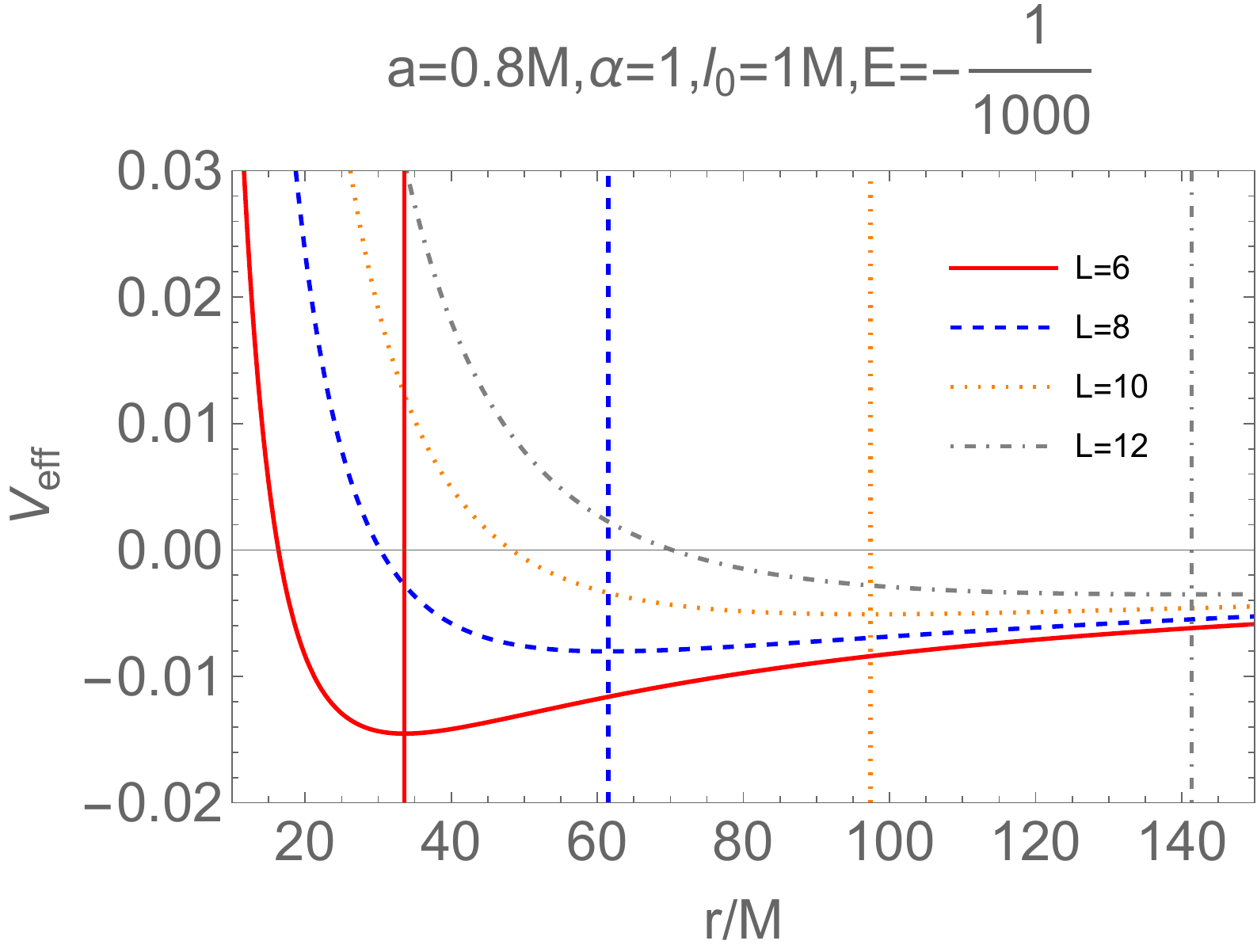}\hspace{0.5cm}
\caption{ Samples of effective potential as the functions  of $r$ with given $E$ and $L$. The vertical lines indicate the positions of minimal effective potential in each case.}\label{fig:Veff}	}
\end{figure}
Consequently, the effective potential of the equatorial timelike geodesics is 
\begin{eqnarray}
V_{eff}(r)&=&E-\frac{1}{2}(p^r)^2\nonumber\\
&=&-\frac{M}{r}+\frac{1}{2}e^{\frac{2 r}{l_0-2 M}}\alpha+\frac{L^2-a^2(\mathcal{E}^2-1)+e^{\frac{2 r}{l_0-2 M}}\alpha(L-a\mathcal{E})^2}{2r^2}-\frac{M(L-a\mathcal{E})^2}{r^3}
\end{eqnarray}
where $E=\frac{1}{2}(\mathcal{E}^2-1)$ is the total relativistic energy of the test particle.
For the stable bound orbits, the total energy should not be smaller than the minimal effective potential which is determined by 
\begin{eqnarray}
\frac{dV_{eff}}{dr}\mid_{r_m}=0,~~~~\frac{d^2V_{eff}}{dr^2}\mid_{r_m}>0.
\end{eqnarray}
It is difficult to give the analytical expression of $r_m$ and the minimal effective potential $V_{min}$, so the we show a sample of $V_{eff}$ in FIG. \ref{fig:Veff} with given $E=-1/1000$ and various $L$. It is obvious that both $r_m$ and $V_{min}$ becomes larger for larger $L$, similar to that found in Kerr spacetime \cite{Bambhaniya:2020zno}. The hairy parameters affect $V_{min}$ and its location. 

\begin{figure}[H]
\center{
\includegraphics[scale=0.6]{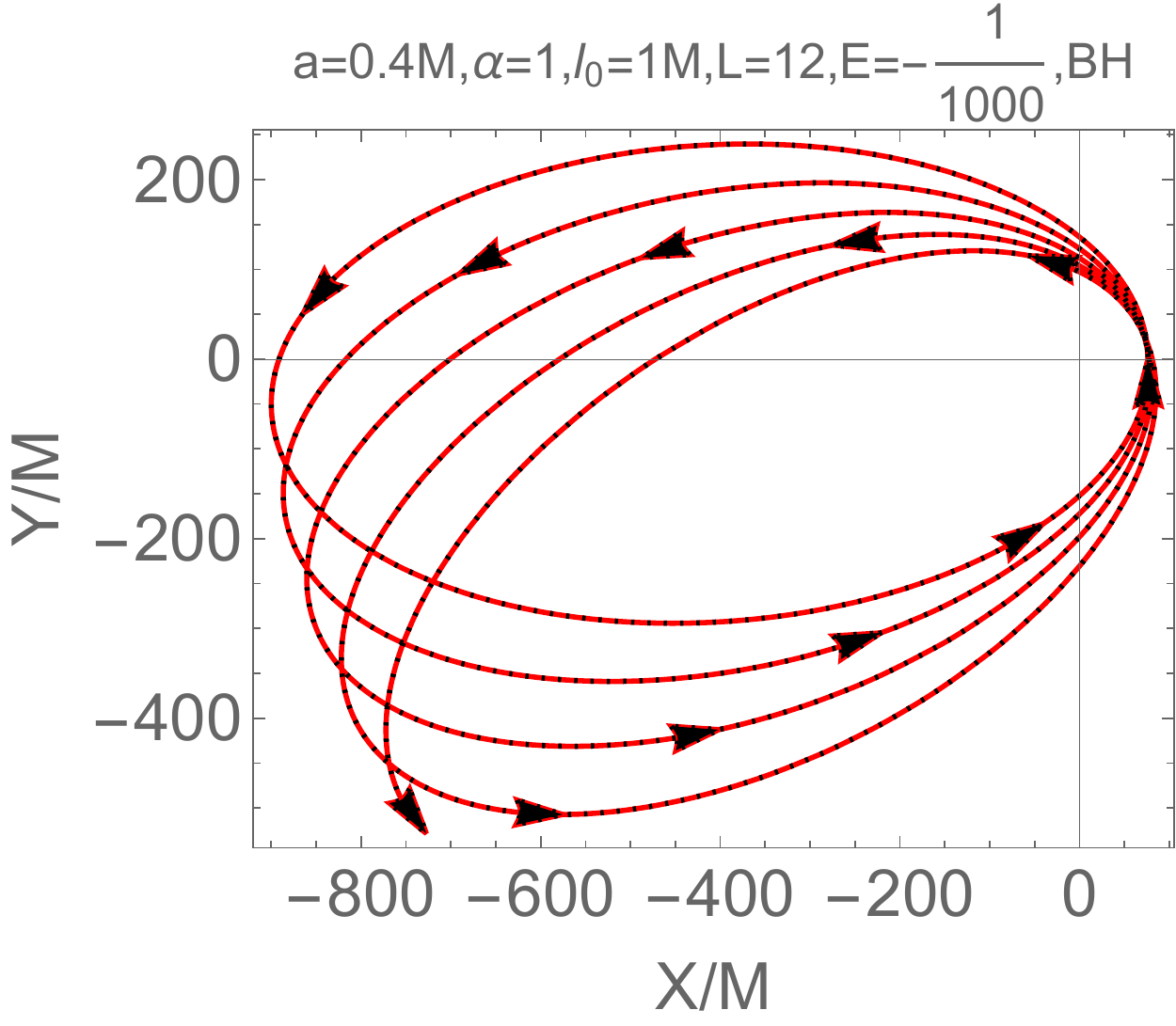}\hspace{0.5cm}
\includegraphics[scale=0.6]{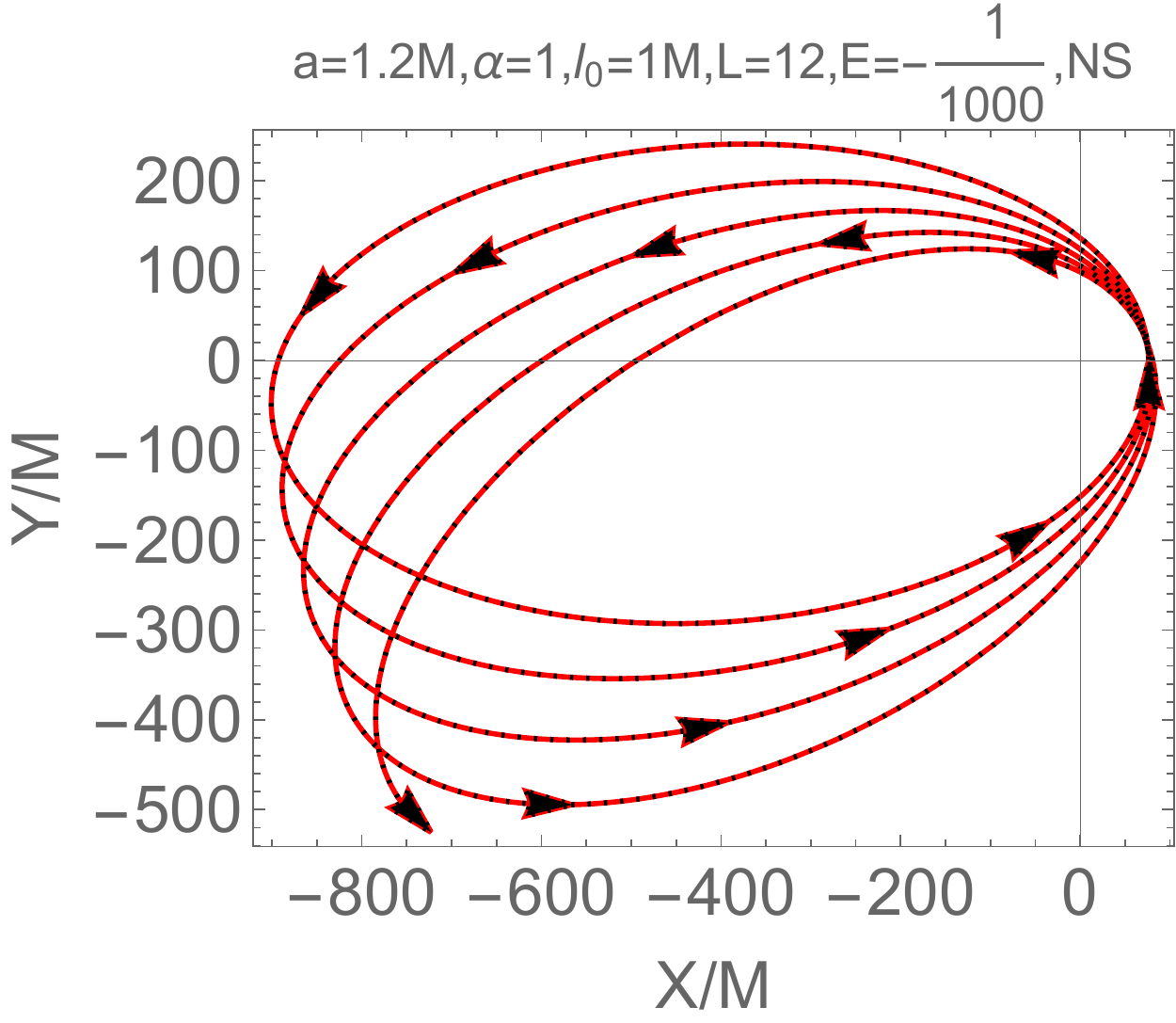}
\caption{The shape of the bound orbit of a test particle freely falling in the hairy Kerr black hole (left panel) and naked singularity spacetime (right panel), respectively.  The black dotted curves indicate the Kerr case with $\alpha=0$. }\label{fig:orbit}}	
\end{figure}

Using the bound orbit condition $V_{min}\leq E<0$ \cite{Bardeen:1972fi}, we will explicitly show the shape of the orbit.
The geodesic could give us how $u=1/r$ changes with respect to $\phi$ in the way 
\begin{eqnarray}
\frac{du}{d\phi}=-u^2\frac{p^r}{p^\phi}=\frac{1+e^{\frac{2}{(l_0-2M)u}}\alpha-2Mu+a^2u^2}{-L+e^{\frac{2}{(l_0-2M)u}}(a\mathcal{E}-L)\alpha+2(L-a\mathcal{E})Mu}X_u,
\end{eqnarray}
with
\begin{eqnarray}
    X_u&=&\sqrt{\mathcal{E}^2-1-e^{\frac{2}{(l_0-2M)u}}\alpha+u(2M+Y_u)},\\
    Y_u&=&u\left(a^2(\mathcal{E}^2-1)-L^2-e^{\frac{2}{(l_0-2M)u}}(L-a\mathcal{E})^2\alpha+2(L-a\mathcal{E})^2Mu\right),
\end{eqnarray}
such that 
\begin{equation}
\frac{d^2u}{d\phi^2}=\left(4(a\mathcal{E}-L)A_uC_u^-+4(C_u^-+a^2(l_0-2M)u^3)B_uX_u^2+Z_u\right)\frac{A_u}{2(l_0-2M)B_u^3}
\end{equation}
with
\begin{eqnarray}
    Z_u&=&B_u A_u(2e^{\frac{2}{(l_0-2M)u}}\alpha+(l_0-2M)u^2(2M+Y_u)+u^2((l_0-2M)Y_u+2(a\mathcal{E}-L)^2C_u^+)),\\
    A_u&=&1+e^{\frac{2}{(l_0-2M)u}}\alpha-2Mu+a^2u^2,\\
    B_u&=&L-e^{\frac{2}{(l_0-2M)u}}\alpha(a\mathcal{E}-L)+2(a\mathcal{E}-L)Mu,\\
    C_u^{\pm}&=&\pm e^{\frac{2}{(l_0-2M)u}}\alpha+(l_0-2M)Mu^2.
\end{eqnarray}
Then by numerically integrating the above orbital equation, we can figure out the shape of the bound orbit of a test particle freely falling in the hairy Kerr BH and NS spacetime. 
We show the bound orbits in Kerr hairy black hole ($a=0.4M$) and naked singularity ($a=1.2M$) in FIG. \ref{fig:orbit}, where we fix $\alpha=1$, $l_0=1M$, $L=12$ and $E=-1/1000$.  In the figure, we also simultaneously plot the corresponding results for Kerr spacetime denoted by black dotted curves, namely with $\alpha=0$. The differences between BH and NS are slight both in hairy Kerr and Kerr spacetimes. 

An interesting type of bound orbit is the circular or spherical orbit which satisfies 
\begin{eqnarray}\label{eq:circular}
V_{eff}\mid_{r_{c}}=0,~~~~\frac{dV_{eff}}{dr}\mid_{r_{c}}=0,
\end{eqnarray}
with $r_c$ the radius of the circular orbit. The bound circular orbit could  either be stable or unstable depending on the sign of $\frac{d^2V_{eff}}{dr^2}$. $\frac{d^2V_{eff}}{dr^2}> 0$ means that the orbit is stable while for $\frac{d^2V_{eff}}{dr^2}<0$ it is unstable. 
Thus, one can define the innermost stable circular orbit (ISCO),  known as the smallest stable marginally bound circular orbit which  satisfies the above conditions accompanying with the vanishing second order derivative  \cite{Bardeen:1972fi}, meaning that the orbital radius, $r_{ISCO}$, is determined by 
\begin{eqnarray}
V_{eff}\mid_{r_{ISCO}}=0,~~~\frac{dV_{eff}}{dr}\mid_{r_{ISCO}}=0,~~~~\frac{d^2V_{eff}}{dr^2}\mid_{r_{ISCO}}= 0.
\end{eqnarray}
 \begin{figure}[H]
\center{
\includegraphics[scale=0.4]{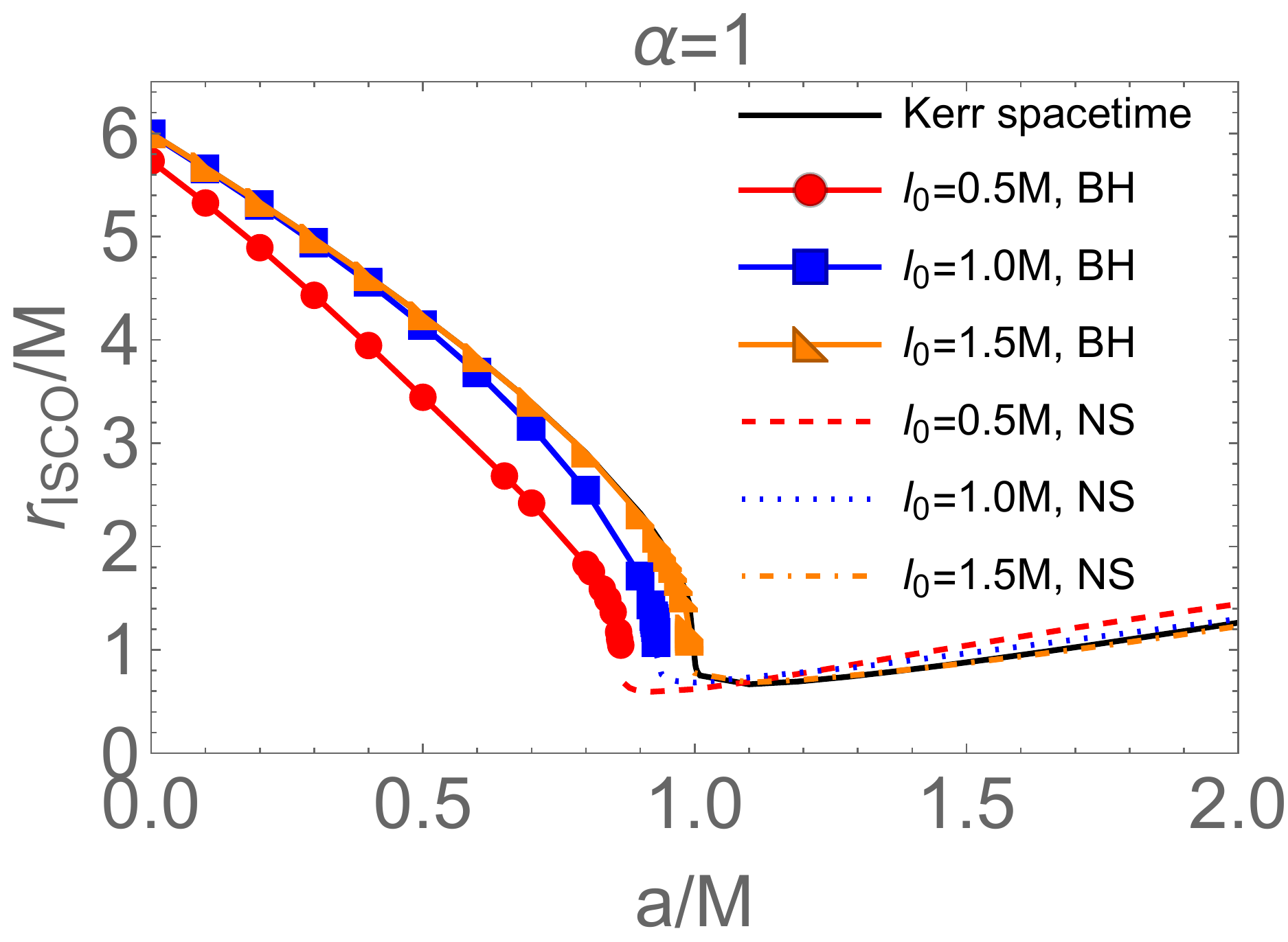}\hspace{0.5cm}
\includegraphics[scale=0.4]{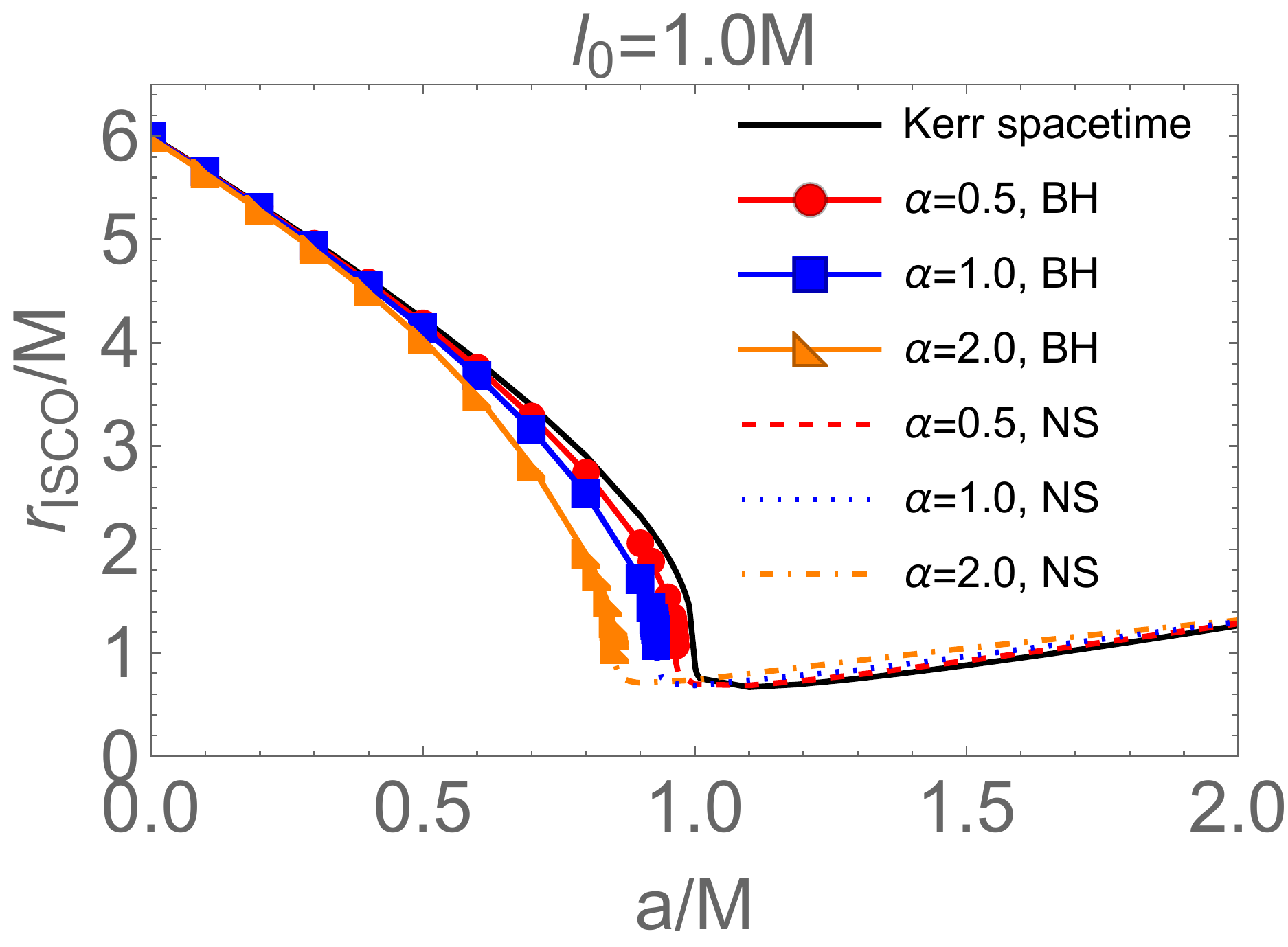}
\caption{ The ISCO of the hairy Kerr spacetime as a function of the spinning parameter. In the left  plot, we fix $\alpha=1$ and focus on the effect of $l_0$, while in the right plot, we fix $l_0=1M$ and tune $\alpha$. In the plots, the solid curves with different shapes are for hairy Kerr black hole, while the dashed, dotted and dotdashed curves are all for naked singularity. }\label{fig:ISCO}	}
\end{figure}
The exact formula of $r_{ISCO}$ for Kerr spacetime was solved out in  \cite{Bardeen:1972fi}, and it was $r_{ISCO}=6M$ as $a=0$ for Schwarzschild spacetime. Again, due to the existence of exponential term, the expression of the  $r_{ISCO}$ for hairy Kerr spacetime becomes difficult, so we  numerically obtain  $r_{ISCO}$ of the hairy Kerr BH and  NS. The values of $r_{ISCO}$ as a function of spin parameter are depicted  in FIG.\ref{fig:ISCO}, from which we see that $r_{ISCO}$  is smaller for faster spinning hairy Kerr BH while it is larger for fast spinning hairy Kerr NS, which is similar to those occur in Kerr spacetime indicated by black curve in each plot. In addition,  $r_{ISCO}$ for hairy Kerr BH increases as $l_0$ increases, but decreases as $\alpha$ increases; while  the dependence of  $r_{ISCO}$ for the hairy NS on the hairy parameters is closely determined by the spinning of the central object.


\subsection{LT precession and periastron precession}
We move on to study the LT precession frequency and the periastron precession of the circular orbit by perturbing the geodesic equation of the massive particle, which could disclose important features of accretion disk  around the central bodies, such that testify the strong gravity of the hairy Kerr spacetime. 

To proceed, we have to model the three fundamental frequencies which are very important for accretion disk physics around the hairy Kerr spacetime. For a test massive particle moving along a circle in the equatorial plane of the metric \eqref{eq-metric}, the orbital angular frequency, $\Omega_{\phi}$, is 
\begin{equation}
 \Omega_{\phi}=\frac{d\phi}{dt}=\frac{-g_{t\phi,r}+\sqrt{(g_{t\phi,r})^2-g_{tt,r} g_{\phi\phi,r}}} {g_{\phi\phi,r}}. 
\end{equation}
If the particle is perturbed, then it oscillates with some characteristic epicyclic frequencies $\Omega_r$ and $\Omega_\theta$ in the radial or vertical direction, respectively, which can be obtained by perturbing the geodesic equation as \cite{Ryan:1995wh,Doneva:2014uma} 
\begin{eqnarray}
\Omega_r=\left\{\frac{1}{2g_{rr}}\left[X^2\partial_r^2\left(\frac{g_{\phi\phi}}{g_{tt}g_{\phi\phi}-g_{t\phi}^2}\right)-2XY\partial_r^2\left(\frac{g_{t\phi}}{g_{tt}g_{\phi\phi}-g_{t\phi}^2}\right)+Y^2\partial_r^2\left(\frac{g_{tt}}{g_{tt}g_{\phi\phi}-g_{t\phi}^2}\right)\right]\right\}^{1/2}, \\
\Omega_\theta=\left\{\frac{1}{2g_{\theta\theta}}\left[X^2\partial_\theta^2\left(\frac{g_{\phi\phi}}{g_{tt}g_{\phi\phi}-g_{t\phi}^2}\right)-2XY\partial_\theta^2\left(\frac{g_{t\phi}}{g_{tt}g_{\phi\phi}-g_{t\phi}^2}\right)+Y^2\partial_\theta^2\left(\frac{g_{tt}}{g_{tt}g_{\phi\phi}-g_{t\phi}^2}\right)\right]\right\}^{1/2} 
\end{eqnarray}
with 
\begin{eqnarray}
X=g_{tt}+g_{t\phi}\Omega_\phi,~~~~~Y=g_{t\phi}+g_{\phi\phi}\Omega_\phi.
\end{eqnarray}
Subsequently, we can extract the nodal  precession frequency $\Omega_{nod}$ and  periastron precession  frequency $\Omega_{per}$ as \cite{Motta:2013wga}
\begin{eqnarray}
\Omega_{nod}=\Omega_{\phi}-\Omega_{\theta}, \\
\Omega_{per}=\Omega_{\phi}-\Omega_{r} 
\end{eqnarray}
which measure the precession of orbital plane  and orbit of the accretion disk, respectively. The nodal precession frequency is also known as LT precession frequency. 

\begin{figure}[H]
\center{
\includegraphics[scale=0.28]{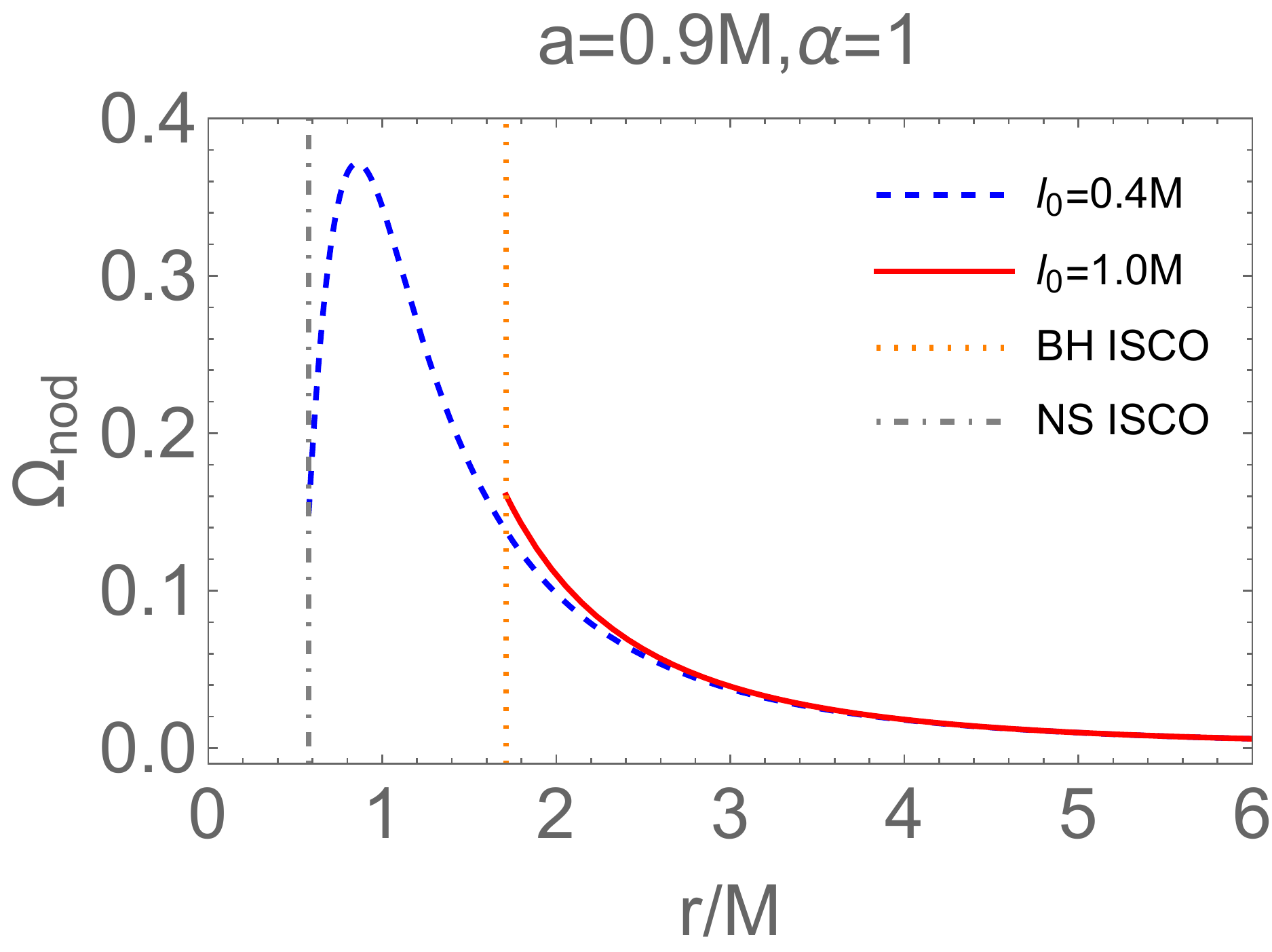}\hspace{0.5cm}
\includegraphics[scale=0.28]{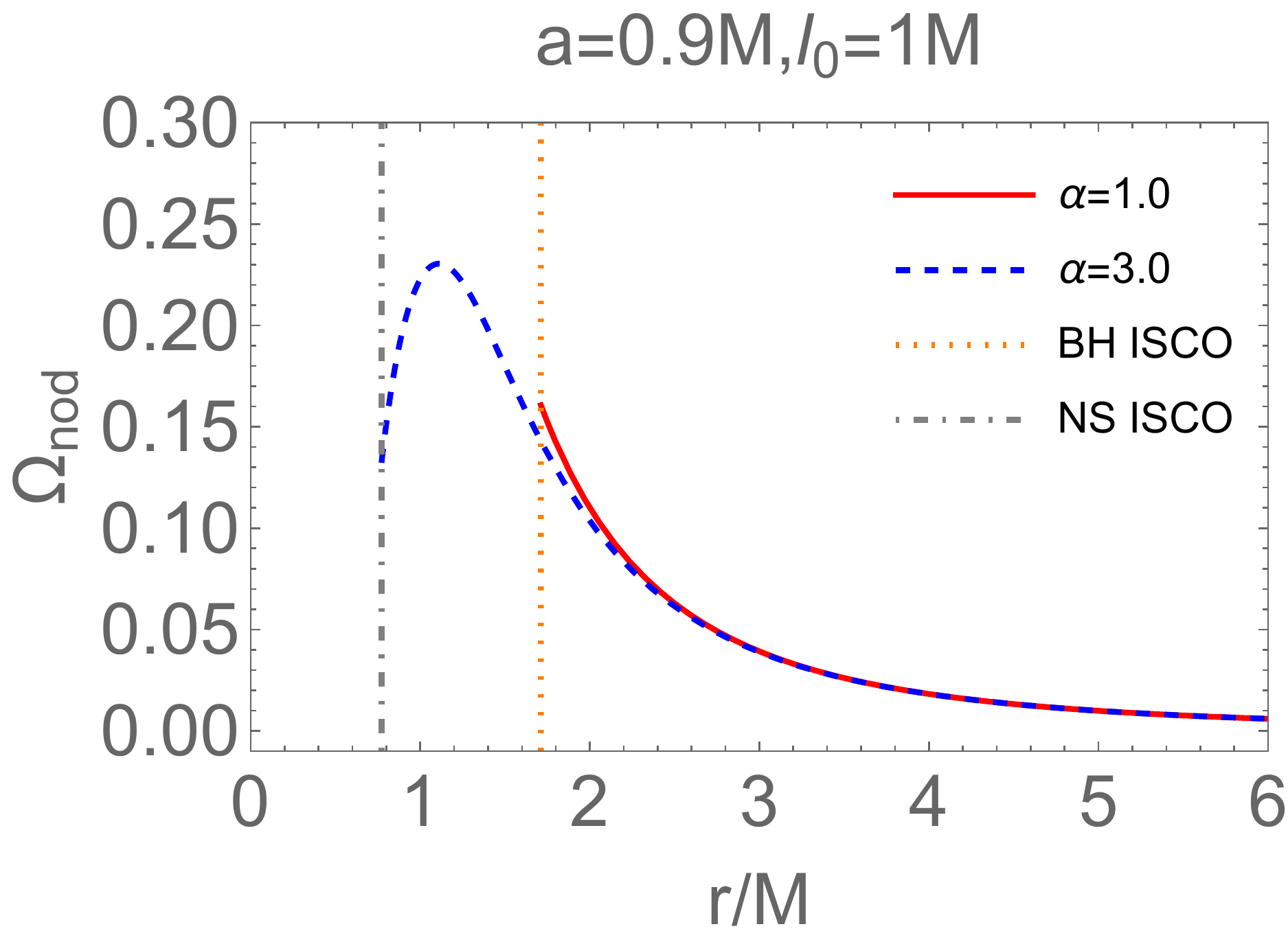}\hspace{0.5cm}
\includegraphics[scale=0.3]{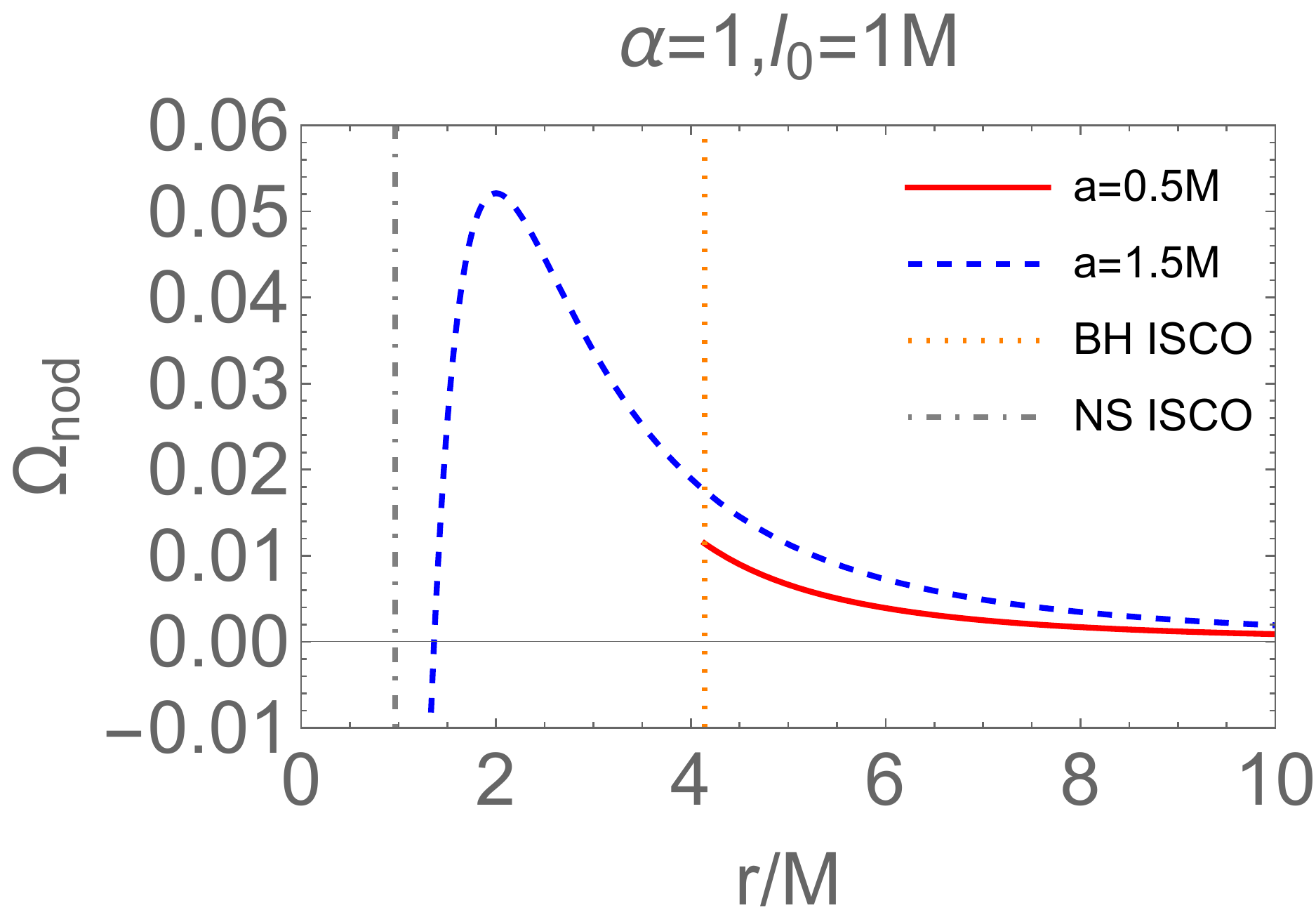}
\includegraphics[scale=0.28]{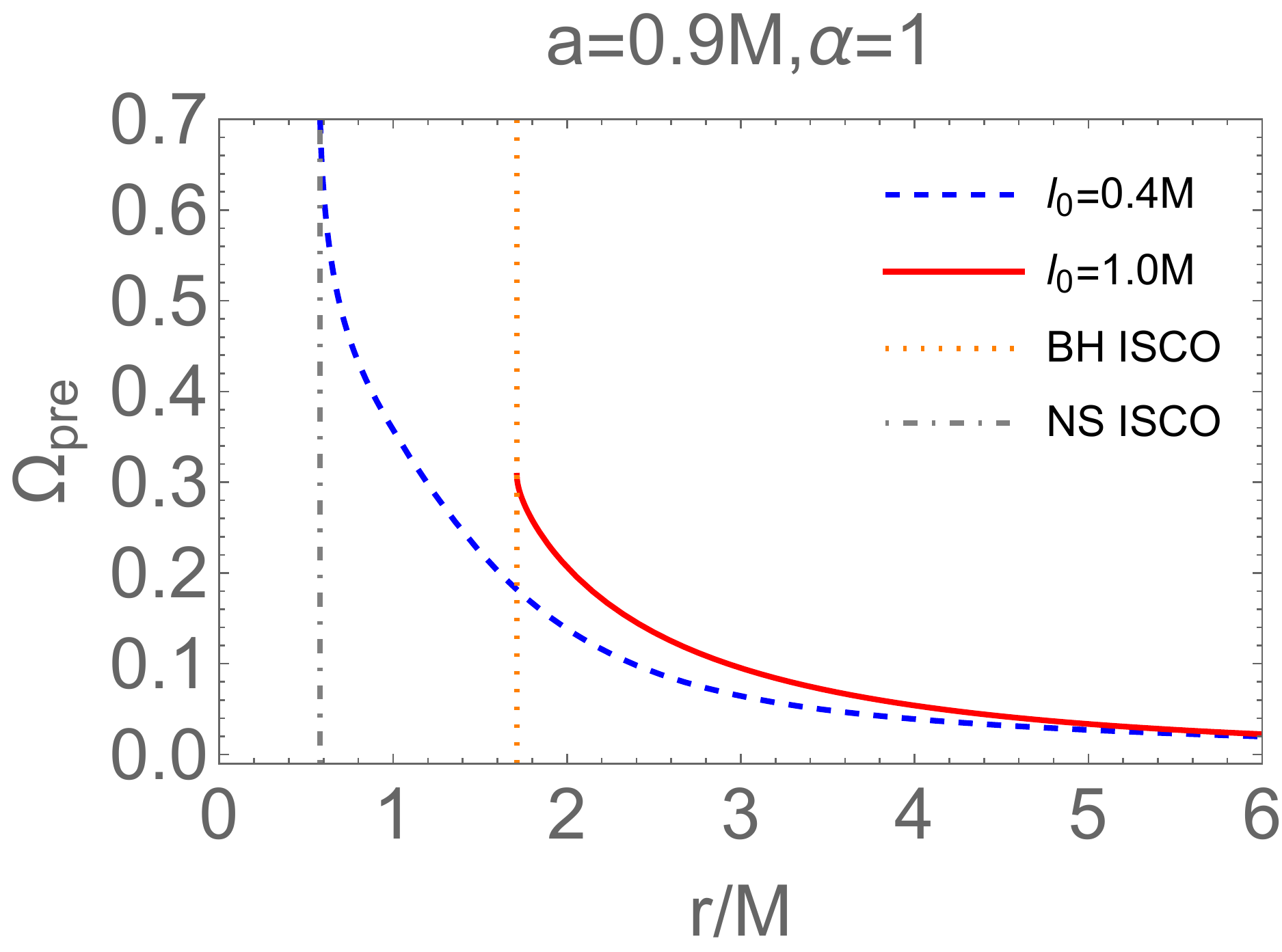}\hspace{0.5cm}
\includegraphics[scale=0.28]{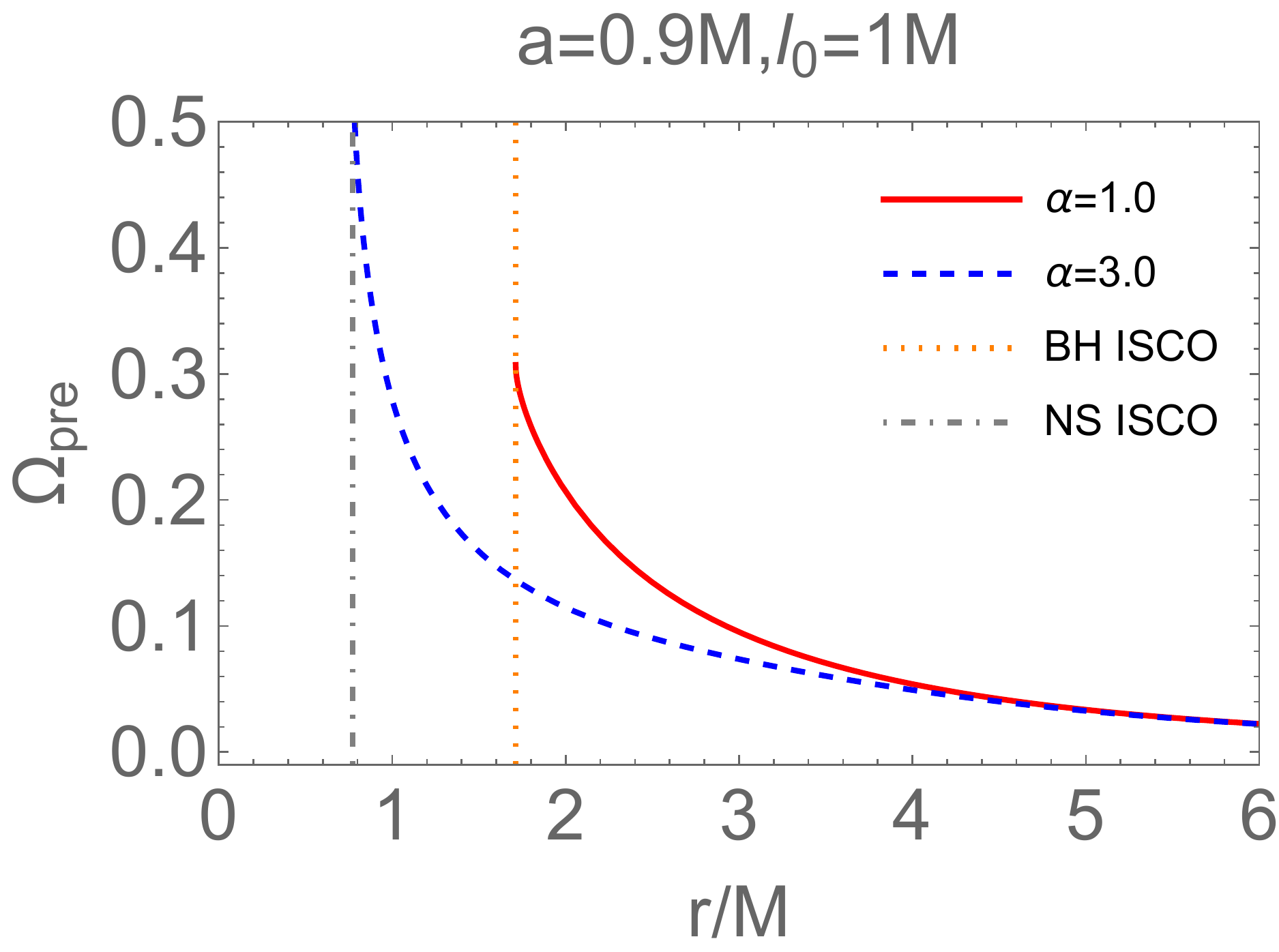}\hspace{0.5cm}
\includegraphics[scale=0.28]{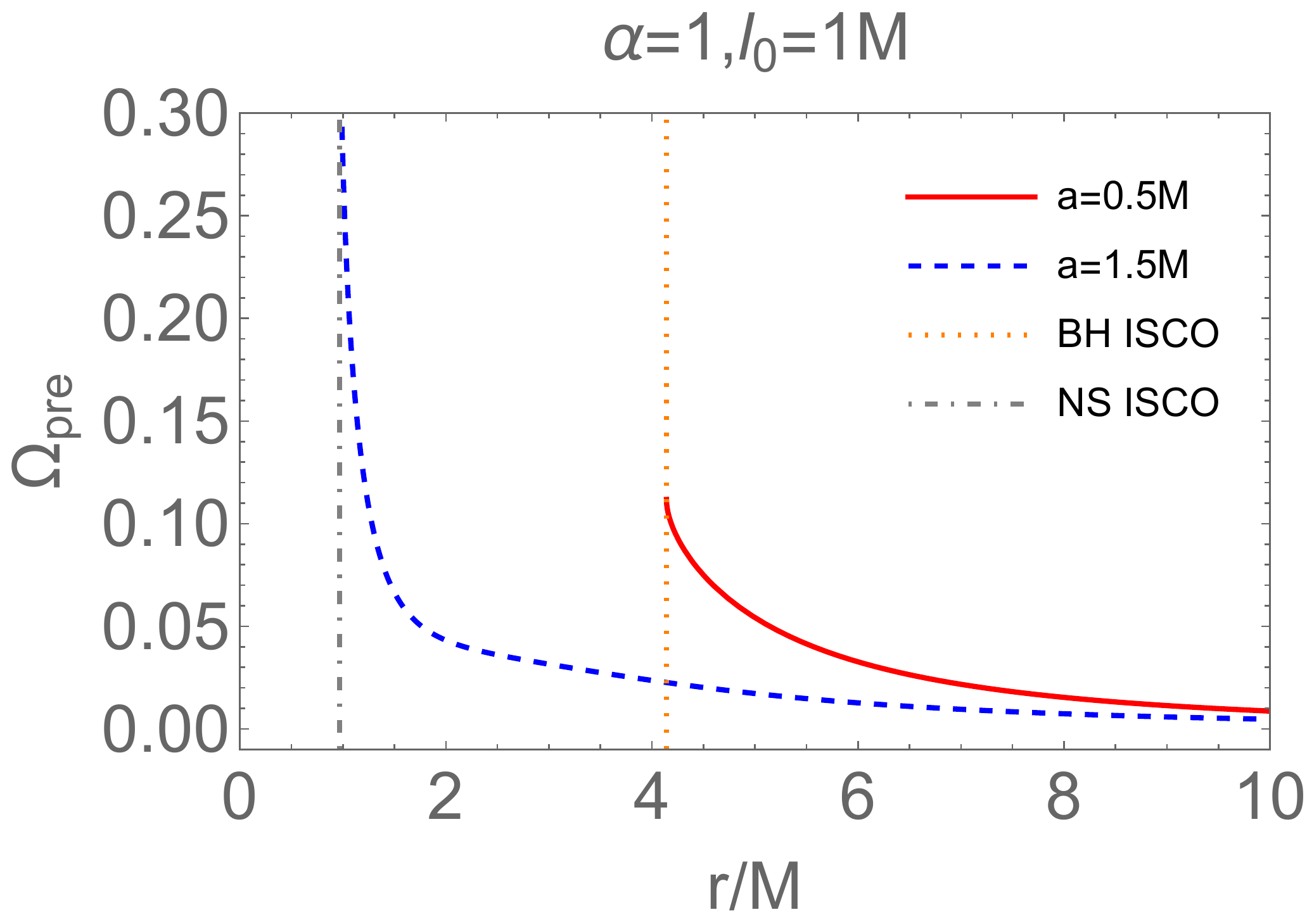}
\caption{The behavior of $\Omega_{nod}$ (upper row) and $\Omega_{pre}$ (bottom row) as a function of $r$ in hairy Kerr black hole and naked singularity, respectively. In all plots, the blue curve is for naked singularity case while the red curve is for black hole case, and the vertical lines indicate the location of the corresponding ISCO. From left to right, the hairy  spacetime is either black hole or naked singularity, depending on $l_0$, $\alpha$ and $a$, respectively.  }\label{fig:OmegaNod}	}	
\end{figure}

\begin{figure}[H]
\center{
\includegraphics[scale=0.4]{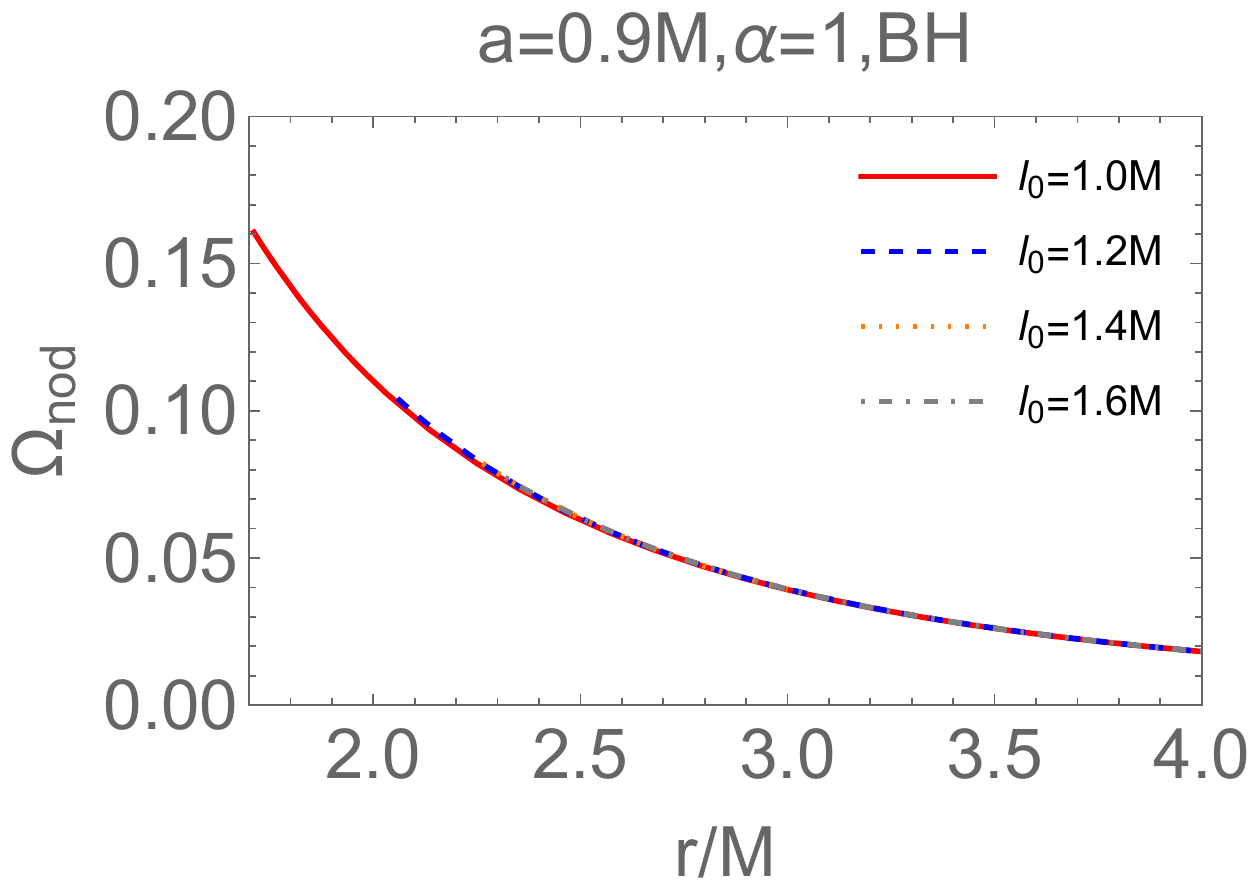}\hspace{0.5cm}
\includegraphics[scale=0.4]{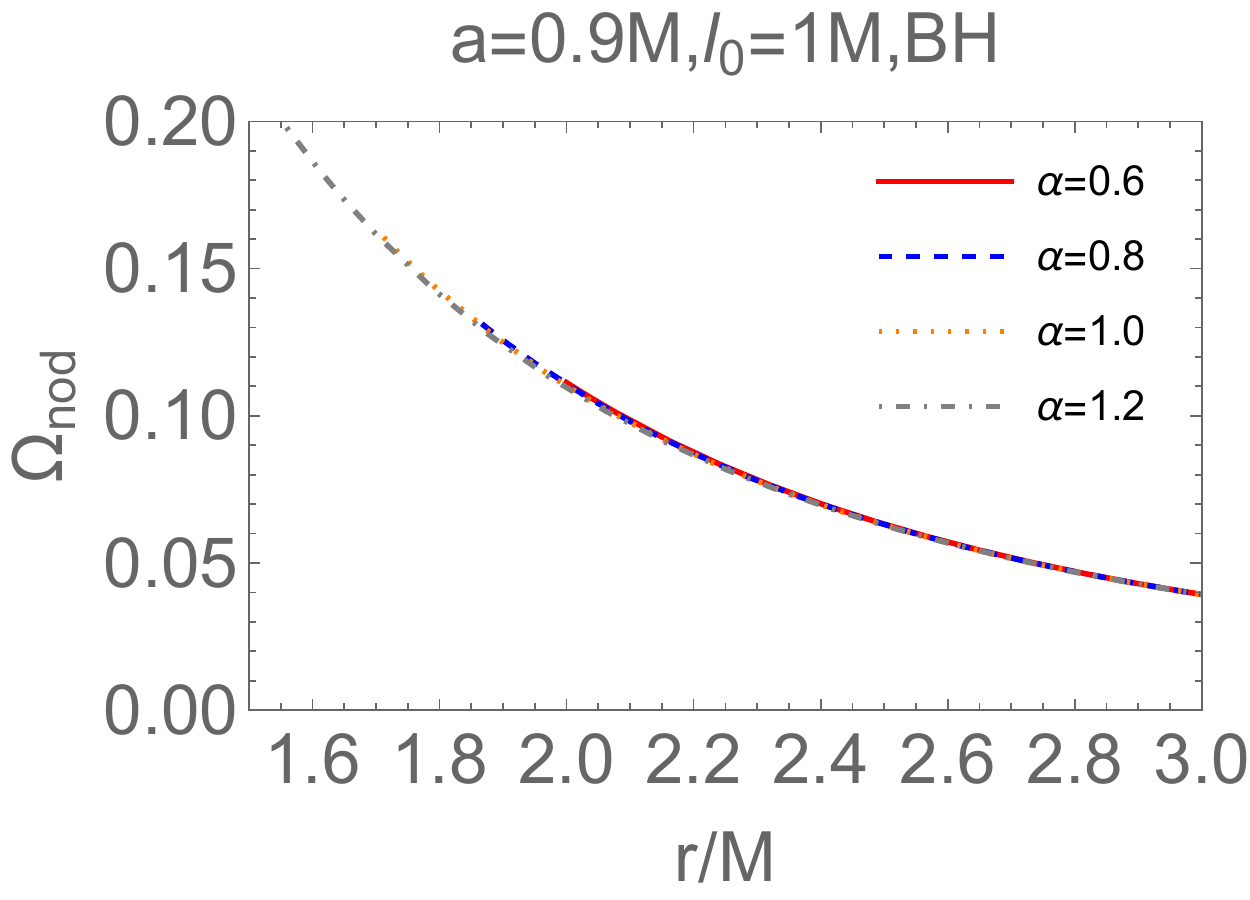}\hspace{0.5cm}
\includegraphics[scale=0.4]{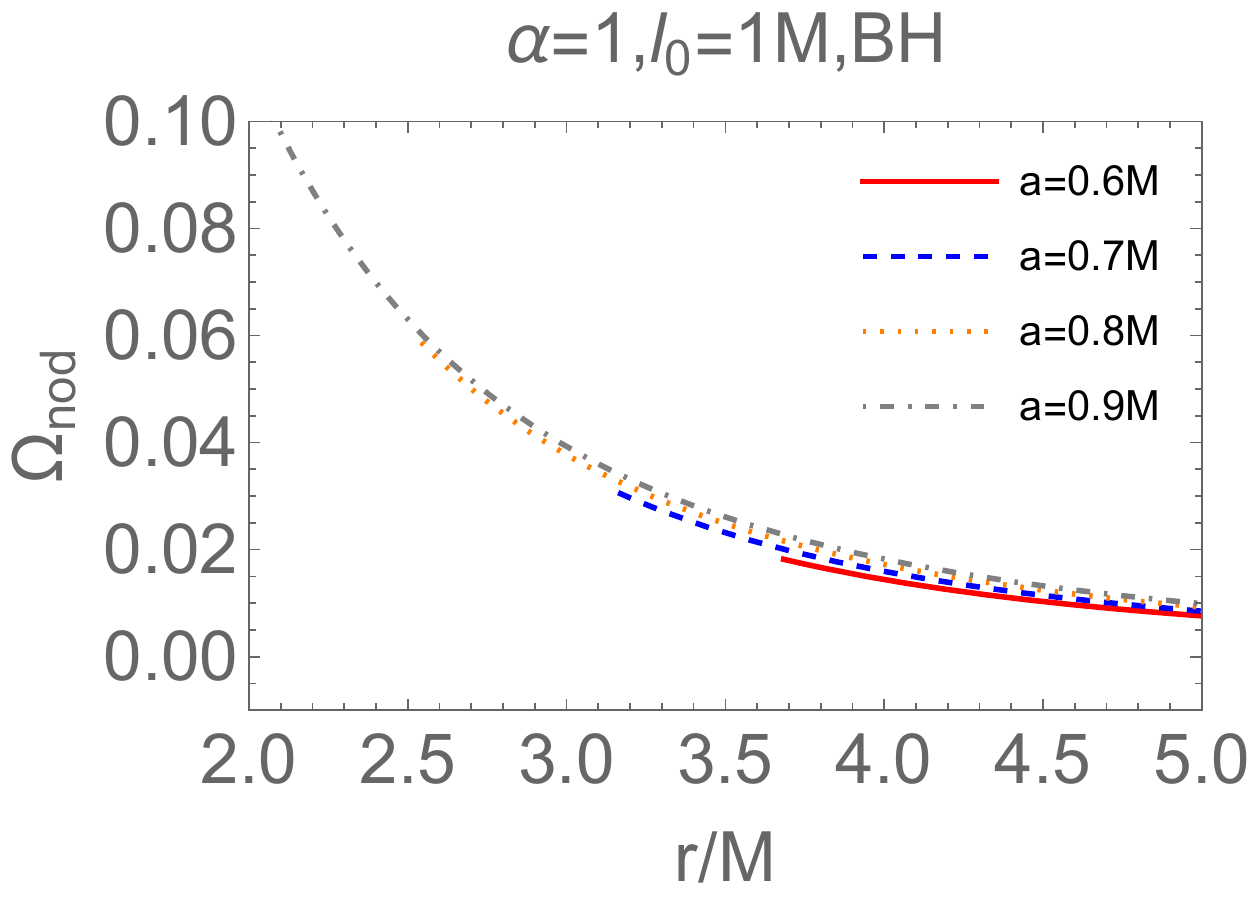}
\includegraphics[scale=0.4]{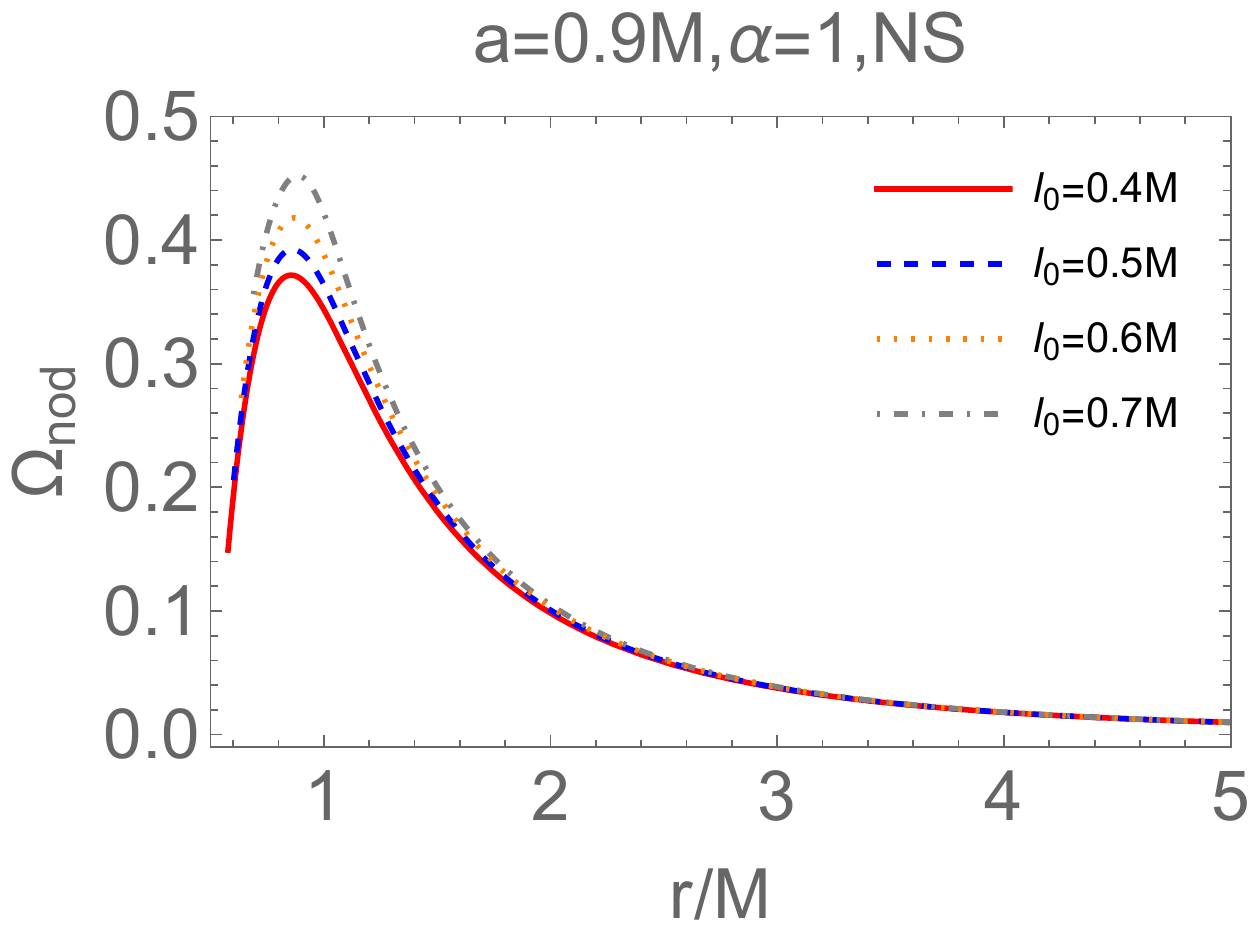}\hspace{0.5cm}
\includegraphics[scale=0.4]{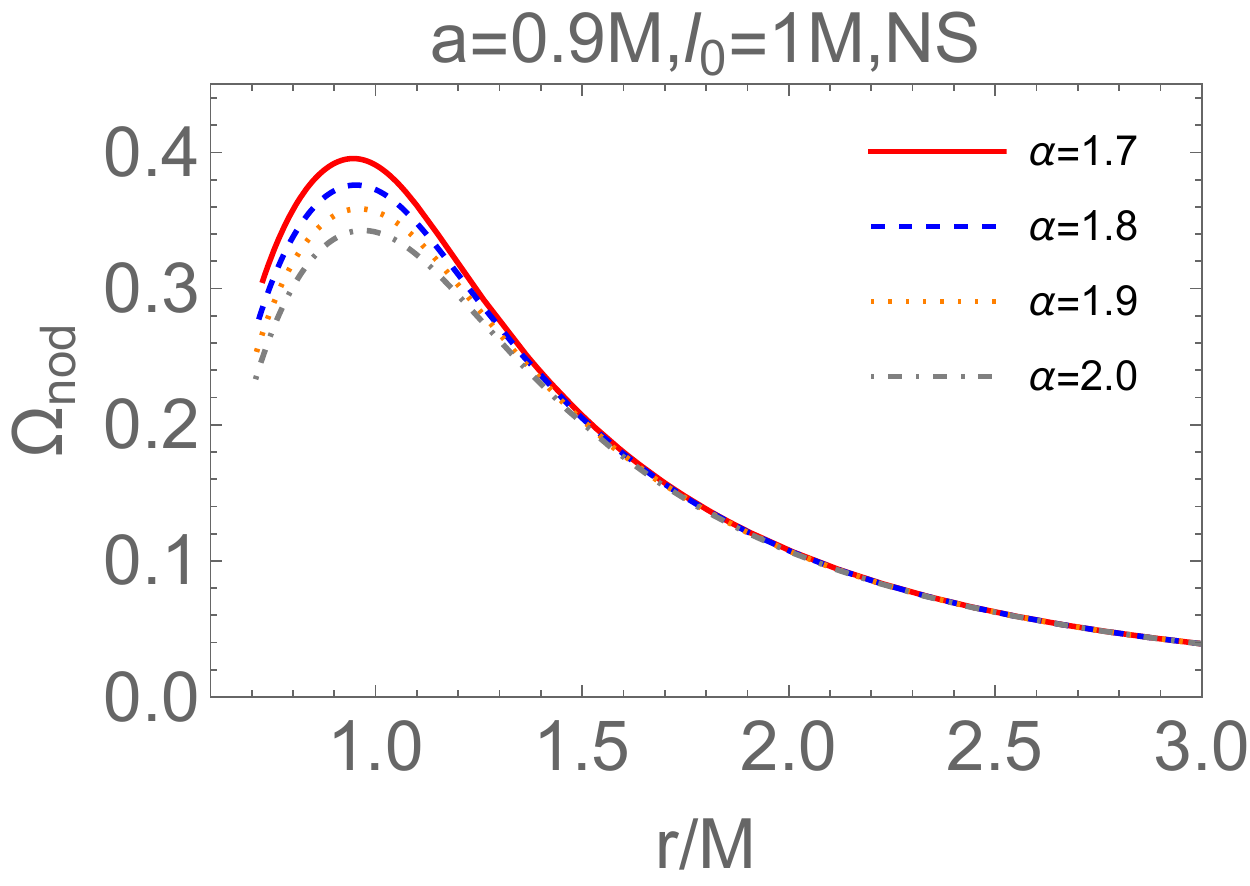}\hspace{0.5cm}
\includegraphics[scale=0.4]{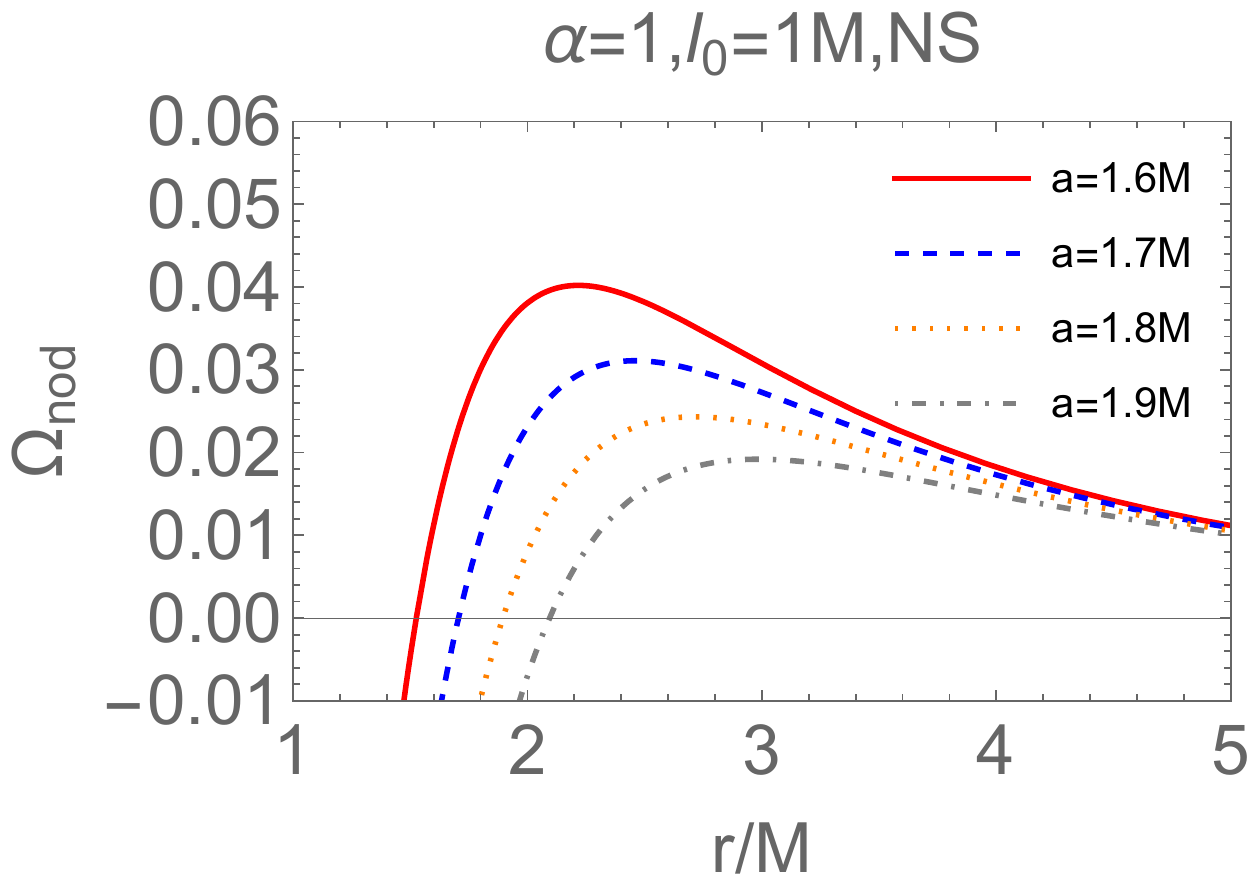}
\caption{The $\Omega_{nod}$  as a function of $r$ for various model parameters. From the plots in the upper row, we can see the effect of $l_0$ (left), $\alpha$ (middle) and $a$ (right) on  $\Omega_{nod}$  in the hairy Kerr black hole. In the bottom row, we see their effects on  $\Omega_{nod}$  in hairy naked singularity.  }\label{fig:OmegaNodBHNS1}	}	
\end{figure}

Focusing on the equatorial plane with $\theta=\pi/2$, we plot $\Omega_{nod}$  and  $\Omega_{pre}$ for hairy Kerr BH and NS with samples of parameters in  FIG.\ref{fig:OmegaNod}, in which the red curves are for NS while the blue curves are for BH, and the vertical lines correspond to the corresponding  location of ISCO. It is obvious that the LT precession frequency increases monotonously as the orbit approaches the ISCO of the hairy Kerr BH. While in the hairy NS spacetime, as the orbit moves towards the ISCO,  $\Omega_{nod}$ first increases to certain peak and then decreases. In addition, $\Omega_{nod}$ at the ISCO of NS is always smaller than that at ISCO of BH, and it even can be negative indicating a reversion of the precession direction.
The periastron precession frequency $\Omega_{per}$ in hairy Kerr BH and NS spacetimes has similar behavior that is increasing with the decrease of $r$. The value of  $\Omega_{per}$ for hairy Kerr BH is always larger than that for NS, and their difference is more significant as the orbit becomes smaller. The effect of different model
parameters on the LT precession frequency are shown in FIG.\ref{fig:OmegaNodBHNS1} which indicates that the  parameters indeed have significant imprint on the LT  frequencies.


\section{Conclusion and discussion}\label{sec:conclusion}
 Gravitational wave and shadow  are two important observations to test strong gravity physics, such that they  allow
us to probe the structure of black holes.  Thus, they could also be used to reveal the scalar fields provided that they leave an
imprint on BH. The well known no-hair theorem in Einstein's GR predicts that  the rotating black holes are described
by the Kerr metric. However, beyond GR with additional surrounding sources, the  hairy rotating black holes should be described by a deformed Kerr metric, including extra hairy parameter.
A hairy Kerr black hole was recently constructed using  the gravitational decoupling approach,  describing deformations of Kerr black hole due to including  additional sources \cite{Contreras:2021yxe},
 Observational investigations related with gravitational wave and shadow  of the hairy Kerr black hole have been studied  in \cite{Cavalcanti:2022cga,Yang:2022ifo,Li:2022hkq,Islam:2021dyk,Afrin:2021imp}, which indicates that the hairy Kerr black hole could not be ruled out by the current observations. In this paper,
 we focused on the Lense-Thirring effect, another important observable effect, to differentiate the hairy Kerr black hole from  naked singularity.

Firstly, we analyzed the spin precession of a test gyro attached to a stationary observer in the hairy Kerr spacetime. When the observer is static with respect to a fixed star, i.e, the the angular velocity vanishes, we calculated the LT precession frequency. It was found that the LT precession frequency diverges as the observer approaches the ergosphere of the hairy Kerr BH  along any direction, while  it keeps finite in the whole region of the hairy NS,  except at the ring singularity. Then, we parameterized the range of angular velocity of a stationary observer by $k$, and systematically studied the general spin precession frequency. The general spin precession frequency diverges as the observer approaches the horizon of the hairy Kerr BH, but it is finite when $k=0.5$ defining ZAMO observer because in this case the test gyro has no rotation with respect to the local geometry. For the hairy NS, it is always finite unless the observer reaches the ring  along the direction $\theta=\pi/2$.  We also obtained  the geodetic precession for observers in a hairy static black hole. The general spin frequency, LT frequency and geodetic frequency all decrease as the parameter $\alpha ~(l_0)$  increases (decreases) in hairy Kerr black hole. And $\alpha$ and $l_0$ have similar effect on the LT frequency in hairy NS as that in BH case, but their effects on general spin frequency in NS depend on the spinning.

Then, we investigated  the quasiperiodic oscillations (QPOs) phenomena as the accretion disk approaches the hairy Kerr BH and NS, which also show difference.  To this end, we first analyzed the orbital precession of bound orbits, and ISCO of a test massive particle orbiting in the equatorial plane of the hairy Kerr BH and NS spacetime, respectively.  Then we perturbed the stable circular orbit and computed the three fundamental frequencies related with QPOs phenomena. Accordingly, our results show that as the orbit moves towards the ISCO, the LT  frequency increases monotonously in hairy Kerr BH,  while it first increases to certain peak and then decreases in hair NS;  the periastron frequency increases in both hairy Kerr BH and NS as the orbit approaches the corresponding ISCO. Moreover, the hairy parameters indeed have  effects on the LT and periastron frequencies.

In conclusion, we do theoretical evaluation on various precession frequencies caused by the frame-dragging effect of the central sources, which differentiate the hairy Kerr BH from NS spacetime.  We expect that our theoretical studies could shed light on  astrophysical observations on distinguishing hairy theory from GR, distinguishing BH  from NS and even further constraining the hairy parameters.

\begin{acknowledgments}
This work is partly supported by Natural Science Foundation of Jiangsu Province under Grant No.BK20211601, Fok Ying Tung Education Foundation under Grant No. 171006.
\end{acknowledgments}

\bibliography{ref}

\begin{thebibliography}{71}
\expandafter\ifx\csname natexlab\endcsname\relax\def\natexlab#1{#1}\fi
\expandafter\ifx\csname bibnamefont\endcsname\relax
  \def\bibnamefont#1{#1}\fi
\expandafter\ifx\csname bibfnamefont\endcsname\relax
  \def\bibfnamefont#1{#1}\fi
\expandafter\ifx\csname citenamefont\endcsname\relax
  \def\citenamefont#1{#1}\fi
\expandafter\ifx\csname url\endcsname\relax
  \def\url#1{\texttt{#1}}\fi
\expandafter\ifx\csname urlprefix\endcsname\relax\def\urlprefix{URL }\fi
\providecommand{\bibinfo}[2]{#2}
\providecommand{\eprint}[2][]{\url{#2}}

\bibitem[{\citenamefont{Abbott et~al.}(2016)}]{LIGOScientific:2016aoc}
\bibinfo{author}{\bibfnamefont{B.~P.} \bibnamefont{Abbott}}
  \bibnamefont{et~al.} (\bibinfo{collaboration}{LIGO Scientific, Virgo}),
  \bibinfo{journal}{Phys. Rev. Lett.} \textbf{\bibinfo{volume}{116}},
  \bibinfo{pages}{061102} (\bibinfo{year}{2016}), \eprint{1602.03837}.

\bibitem[{\citenamefont{Abbott et~al.}(2019)}]{LIGOScientific:2018mvr}
\bibinfo{author}{\bibfnamefont{B.~P.} \bibnamefont{Abbott}}
  \bibnamefont{et~al.} (\bibinfo{collaboration}{LIGO Scientific, Virgo}),
  \bibinfo{journal}{Phys. Rev. X} \textbf{\bibinfo{volume}{9}},
  \bibinfo{pages}{031040} (\bibinfo{year}{2019}), \eprint{1811.12907}.

\bibitem[{\citenamefont{Abbott et~al.}(2020)}]{LIGOScientific:2020aai}
\bibinfo{author}{\bibfnamefont{B.~P.} \bibnamefont{Abbott}}
  \bibnamefont{et~al.} (\bibinfo{collaboration}{LIGO Scientific, Virgo}),
  \bibinfo{journal}{Astrophys. J. Lett.} \textbf{\bibinfo{volume}{892}},
  \bibinfo{pages}{L3} (\bibinfo{year}{2020}), \eprint{2001.01761}.

\bibitem[{\citenamefont{Akiyama
  et~al.}(2019{\natexlab{a}})}]{EventHorizonTelescope:2019dse}
\bibinfo{author}{\bibfnamefont{K.}~\bibnamefont{Akiyama}} \bibnamefont{et~al.}
  (\bibinfo{collaboration}{Event Horizon Telescope}),
  \bibinfo{journal}{Astrophys. J. Lett.} \textbf{\bibinfo{volume}{875}},
  \bibinfo{pages}{L1} (\bibinfo{year}{2019}{\natexlab{a}}),
  \eprint{1906.11238}.

\bibitem[{\citenamefont{Akiyama
  et~al.}(2019{\natexlab{b}})}]{EventHorizonTelescope:2019ths}
\bibinfo{author}{\bibfnamefont{K.}~\bibnamefont{Akiyama}} \bibnamefont{et~al.}
  (\bibinfo{collaboration}{Event Horizon Telescope}),
  \bibinfo{journal}{Astrophys. J. Lett.} \textbf{\bibinfo{volume}{875}},
  \bibinfo{pages}{L4} (\bibinfo{year}{2019}{\natexlab{b}}),
  \eprint{1906.11241}.

\bibitem[{\citenamefont{Akiyama
  et~al.}(2019{\natexlab{c}})}]{EventHorizonTelescope:2019pgp}
\bibinfo{author}{\bibfnamefont{K.}~\bibnamefont{Akiyama}} \bibnamefont{et~al.}
  (\bibinfo{collaboration}{Event Horizon Telescope}),
  \bibinfo{journal}{Astrophys. J. Lett.} \textbf{\bibinfo{volume}{875}},
  \bibinfo{pages}{L5} (\bibinfo{year}{2019}{\natexlab{c}}),
  \eprint{1906.11242}.

\bibitem[{\citenamefont{Akiyama
  et~al.}(2022{\natexlab{a}})}]{EventHorizonTelescope:2022xnr}
\bibinfo{author}{\bibfnamefont{K.}~\bibnamefont{Akiyama}} \bibnamefont{et~al.}
  (\bibinfo{collaboration}{Event Horizon Telescope}),
  \bibinfo{journal}{Astrophys. J. Lett.} \textbf{\bibinfo{volume}{930}},
  \bibinfo{pages}{L12} (\bibinfo{year}{2022}{\natexlab{a}}).

\bibitem[{\citenamefont{Akiyama
  et~al.}(2022{\natexlab{b}})}]{EventHorizonTelescope:2022xqj}
\bibinfo{author}{\bibfnamefont{K.}~\bibnamefont{Akiyama}} \bibnamefont{et~al.}
  (\bibinfo{collaboration}{Event Horizon Telescope}),
  \bibinfo{journal}{Astrophys. J. Lett.} \textbf{\bibinfo{volume}{930}},
  \bibinfo{pages}{L17} (\bibinfo{year}{2022}{\natexlab{b}}).

\bibitem[{\citenamefont{{D. James et al.}}(2019)}]{2019clrp.2020...32D}
\bibinfo{author}{\bibnamefont{{D. James et al.}}}, in
  \emph{\bibinfo{booktitle}{Canadian Long Range Plan for Astronomy and
  Astrophysics White Papers}} (\bibinfo{year}{2019}), vol.
  \bibinfo{volume}{2020}, p.~\bibinfo{pages}{32}, \eprint{1911.01517}.

\bibitem[{\citenamefont{Skidmore et~al.}(2015)}]{TMT:2015pvw}
\bibinfo{author}{\bibfnamefont{W.}~\bibnamefont{Skidmore}} \bibnamefont{et~al.}
  (\bibinfo{collaboration}{TMT International Science Development Teams \& TMT
  Science Advisory Committee}), \bibinfo{journal}{Res. Astron. Astrophys.}
  \textbf{\bibinfo{volume}{15}}, \bibinfo{pages}{1945} (\bibinfo{year}{2015}),
  \eprint{1505.01195}.

\bibitem[{\citenamefont{Ovalle et~al.}(2021)\citenamefont{Ovalle, Casadio,
  Contreras, and Sotomayor}}]{Ovalle:2020kpd}
\bibinfo{author}{\bibfnamefont{J.}~\bibnamefont{Ovalle}},
  \bibinfo{author}{\bibfnamefont{R.}~\bibnamefont{Casadio}},
  \bibinfo{author}{\bibfnamefont{E.}~\bibnamefont{Contreras}},
  \bibnamefont{and}
  \bibinfo{author}{\bibfnamefont{A.}~\bibnamefont{Sotomayor}},
  \bibinfo{journal}{Phys. Dark Univ.} \textbf{\bibinfo{volume}{31}},
  \bibinfo{pages}{100744} (\bibinfo{year}{2021}), \eprint{2006.06735}.

\bibitem[{\citenamefont{Contreras et~al.}(2021)\citenamefont{Contreras, Ovalle,
  and Casadio}}]{Contreras:2021yxe}
\bibinfo{author}{\bibfnamefont{E.}~\bibnamefont{Contreras}},
  \bibinfo{author}{\bibfnamefont{J.}~\bibnamefont{Ovalle}}, \bibnamefont{and}
  \bibinfo{author}{\bibfnamefont{R.}~\bibnamefont{Casadio}},
  \bibinfo{journal}{Phys. Rev. D} \textbf{\bibinfo{volume}{103}},
  \bibinfo{pages}{044020} (\bibinfo{year}{2021}), \eprint{2101.08569}.

\bibitem[{\citenamefont{Mahapatra and Banerjee}(2022)}]{Mahapatra:2022xea}
\bibinfo{author}{\bibfnamefont{S.}~\bibnamefont{Mahapatra}} \bibnamefont{and}
  \bibinfo{author}{\bibfnamefont{I.}~\bibnamefont{Banerjee}}
  (\bibinfo{year}{2022}), \eprint{2208.05796}.

\bibitem[{\citenamefont{Cavalcanti et~al.}(2022)\citenamefont{Cavalcanti,
  de~Paiva, and da~Rocha}}]{Cavalcanti:2022cga}
\bibinfo{author}{\bibfnamefont{R.~T.} \bibnamefont{Cavalcanti}},
  \bibinfo{author}{\bibfnamefont{R.~C.} \bibnamefont{de~Paiva}},
  \bibnamefont{and} \bibinfo{author}{\bibfnamefont{R.}~\bibnamefont{da~Rocha}},
  \bibinfo{journal}{Eur. Phys. J. Plus} \textbf{\bibinfo{volume}{137}},
  \bibinfo{pages}{1185} (\bibinfo{year}{2022}), \eprint{2203.08740}.

\bibitem[{\citenamefont{Yang et~al.}(2022)\citenamefont{Yang, Liu, \"Ovg\"un,
  Long, and Xu}}]{Yang:2022ifo}
\bibinfo{author}{\bibfnamefont{Y.}~\bibnamefont{Yang}},
  \bibinfo{author}{\bibfnamefont{D.}~\bibnamefont{Liu}},
  \bibinfo{author}{\bibfnamefont{A.}~\bibnamefont{\"Ovg\"un}},
  \bibinfo{author}{\bibfnamefont{Z.-W.} \bibnamefont{Long}}, \bibnamefont{and}
  \bibinfo{author}{\bibfnamefont{Z.}~\bibnamefont{Xu}} (\bibinfo{year}{2022}),
  \eprint{2203.11551}.

\bibitem[{\citenamefont{Li}(2022)}]{Li:2022hkq}
\bibinfo{author}{\bibfnamefont{Z.}~\bibnamefont{Li}} (\bibinfo{year}{2022}),
  \eprint{2212.08112}.

\bibitem[{\citenamefont{Islam and Ghosh}(2021)}]{Islam:2021dyk}
\bibinfo{author}{\bibfnamefont{S.~U.} \bibnamefont{Islam}} \bibnamefont{and}
  \bibinfo{author}{\bibfnamefont{S.~G.} \bibnamefont{Ghosh}},
  \bibinfo{journal}{Phys. Rev. D} \textbf{\bibinfo{volume}{103}},
  \bibinfo{pages}{124052} (\bibinfo{year}{2021}), \eprint{2102.08289}.

\bibitem[{\citenamefont{Afrin et~al.}(2021)\citenamefont{Afrin, Kumar, and
  Ghosh}}]{Afrin:2021imp}
\bibinfo{author}{\bibfnamefont{M.}~\bibnamefont{Afrin}},
  \bibinfo{author}{\bibfnamefont{R.}~\bibnamefont{Kumar}}, \bibnamefont{and}
  \bibinfo{author}{\bibfnamefont{S.~G.} \bibnamefont{Ghosh}},
  \bibinfo{journal}{Mon. Not. Roy. Astron. Soc.}
  \textbf{\bibinfo{volume}{504}}, \bibinfo{pages}{5927} (\bibinfo{year}{2021}),
  \eprint{2103.11417}.

\bibitem[{\citenamefont{Joshi et~al.}(2011)\citenamefont{Joshi, Malafarina, and
  Narayan}}]{Joshi:2011zm}
\bibinfo{author}{\bibfnamefont{P.~S.} \bibnamefont{Joshi}},
  \bibinfo{author}{\bibfnamefont{D.}~\bibnamefont{Malafarina}},
  \bibnamefont{and} \bibinfo{author}{\bibfnamefont{R.}~\bibnamefont{Narayan}},
  \bibinfo{journal}{Class. Quant. Grav.} \textbf{\bibinfo{volume}{28}},
  \bibinfo{pages}{235018} (\bibinfo{year}{2011}), \eprint{1106.5438}.

\bibitem[{\citenamefont{Shaikh et~al.}(2019)\citenamefont{Shaikh, Kocherlakota,
  Narayan, and Joshi}}]{Shaikh:2018lcc}
\bibinfo{author}{\bibfnamefont{R.}~\bibnamefont{Shaikh}},
  \bibinfo{author}{\bibfnamefont{P.}~\bibnamefont{Kocherlakota}},
  \bibinfo{author}{\bibfnamefont{R.}~\bibnamefont{Narayan}}, \bibnamefont{and}
  \bibinfo{author}{\bibfnamefont{P.~S.} \bibnamefont{Joshi}},
  \bibinfo{journal}{Mon. Not. Roy. Astron. Soc.}
  \textbf{\bibinfo{volume}{482}}, \bibinfo{pages}{52} (\bibinfo{year}{2019}),
  \eprint{1802.08060}.

\bibitem[{\citenamefont{Joshi et~al.}(2020)\citenamefont{Joshi, Dey, Joshi, and
  Bambhaniya}}]{Joshi:2020tlq}
\bibinfo{author}{\bibfnamefont{A.~B.} \bibnamefont{Joshi}},
  \bibinfo{author}{\bibfnamefont{D.}~\bibnamefont{Dey}},
  \bibinfo{author}{\bibfnamefont{P.~S.} \bibnamefont{Joshi}}, \bibnamefont{and}
  \bibinfo{author}{\bibfnamefont{P.}~\bibnamefont{Bambhaniya}},
  \bibinfo{journal}{Phys. Rev. D} \textbf{\bibinfo{volume}{102}},
  \bibinfo{pages}{024022} (\bibinfo{year}{2020}), \eprint{2004.06525}.

\bibitem[{\citenamefont{Dey et~al.}(2020)\citenamefont{Dey, Shaikh, and
  Joshi}}]{Dey:2020haf}
\bibinfo{author}{\bibfnamefont{D.}~\bibnamefont{Dey}},
  \bibinfo{author}{\bibfnamefont{R.}~\bibnamefont{Shaikh}}, \bibnamefont{and}
  \bibinfo{author}{\bibfnamefont{P.~S.} \bibnamefont{Joshi}},
  \bibinfo{journal}{Phys. Rev. D} \textbf{\bibinfo{volume}{102}},
  \bibinfo{pages}{044042} (\bibinfo{year}{2020}), \eprint{2003.06810}.

\bibitem[{\citenamefont{Dey et~al.}(2021)\citenamefont{Dey, Shaikh, and
  Joshi}}]{Dey:2020bgo}
\bibinfo{author}{\bibfnamefont{D.}~\bibnamefont{Dey}},
  \bibinfo{author}{\bibfnamefont{R.}~\bibnamefont{Shaikh}}, \bibnamefont{and}
  \bibinfo{author}{\bibfnamefont{P.~S.} \bibnamefont{Joshi}},
  \bibinfo{journal}{Phys. Rev. D} \textbf{\bibinfo{volume}{103}},
  \bibinfo{pages}{024015} (\bibinfo{year}{2021}), \eprint{2009.07487}.

\bibitem[{\citenamefont{Pugliese et~al.}(2011)\citenamefont{Pugliese, Quevedo,
  and Ruffini}}]{Pugliese:2011xn}
\bibinfo{author}{\bibfnamefont{D.}~\bibnamefont{Pugliese}},
  \bibinfo{author}{\bibfnamefont{H.}~\bibnamefont{Quevedo}}, \bibnamefont{and}
  \bibinfo{author}{\bibfnamefont{R.}~\bibnamefont{Ruffini}},
  \bibinfo{journal}{Phys. Rev. D} \textbf{\bibinfo{volume}{84}},
  \bibinfo{pages}{044030} (\bibinfo{year}{2011}), \eprint{1105.2959}.

\bibitem[{\citenamefont{Mart\'\i{}nez et~al.}(2019)\citenamefont{Mart\'\i{}nez,
  Parra, Vald\'es, and Zanelli}}]{Martinez:2019nor}
\bibinfo{author}{\bibfnamefont{C.}~\bibnamefont{Mart\'\i{}nez}},
  \bibinfo{author}{\bibfnamefont{N.}~\bibnamefont{Parra}},
  \bibinfo{author}{\bibfnamefont{N.}~\bibnamefont{Vald\'es}}, \bibnamefont{and}
  \bibinfo{author}{\bibfnamefont{J.}~\bibnamefont{Zanelli}},
  \bibinfo{journal}{Phys. Rev. D} \textbf{\bibinfo{volume}{100}},
  \bibinfo{pages}{024026} (\bibinfo{year}{2019}), \eprint{1902.00145}.

\bibitem[{\citenamefont{Hackmann et~al.}(2014)\citenamefont{Hackmann,
  L\"ammerzahl, Obukhov, Puetzfeld, and Schaffer}}]{Hackmann:2014tga}
\bibinfo{author}{\bibfnamefont{E.}~\bibnamefont{Hackmann}},
  \bibinfo{author}{\bibfnamefont{C.}~\bibnamefont{L\"ammerzahl}},
  \bibinfo{author}{\bibfnamefont{Y.~N.} \bibnamefont{Obukhov}},
  \bibinfo{author}{\bibfnamefont{D.}~\bibnamefont{Puetzfeld}},
  \bibnamefont{and} \bibinfo{author}{\bibfnamefont{I.}~\bibnamefont{Schaffer}},
  \bibinfo{journal}{Phys. Rev. D} \textbf{\bibinfo{volume}{90}},
  \bibinfo{pages}{064035} (\bibinfo{year}{2014}), \eprint{1408.1773}.

\bibitem[{\citenamefont{Potashov et~al.}(2019)\citenamefont{Potashov,
  Tchemarina, and Tsirulev}}]{Potashov:2019kxq}
\bibinfo{author}{\bibfnamefont{I.~M.} \bibnamefont{Potashov}},
  \bibinfo{author}{\bibfnamefont{J.~V.} \bibnamefont{Tchemarina}},
  \bibnamefont{and} \bibinfo{author}{\bibfnamefont{A.~N.}
  \bibnamefont{Tsirulev}}, \bibinfo{journal}{Eur. Phys. J. C}
  \textbf{\bibinfo{volume}{79}}, \bibinfo{pages}{709} (\bibinfo{year}{2019}),
  \eprint{1908.03700}.

\bibitem[{\citenamefont{Bhattacharya et~al.}(2020)\citenamefont{Bhattacharya,
  Dey, Mazumdar, and Sarkar}}]{Bhattacharya:2017chr}
\bibinfo{author}{\bibfnamefont{K.}~\bibnamefont{Bhattacharya}},
  \bibinfo{author}{\bibfnamefont{D.}~\bibnamefont{Dey}},
  \bibinfo{author}{\bibfnamefont{A.}~\bibnamefont{Mazumdar}}, \bibnamefont{and}
  \bibinfo{author}{\bibfnamefont{T.}~\bibnamefont{Sarkar}},
  \bibinfo{journal}{Phys. Rev. D} \textbf{\bibinfo{volume}{101}},
  \bibinfo{pages}{043005} (\bibinfo{year}{2020}), \eprint{1709.03798}.

\bibitem[{\citenamefont{Bambhaniya et~al.}(2019)\citenamefont{Bambhaniya,
  Joshi, Dey, and Joshi}}]{Bambhaniya:2019pbr}
\bibinfo{author}{\bibfnamefont{P.}~\bibnamefont{Bambhaniya}},
  \bibinfo{author}{\bibfnamefont{A.~B.} \bibnamefont{Joshi}},
  \bibinfo{author}{\bibfnamefont{D.}~\bibnamefont{Dey}}, \bibnamefont{and}
  \bibinfo{author}{\bibfnamefont{P.~S.} \bibnamefont{Joshi}},
  \bibinfo{journal}{Phys. Rev. D} \textbf{\bibinfo{volume}{100}},
  \bibinfo{pages}{124020} (\bibinfo{year}{2019}), \eprint{1908.07171}.

\bibitem[{\citenamefont{Deng}(2020)}]{Deng:2020hxw}
\bibinfo{author}{\bibfnamefont{X.-M.} \bibnamefont{Deng}},
  \bibinfo{journal}{Eur. Phys. J. C} \textbf{\bibinfo{volume}{80}},
  \bibinfo{pages}{489} (\bibinfo{year}{2020}).

\bibitem[{\citenamefont{Lin and Deng}(2021)}]{Lin:2021noq}
\bibinfo{author}{\bibfnamefont{H.-Y.} \bibnamefont{Lin}} \bibnamefont{and}
  \bibinfo{author}{\bibfnamefont{X.-M.} \bibnamefont{Deng}},
  \bibinfo{journal}{Phys. Dark Univ.} \textbf{\bibinfo{volume}{31}},
  \bibinfo{pages}{100745} (\bibinfo{year}{2021}).

\bibitem[{\citenamefont{Bambhaniya
  et~al.}(2021{\natexlab{a}})\citenamefont{Bambhaniya, Verma, Dey, Joshi, and
  Joshi}}]{Bambhaniya:2021jum}
\bibinfo{author}{\bibfnamefont{P.}~\bibnamefont{Bambhaniya}},
  \bibinfo{author}{\bibfnamefont{J.~S.} \bibnamefont{Verma}},
  \bibinfo{author}{\bibfnamefont{D.}~\bibnamefont{Dey}},
  \bibinfo{author}{\bibfnamefont{P.~S.} \bibnamefont{Joshi}}, \bibnamefont{and}
  \bibinfo{author}{\bibfnamefont{A.~B.} \bibnamefont{Joshi}}
  (\bibinfo{year}{2021}{\natexlab{a}}), \eprint{2109.11137}.

\bibitem[{\citenamefont{Ota et~al.}(2022)\citenamefont{Ota, Kobayashi, and
  Nakashi}}]{Ota:2021mub}
\bibinfo{author}{\bibfnamefont{K.}~\bibnamefont{Ota}},
  \bibinfo{author}{\bibfnamefont{S.}~\bibnamefont{Kobayashi}},
  \bibnamefont{and} \bibinfo{author}{\bibfnamefont{K.}~\bibnamefont{Nakashi}},
  \bibinfo{journal}{Phys. Rev. D} \textbf{\bibinfo{volume}{105}},
  \bibinfo{pages}{024037} (\bibinfo{year}{2022}), \eprint{2110.07503}.

\bibitem[{\citenamefont{Baub\"ock et~al.}(2020)}]{GRAVITY:2020lpa}
\bibinfo{author}{\bibfnamefont{M.}~\bibnamefont{Baub\"ock}}
  \bibnamefont{et~al.} (\bibinfo{collaboration}{GRAVITY}),
  \bibinfo{journal}{Astron. Astrophys.} \textbf{\bibinfo{volume}{635}},
  \bibinfo{pages}{A143} (\bibinfo{year}{2020}), \eprint{2002.08374}.

\bibitem[{\citenamefont{Eisenhauer et~al.}(2005)}]{Eisenhauer:2005cv}
\bibinfo{author}{\bibfnamefont{F.}~\bibnamefont{Eisenhauer}}
  \bibnamefont{et~al.}, \bibinfo{journal}{Astrophys. J.}
  \textbf{\bibinfo{volume}{628}}, \bibinfo{pages}{246} (\bibinfo{year}{2005}),
  \eprint{astro-ph/0502129}.

\bibitem[{\citenamefont{de~Sitter}(1916)}]{deSitter:1916zz}
\bibinfo{author}{\bibfnamefont{W.}~\bibnamefont{de~Sitter}},
  \bibinfo{journal}{Mon. Not. Roy. Astron. Soc.} \textbf{\bibinfo{volume}{77}},
  \bibinfo{pages}{155} (\bibinfo{year}{1916}).

\bibitem[{\citenamefont{Lense and Thirring}(1918)}]{Lense:1918LT}
\bibinfo{author}{\bibfnamefont{J.}~\bibnamefont{Lense}} \bibnamefont{and}
  \bibinfo{author}{\bibfnamefont{H.}~\bibnamefont{Thirring}},
  \bibinfo{journal}{Physikalische Zeitschrift} \textbf{\bibinfo{volume}{19}},
  \bibinfo{pages}{156} (\bibinfo{year}{1918}).

\bibitem[{\citenamefont{Everitt et~al.}(2011)}]{Everitt:2011hp}
\bibinfo{author}{\bibfnamefont{C.~W.~F.} \bibnamefont{Everitt}}
  \bibnamefont{et~al.}, \bibinfo{journal}{Phys. Rev. Lett.}
  \textbf{\bibinfo{volume}{106}}, \bibinfo{pages}{221101}
  (\bibinfo{year}{2011}), \eprint{1105.3456}.

\bibitem[{\citenamefont{Sakina and Chiba}(1979)}]{sakina1979parallel}
\bibinfo{author}{\bibfnamefont{K.-i.} \bibnamefont{Sakina}} \bibnamefont{and}
  \bibinfo{author}{\bibfnamefont{J.}~\bibnamefont{Chiba}},
  \bibinfo{journal}{Physical Review D} \textbf{\bibinfo{volume}{19}},
  \bibinfo{pages}{2280} (\bibinfo{year}{1979}).

\bibitem[{\citenamefont{Hartle}(2009)}]{2009Hartle}
\bibinfo{author}{\bibfnamefont{J.~B.} \bibnamefont{Hartle}},
  \emph{\bibinfo{title}{{Gravity: An introduction to Einstein¡¯s General
  relativity}}} (\bibinfo{publisher}{Pearson}, \bibinfo{year}{2009}).

\bibitem[{\citenamefont{Chakraborty and Majumdar}(2014)}]{Chakraborty:2013naa}
\bibinfo{author}{\bibfnamefont{C.}~\bibnamefont{Chakraborty}} \bibnamefont{and}
  \bibinfo{author}{\bibfnamefont{P.}~\bibnamefont{Majumdar}},
  \bibinfo{journal}{Class. Quant. Grav.} \textbf{\bibinfo{volume}{31}},
  \bibinfo{pages}{075006} (\bibinfo{year}{2014}), \eprint{1304.6936}.

\bibitem[{\citenamefont{Bini et~al.}(2016)\citenamefont{Bini, Geralico, and
  Jantzen}}]{Bini:2016iym}
\bibinfo{author}{\bibfnamefont{D.}~\bibnamefont{Bini}},
  \bibinfo{author}{\bibfnamefont{A.}~\bibnamefont{Geralico}}, \bibnamefont{and}
  \bibinfo{author}{\bibfnamefont{R.~T.} \bibnamefont{Jantzen}},
  \bibinfo{journal}{Phys. Rev. D} \textbf{\bibinfo{volume}{94}},
  \bibinfo{pages}{064066} (\bibinfo{year}{2016}), \eprint{1607.08427}.

\bibitem[{\citenamefont{Chakraborty and Pradhan}(2013)}]{Chakraborty:2012wv}
\bibinfo{author}{\bibfnamefont{C.}~\bibnamefont{Chakraborty}} \bibnamefont{and}
  \bibinfo{author}{\bibfnamefont{P.~P.} \bibnamefont{Pradhan}},
  \bibinfo{journal}{Eur. Phys. J. C} \textbf{\bibinfo{volume}{73}},
  \bibinfo{pages}{2536} (\bibinfo{year}{2013}), \eprint{1209.1945}.

\bibitem[{\citenamefont{Chakraborty and Pradhan}(2017)}]{Chakraborty:2016oja}
\bibinfo{author}{\bibfnamefont{C.}~\bibnamefont{Chakraborty}} \bibnamefont{and}
  \bibinfo{author}{\bibfnamefont{P.}~\bibnamefont{Pradhan}},
  \bibinfo{journal}{JCAP} \textbf{\bibinfo{volume}{03}}, \bibinfo{pages}{035}
  (\bibinfo{year}{2017}), \eprint{1603.09683}.

\bibitem[{\citenamefont{Chakraborty et~al.}(2014)\citenamefont{Chakraborty,
  Modak, and Bandyopadhyay}}]{Chakraborty:2014qba}
\bibinfo{author}{\bibfnamefont{C.}~\bibnamefont{Chakraborty}},
  \bibinfo{author}{\bibfnamefont{K.~P.} \bibnamefont{Modak}}, \bibnamefont{and}
  \bibinfo{author}{\bibfnamefont{D.}~\bibnamefont{Bandyopadhyay}},
  \bibinfo{journal}{Astrophys. J.} \textbf{\bibinfo{volume}{790}},
  \bibinfo{pages}{2} (\bibinfo{year}{2014}), \eprint{1402.6108}.

\bibitem[{\citenamefont{Chakraborty
  et~al.}(2017{\natexlab{a}})\citenamefont{Chakraborty, Patil, Kocherlakota,
  Bhattacharyya, Joshi, and Kr\'olak}}]{Chakraborty:2016mhx}
\bibinfo{author}{\bibfnamefont{C.}~\bibnamefont{Chakraborty}},
  \bibinfo{author}{\bibfnamefont{M.}~\bibnamefont{Patil}},
  \bibinfo{author}{\bibfnamefont{P.}~\bibnamefont{Kocherlakota}},
  \bibinfo{author}{\bibfnamefont{S.}~\bibnamefont{Bhattacharyya}},
  \bibinfo{author}{\bibfnamefont{P.~S.} \bibnamefont{Joshi}}, \bibnamefont{and}
  \bibinfo{author}{\bibfnamefont{A.}~\bibnamefont{Kr\'olak}},
  \bibinfo{journal}{Phys. Rev. D} \textbf{\bibinfo{volume}{95}},
  \bibinfo{pages}{084024} (\bibinfo{year}{2017}{\natexlab{a}}),
  \eprint{1611.08808}.

\bibitem[{\citenamefont{Rizwan et~al.}(2018)\citenamefont{Rizwan, Jamil, and
  Wang}}]{Rizwan:2018lht}
\bibinfo{author}{\bibfnamefont{M.}~\bibnamefont{Rizwan}},
  \bibinfo{author}{\bibfnamefont{M.}~\bibnamefont{Jamil}}, \bibnamefont{and}
  \bibinfo{author}{\bibfnamefont{A.}~\bibnamefont{Wang}},
  \bibinfo{journal}{Phys. Rev. D} \textbf{\bibinfo{volume}{98}},
  \bibinfo{pages}{024015} (\bibinfo{year}{2018}), \bibinfo{note}{[Erratum:
  Phys.Rev.D 100, 029902 (2019)]}, \eprint{1802.04301}.

\bibitem[{\citenamefont{Rizwan et~al.}(2019)\citenamefont{Rizwan, Jamil, and
  Jusufi}}]{Rizwan:2018rgs}
\bibinfo{author}{\bibfnamefont{M.}~\bibnamefont{Rizwan}},
  \bibinfo{author}{\bibfnamefont{M.}~\bibnamefont{Jamil}}, \bibnamefont{and}
  \bibinfo{author}{\bibfnamefont{K.}~\bibnamefont{Jusufi}},
  \bibinfo{journal}{Phys. Rev. D} \textbf{\bibinfo{volume}{99}},
  \bibinfo{pages}{024050} (\bibinfo{year}{2019}), \eprint{1812.01331}.

\bibitem[{\citenamefont{Pradhan}(2020)}]{Pradhan:2020nno}
\bibinfo{author}{\bibfnamefont{P.}~\bibnamefont{Pradhan}}
  (\bibinfo{year}{2020}), \eprint{2007.01347}.

\bibitem[{\citenamefont{Solanki et~al.}(2022)\citenamefont{Solanki, Bambhaniya,
  Dey, Joshi, and Pathak}}]{Solanki:2021mkt}
\bibinfo{author}{\bibfnamefont{D.~N.} \bibnamefont{Solanki}},
  \bibinfo{author}{\bibfnamefont{P.}~\bibnamefont{Bambhaniya}},
  \bibinfo{author}{\bibfnamefont{D.}~\bibnamefont{Dey}},
  \bibinfo{author}{\bibfnamefont{P.~S.} \bibnamefont{Joshi}}, \bibnamefont{and}
  \bibinfo{author}{\bibfnamefont{K.~N.} \bibnamefont{Pathak}},
  \bibinfo{journal}{Eur. Phys. J. C} \textbf{\bibinfo{volume}{82}},
  \bibinfo{pages}{77} (\bibinfo{year}{2022}), \eprint{2109.14937}.

\bibitem[{\citenamefont{Torok et~al.}(2011)\citenamefont{Torok, Kotrlova,
  Sramkova, and Stuchlik}}]{Torok:2011qy}
\bibinfo{author}{\bibfnamefont{G.}~\bibnamefont{Torok}},
  \bibinfo{author}{\bibfnamefont{A.}~\bibnamefont{Kotrlova}},
  \bibinfo{author}{\bibfnamefont{E.}~\bibnamefont{Sramkova}}, \bibnamefont{and}
  \bibinfo{author}{\bibfnamefont{Z.}~\bibnamefont{Stuchlik}},
  \bibinfo{journal}{Astron. Astrophys.} \textbf{\bibinfo{volume}{531}},
  \bibinfo{pages}{A59} (\bibinfo{year}{2011}), \eprint{1103.2438}.

\bibitem[{\citenamefont{Bambi}(2012)}]{Bambi:2012ku}
\bibinfo{author}{\bibfnamefont{C.}~\bibnamefont{Bambi}},
  \bibinfo{journal}{Phys. Rev. D} \textbf{\bibinfo{volume}{85}},
  \bibinfo{pages}{043002} (\bibinfo{year}{2012}), \eprint{1201.1638}.

\bibitem[{\citenamefont{Tripathi et~al.}(2019)}]{Tripathi:2019bya}
\bibinfo{author}{\bibfnamefont{A.}~\bibnamefont{Tripathi}}
  \bibnamefont{et~al.}, \bibinfo{journal}{Astrophys. J.}
  \textbf{\bibinfo{volume}{874}}, \bibinfo{pages}{135} (\bibinfo{year}{2019}),
  \eprint{1901.03064}.

\bibitem[{\citenamefont{van~der Klisin}(2006)}]{Klisin2006}
\bibinfo{author}{\bibfnamefont{M.}~\bibnamefont{van~der Klisin}},
  \emph{\bibinfo{title}{{Compact Stellar X-ray Sources (Cambridge
  Astrophysics)}}} (\bibinfo{publisher}{Cambridge University Press},
  \bibinfo{year}{2006}).

\bibitem[{\citenamefont{Aliev et~al.}(2013)\citenamefont{Aliev, Esmer, and
  Talazan}}]{Aliev:2013jqz}
\bibinfo{author}{\bibfnamefont{A.~N.} \bibnamefont{Aliev}},
  \bibinfo{author}{\bibfnamefont{G.~D.} \bibnamefont{Esmer}}, \bibnamefont{and}
  \bibinfo{author}{\bibfnamefont{P.}~\bibnamefont{Talazan}},
  \bibinfo{journal}{Class. Quant. Grav.} \textbf{\bibinfo{volume}{30}},
  \bibinfo{pages}{045010} (\bibinfo{year}{2013}), \eprint{1205.2838}.

\bibitem[{\citenamefont{Doneva et~al.}(2014)\citenamefont{Doneva, Yazadjiev,
  Stergioulas, Kokkotas, and Athanasiadis}}]{Doneva:2014uma}
\bibinfo{author}{\bibfnamefont{D.~D.} \bibnamefont{Doneva}},
  \bibinfo{author}{\bibfnamefont{S.~S.} \bibnamefont{Yazadjiev}},
  \bibinfo{author}{\bibfnamefont{N.}~\bibnamefont{Stergioulas}},
  \bibinfo{author}{\bibfnamefont{K.~D.} \bibnamefont{Kokkotas}},
  \bibnamefont{and} \bibinfo{author}{\bibfnamefont{T.~M.}
  \bibnamefont{Athanasiadis}}, \bibinfo{journal}{Phys. Rev. D}
  \textbf{\bibinfo{volume}{90}}, \bibinfo{pages}{044004}
  (\bibinfo{year}{2014}), \eprint{1405.6976}.

\bibitem[{\citenamefont{Azreg-A\"\i{}nou
  et~al.}(2020)\citenamefont{Azreg-A\"\i{}nou, Chen, Deng, Jamil, Zhu, Wu, and
  Lim}}]{Azreg-Ainou:2020bfl}
\bibinfo{author}{\bibfnamefont{M.}~\bibnamefont{Azreg-A\"\i{}nou}},
  \bibinfo{author}{\bibfnamefont{Z.}~\bibnamefont{Chen}},
  \bibinfo{author}{\bibfnamefont{B.}~\bibnamefont{Deng}},
  \bibinfo{author}{\bibfnamefont{M.}~\bibnamefont{Jamil}},
  \bibinfo{author}{\bibfnamefont{T.}~\bibnamefont{Zhu}},
  \bibinfo{author}{\bibfnamefont{Q.}~\bibnamefont{Wu}}, \bibnamefont{and}
  \bibinfo{author}{\bibfnamefont{Y.-K.} \bibnamefont{Lim}},
  \bibinfo{journal}{Phys. Rev. D} \textbf{\bibinfo{volume}{102}},
  \bibinfo{pages}{044028} (\bibinfo{year}{2020}), \eprint{2004.02602}.

\bibitem[{\citenamefont{Jusufi et~al.}(2021)\citenamefont{Jusufi,
  Azreg-A\"\i{}nou, Jamil, Wei, Wu, and Wang}}]{Jusufi:2020odz}
\bibinfo{author}{\bibfnamefont{K.}~\bibnamefont{Jusufi}},
  \bibinfo{author}{\bibfnamefont{M.}~\bibnamefont{Azreg-A\"\i{}nou}},
  \bibinfo{author}{\bibfnamefont{M.}~\bibnamefont{Jamil}},
  \bibinfo{author}{\bibfnamefont{S.-W.} \bibnamefont{Wei}},
  \bibinfo{author}{\bibfnamefont{Q.}~\bibnamefont{Wu}}, \bibnamefont{and}
  \bibinfo{author}{\bibfnamefont{A.}~\bibnamefont{Wang}},
  \bibinfo{journal}{Phys. Rev. D} \textbf{\bibinfo{volume}{103}},
  \bibinfo{pages}{024013} (\bibinfo{year}{2021}), \eprint{2008.08450}.

\bibitem[{\citenamefont{Ghasemi-Nodehi
  et~al.}(2020)\citenamefont{Ghasemi-Nodehi, Azreg-A\"\i{}nou, Jusufi, and
  Jamil}}]{Ghasemi-Nodehi:2020oiz}
\bibinfo{author}{\bibfnamefont{M.}~\bibnamefont{Ghasemi-Nodehi}},
  \bibinfo{author}{\bibfnamefont{M.}~\bibnamefont{Azreg-A\"\i{}nou}},
  \bibinfo{author}{\bibfnamefont{K.}~\bibnamefont{Jusufi}}, \bibnamefont{and}
  \bibinfo{author}{\bibfnamefont{M.}~\bibnamefont{Jamil}},
  \bibinfo{journal}{Phys. Rev. D} \textbf{\bibinfo{volume}{102}},
  \bibinfo{pages}{104032} (\bibinfo{year}{2020}), \eprint{2011.02276}.

\bibitem[{\citenamefont{Motta et~al.}(2018)\citenamefont{Motta, Franchini,
  Lodato, and Mastroserio}}]{Motta:2017ijc}
\bibinfo{author}{\bibfnamefont{S.~E.} \bibnamefont{Motta}},
  \bibinfo{author}{\bibfnamefont{A.}~\bibnamefont{Franchini}},
  \bibinfo{author}{\bibfnamefont{G.}~\bibnamefont{Lodato}}, \bibnamefont{and}
  \bibinfo{author}{\bibfnamefont{G.}~\bibnamefont{Mastroserio}},
  \bibinfo{journal}{Mon. Not. Roy. Astron. Soc.}
  \textbf{\bibinfo{volume}{473}}, \bibinfo{pages}{431} (\bibinfo{year}{2018}),
  \eprint{1709.02608}.

\bibitem[{\citenamefont{Israel}(1967)}]{Israel:1967wq}
\bibinfo{author}{\bibfnamefont{W.}~\bibnamefont{Israel}},
  \bibinfo{journal}{Phys. Rev.} \textbf{\bibinfo{volume}{164}},
  \bibinfo{pages}{1776} (\bibinfo{year}{1967}).

\bibitem[{\citenamefont{Hawking}(1972)}]{Hawking:1971vc}
\bibinfo{author}{\bibfnamefont{S.~W.} \bibnamefont{Hawking}},
  \bibinfo{journal}{Commun. Math. Phys.} \textbf{\bibinfo{volume}{25}},
  \bibinfo{pages}{152} (\bibinfo{year}{1972}).

\bibitem[{\citenamefont{Mazur}(1982)}]{Mazur:1982db}
\bibinfo{author}{\bibfnamefont{P.~O.} \bibnamefont{Mazur}},
  \bibinfo{journal}{J. Phys. A} \textbf{\bibinfo{volume}{15}},
  \bibinfo{pages}{3173} (\bibinfo{year}{1982}).

\bibitem[{\citenamefont{Carter}(1968)}]{Carter:1968rr}
\bibinfo{author}{\bibfnamefont{B.}~\bibnamefont{Carter}},
  \bibinfo{journal}{Phys. Rev.} \textbf{\bibinfo{volume}{174}},
  \bibinfo{pages}{1559} (\bibinfo{year}{1968}).

\bibitem[{\citenamefont{{Straumann}}(2004)}]{2004graa.book}
\bibinfo{author}{\bibfnamefont{N.}~\bibnamefont{{Straumann}}},
  \emph{\bibinfo{title}{{General relativity with applications to
  astrophysics}}} (\bibinfo{publisher}{Springer}, \bibinfo{year}{2004}).

\bibitem[{\citenamefont{Chakraborty
  et~al.}(2017{\natexlab{b}})\citenamefont{Chakraborty, Kocherlakota, and
  Joshi}}]{Chakraborty:2016ipk}
\bibinfo{author}{\bibfnamefont{C.}~\bibnamefont{Chakraborty}},
  \bibinfo{author}{\bibfnamefont{P.}~\bibnamefont{Kocherlakota}},
  \bibnamefont{and} \bibinfo{author}{\bibfnamefont{P.~S.} \bibnamefont{Joshi}},
  \bibinfo{journal}{Phys. Rev. D} \textbf{\bibinfo{volume}{95}},
  \bibinfo{pages}{044006} (\bibinfo{year}{2017}{\natexlab{b}}),
  \eprint{1605.00600}.

\bibitem[{\citenamefont{Glendenning and Weber}(1994)}]{Glendenning:1993di}
\bibinfo{author}{\bibfnamefont{N.~K.} \bibnamefont{Glendenning}}
  \bibnamefont{and} \bibinfo{author}{\bibfnamefont{F.}~\bibnamefont{Weber}},
  \bibinfo{journal}{Phys. Rev. D} \textbf{\bibinfo{volume}{50}},
  \bibinfo{pages}{3836} (\bibinfo{year}{1994}).

\bibitem[{\citenamefont{Bardeen et~al.}(1972)\citenamefont{Bardeen, Press, and
  Teukolsky}}]{Bardeen:1972fi}
\bibinfo{author}{\bibfnamefont{J.~M.} \bibnamefont{Bardeen}},
  \bibinfo{author}{\bibfnamefont{W.~H.} \bibnamefont{Press}}, \bibnamefont{and}
  \bibinfo{author}{\bibfnamefont{S.~A.} \bibnamefont{Teukolsky}},
  \bibinfo{journal}{Astrophys. J.} \textbf{\bibinfo{volume}{178}},
  \bibinfo{pages}{347} (\bibinfo{year}{1972}).

\bibitem[{\citenamefont{Bambhaniya
  et~al.}(2021{\natexlab{b}})\citenamefont{Bambhaniya, Solanki, Dey, Joshi,
  Joshi, and Patel}}]{Bambhaniya:2020zno}
\bibinfo{author}{\bibfnamefont{P.}~\bibnamefont{Bambhaniya}},
  \bibinfo{author}{\bibfnamefont{D.~N.} \bibnamefont{Solanki}},
  \bibinfo{author}{\bibfnamefont{D.}~\bibnamefont{Dey}},
  \bibinfo{author}{\bibfnamefont{A.~B.} \bibnamefont{Joshi}},
  \bibinfo{author}{\bibfnamefont{P.~S.} \bibnamefont{Joshi}}, \bibnamefont{and}
  \bibinfo{author}{\bibfnamefont{V.}~\bibnamefont{Patel}},
  \bibinfo{journal}{Eur. Phys. J. C} \textbf{\bibinfo{volume}{81}},
  \bibinfo{pages}{205} (\bibinfo{year}{2021}{\natexlab{b}}),
  \eprint{2007.12086}.

\bibitem[{\citenamefont{Ryan}(1995)}]{Ryan:1995wh}
\bibinfo{author}{\bibfnamefont{F.~D.} \bibnamefont{Ryan}},
  \bibinfo{journal}{Phys. Rev. D} \textbf{\bibinfo{volume}{52}},
  \bibinfo{pages}{5707} (\bibinfo{year}{1995}).

\bibitem[{\citenamefont{Motta et~al.}(2014)\citenamefont{Motta, Belloni,
  Stella, Mu\~noz Darias, and Fender}}]{Motta:2013wga}
\bibinfo{author}{\bibfnamefont{S.~E.} \bibnamefont{Motta}},
  \bibinfo{author}{\bibfnamefont{T.~M.} \bibnamefont{Belloni}},
  \bibinfo{author}{\bibfnamefont{L.}~\bibnamefont{Stella}},
  \bibinfo{author}{\bibfnamefont{T.}~\bibnamefont{Mu\~noz Darias}},
  \bibnamefont{and} \bibinfo{author}{\bibfnamefont{R.}~\bibnamefont{Fender}},
  \bibinfo{journal}{Mon. Not. Roy. Astron. Soc.}
  \textbf{\bibinfo{volume}{437}}, \bibinfo{pages}{2554} (\bibinfo{year}{2014}),
  \eprint{1309.3652}.

\end{thebibliography}
\bibliographystyle{apsrev}

\end{document}